\documentclass[floatfix,twocolumn,showpacs,prc,aps,nofootinbib]{revtex4}
\usepackage[dvips]{graphicx}
\begin{document}
\renewcommand{\thefootnote}{\fnsymbol{footnote}}
\title{New potentialities of the Li\`ege intranuclear cascade (INCL)
model for reactions induced by nucleons and light charged
particles 
\footnote[2]{Work partly supported by the ANDES Collaboration, EU 
Contract ANDES-N$^{\circ}$ 249671  (RTD-J5/AC/MC/ap/Ares (2010) 662150 )}}
\renewcommand{\thefootnote}{\arabic{footnote}}
\setcounter{footnote}{0}
\author{A. Boudard$^1$, J. Cugnon$^2$, J.-C. David$^1$, S. Leray$^1$ and D. Mancusi$^1$}
\affiliation{$^1$Irfu/SPhN, CEA/Saclay, F-91191 Gif-sur-Yvette Cedex,
France\\$^2$University of Li\`ege, AGO Department, all\'ee du 6 Ao\^{u}t 17, b\^{a}t.B 5,
B-4000 LIEGE 1, Belgium}

\date{\today} 
\everymath={\displaystyle}   
 
\renewcommand{\baselinestretch}{0.5}
\begin{abstract}\vspace{2cm} 
The new version (INCL4.6) of the Li\`ege intranuclear cascade (INC) model for the description of spallation reactions is presented in detail. Compared to the standard version (INCL4.2), it incorporates several new features, the most important of which are: {\it(i)} the inclusion of cluster production through a dynamical phase space coalescence model, {\it(ii)} the Coulomb deflection for entering and outgoing charged particles, {\it(iii)} the improvement of the treatment of Pauli blocking and of soft collisions, {\it(iv)} the introduction of experimental threshold values for the emission of particles, {\it(v)} the improvement of pion dynamics, {\it(vi)} a detailed procedure for the treatment of light-cluster induced reactions taking care of the effects of binding energy of the nucleons inside the incident cluster and of the possible fusion reaction at low energy. Performances of the new model concerning nucleon-induced reactions are illustrated by a comparison with experimental data covering total reaction cross-sections, neutron, proton, pion and composite double differential cross-sections, neutron multiplicities, residue mass and charge distributions, and residue recoil velocity distributions. Whenever necessary, the INCL4.6 model is coupled to the ABLA07 de-excitation model and the respective merits of the two models are then tentatively disentangled. Good agreement is generally obtained in the 200 MeV-2 GeV range. Below 200 MeV and down to a few tens of MeV, the total reaction cross section is well reproduced and differential cross sections are reasonably well described. The model is also tested for light-ion induced reactions at low energy, below 100 MeV incident energy per nucleon. Beyond presenting the update of the INCL4.2 model, attention has been paid to applications of the new model to three topics for which some particular aspects are discussed for the first time. The first topic is the production of clusters heavier than alpha particle. It is shown that the energy spectra of these produced clusters are consistent with coalescence. The second topic regards the longitudinal residue recoil velocity and its fluctuations. Excellent results are obtained for these quantities. It addition, it is shown that the distributions of these quantities display typical random-walk characterics, at least for not too large mass losses. They are interpreted as a direct consequence of the independence of successive binary collisions occurring during the cascade process. The last topic concerns the total reaction cross section and the residue production cross sections for low energy incident light ions. It is shown that our new model can give a rather satisfactory account of these cross sections, offering so an alternative to fusion models and the advantage of a single model for the progressive change from fusion to pre-equilibrium mechanisms.

\vspace{2cm}  
\end{abstract}
\renewcommand{\baselinestretch}{0.5} 

\pacs{25.40.-h,25.40.Sc,25.45.-z,25.55.-e}
\maketitle

\section{Introduction} 
It is largely accepted that nucleon-induced spallation reactions proceed through a two-stage process: a first stage dominated by hard nucleon-nucleon (NN) collisions emitting fast particles, followed by the de-excitation of a more or less thermalized remnant, akin to evaporation and/or fission. The commonly used tools to describe these reactions result from the coupling of an intra-nuclear cascade (INC) model for the first stage  to an evaporation-fission model for the second stage. There exist many INC and de-excitation models, differing from each other by the ingredients and by the results. In the last years, there has been a very strong development of applications involving spallation reactions, including neutron sources for condensed matter and material studies~\cite{FI10}, transmutation of nuclear waste~\cite{GU99,AB10}, simulation of experimental set-ups in nuclear and particle physics~\cite{Geant4}, production of rare isotopes~\cite{TA89},  protection against radiation near accelerators and in space missions~\cite{DU02}, interaction of cosmic rays in the atmosphere~\cite{LO97} and  cancer hadrontherapy~\cite{KR90}. This activity is at the origin of a strong demand for the improvement of the accuracy of the reaction models and for reliable data to benchmark these models efficiently. Indeed, in Europe, an important effort has been devoted in the last fifteen years to collect and produce high-quality data concerning all emission channels: neutrons, light charged particles (lcp) and residues~\cite{HIN,EUR}. Meanwhile, model developers have constantly attempted to improve their models. We are interested here in the development of the Li\`ege INC model (INCL4). The first standard version (INCL4.2) is described extensively in Ref. \cite{BO02}. Coupled with the ABLA evaporation/fission model~\cite{GA91,JU98,BE98} (actually the KHSv3p version), it is able to describe fairly well a large body of experimental data in the 200 MeV-2 GeV range, namely total reaction cross sections, double differential proton, neutron and pion production cross sections, residue mass and charge spectra, isotopic distributions and, to some extent, recoil energies \cite{BO02}. This was achieved without any parameter adjustment in the INC model and with the use of standard values for the parameters of the ABLA model. More precisely, there is no fitting parameter in the INCL4.2 model. The basic nuclear physics parameters, like the radius and the diffuseness of nuclei, are taken from phenomenology and ``technical parameters'', like  those used to evaluate phase space occupation entering the Pauli blockers, have been fixed once for all.  In addition the stopping time is generated self-consistently by the model itself. Nevertheless, the results obtained with INCL4.2 show some systematic shortcomings and/or discrepancies with experimental data. They have been identified in the frame of the European HINDAS and NUDATRA collaborations~\cite{HIN,EUR} and are shortly presented below. The most important of these shortcomings is the inability of INCL4.2 to produce lcp's (except protons of course).  Since the release of the INCL4.2 model, a constant effort has been made to palliate its deficiencies.

It is the purpose of this paper to set out the present status of the INCL4 model, that is designated as INCL4.6. The various points of improvements are described below. But, this paper is not a mere update of the INCL model. It highlights at least three innovative points concerning spallation reactions and bearing on $\it{(i)}$ production of clusters in the cascade, $\it{(ii)}$ residue recoil velocities and $\it{(iii)}$ the behaviour of the model at low energy, respectively. Concerning the first point, a dynamical coalescence model has been implemented to allow the emission of clusters in the cascade stage. An intermediate version of our model~\cite{BO04}, sometimes denoted as INCL4.3, consisting basically of INCL4.2 plus the treatment of the emission of light clusters, up to alpha particles, has opened the path to the improvement along this line. We report here on the continuation of this effort and we show, for the first time, that  emission of heavier clusters (in practice, up to A=8) can be explained by dynamical coalescence.  Concerning point $\it{(ii)}$, we show below that we are now able to reproduce average values and standard deviations of the residue velocity distributions. Finally, let us come to point $\it{(iii)}$. It is generally stated that INC models cannot be reliable below $\sim$200 MeV incident energy, although in some cases where such models are used, they occasionally give surprisingly satisfactory results. In the last years, we have tentatively but systematically improved the model in this energy range. We  considerably improved the model concerning  reactions induced by nucleons and by light clusters of nucleons, typically from deuterons to alpha particles, and this for incident energies per nucleon down to a few MeV. Our model is now able to produce good total reaction cross sections for both incident nucleons and incident light clusters. In addition, we show below that we are reproducing reasonably well residue production cross sections for cluster-induced reactions at low energy. The developments concerning the above-mentioned three points were motivated in part by studies of thick targets, concerning release of volatile elements, such as $H$ and $He$ isotopes (point $\it{(i)}$), radiation damages (point $\it{(ii)}$) and radiotoxicity (point $\it{(iii)}$). Sometimes, the latter may crucially depend upon secondary reactions induced by light clusters produced in primary collisions~\cite{RA06,RA06}. 

The predictive power of this improved model is considerably better than the one of INCL4.2. This has been verified on an intermediate version (named INCL4.5, not very different from INCL4.6, and defined below) in the course of an intercomparison of spallation codes organized by the IAEA~\cite{IAEA,CU10,IAEA2}. As one may surmise, the passage from INCL4.2 to INCL4.6 (even to INCL4.5) could not have  been done entirely in the same spirit of the building of INCL4.2, i. e. resting on known phenomelogy with well determined parameters. We were forced to introduce less well established features with less solidly determined parameters. The purpose of this paper is to present all the new features that have been introduced in INCL4.6 and to give, whenever possible, the physics motivation for these new features. Although a lot of results with INCL4.5 are available in Ref.~\cite{IAEA}, perhaps not sufficiently commented or detailed, we present here the most important ones, especially to illustrate the physics aspects of the improvements brought into the model.

The paper is divided as follows. In
Section II, we describe the  INCL model with emphasis on the new features of INCL4.6. Section III is devoted to an extensive comparison with a
representative panel of experimental data, for nucleon-induced reactions. In Section IV, we discuss theoretical results for light cluster-induced reactions concerning a few key experiments. Finally, in Section V, we
critically examine the new features and their effects and we present our conclusion.

\section{Description of the INCL4.6 model}
\subsection{A short reminder of the INCL4.2 model }

The INCL4.2 model is  extensively described in Ref.~\cite{BO02}. It is sufficient here to remind the salient features. The INCL4 model is a time-like intranuclear cascade model. In the initial state, all nucleons are prepared in phase space. Target nucleons are given position and momentum at random in agreement with a Saxon-Woods and a Fermi sphere distributions, respectively. They are moving in a constant potential well, the same for protons and neutrons, describing the nuclear mean field. The incident particle (nucleon or pion) is given the appropriate energy and an impact parameter at random. All nucleons are then set into motion and followed in space-time. The cascade process involves binary collisions between nucleons and produced pions and Delta resonances. Particles can be  transmitted through, or reflected on, the surface of the square well potential they feel. Delta resonances  decay according to their lifetime. Details on collisions, utilized cross sections, Pauli blocking, etc, may be found in Ref.~\cite{BO02} and previous publications cited therein. We remind that the stopping time of the cascade is determined self-consistently by the model itself. It can  simply be parametrized (in fm/c) by
\begin{equation}
t_{stop}=29.8 A_T^{0.16},
\label{tstop} 
\end{equation}
for incident nucleons(we denote by $Z_T$ and $A_T$ the charge and mass numbers of the target).

It is useful to remind some other specific features which may be of importance for the rest of the paper. At the beginning of the cascade process, the incident nucleon or pion is located with its own impact parameter on the surface of the ``working sphere'', centered on the target with a radius 
\begin{equation}
R_{max}=R_0 + 8a,
\label{Rmax} 
\end{equation} 
where $R_0$ and $a$ are respectively the radius and the diffuseness of the target nucleus density. Particles are moving along straight-line trajectories between collisions inside the working sphere. They are divided into participants and spectators in the usual sense. When participants leave the working sphere, they are considered as ejectiles and do not interact any more. Nucleons are moving in a  potential well with constant depth and with  a radius which is dependent upon their momentum.  The potential radius for particles with energy larger than the Fermi energy is also taken to be equal to $R_{max}$. Pions do not experience any potential. Motivation and details are given in Ref.~\cite{BO02}. 

The INCL4.2 model can accomodate light clusters (up to $\alpha$-particles) as projectiles. In this case, the nucleons are given initially position and momentum inside the cluster and the latter is positioned at the beginning in such a way that one of its nucleons is touching the working sphere. See Ref.~\cite{BO02} for detail.

We comment a little bit on soft NN collisions. The latter do not contribute very much to the ejection of particles in the cascade. There are several arguments based on nuclear transport theory indicating that these collisions should be disregarded, since they induce slow modifications of the particle density distribution, a feature that is supposedly taken into account by the average nuclear potential~\cite{BO86,BO90,BU85,WE89,CU11}. In INCL4.2, soft NN collisions, with c.m. energy smaller than $\sqrt{s_0}$=1925 MeV, are simply disregarded. This cut-off value may seem to be rather large, but one has to realize that low energy NN collisions are largely Pauli-blocked in spallation reactions. Lowering the value of $\sqrt{s_0}$ does not significantly change the results at high incident energy.  It is no longer the case at low energy, say below $\sim$200 MeV, as discussed in Ref. \cite{CU03}. This matter will be re-examined in this paper.    

\subsection{Shortcomings of the INCL4.2 model }
The shortcomings of the INCL4.2 model have been identified in various places~\cite{BO02,HIN,EUR}.  We just remind them very briefly. Some phenomenological aspects of nuclear physics are neglected. The model cannot accommodate production of clusters in the cascade, i.e. with a kinetic energy definitely larger than the typical evaporation energies, as it can be seen experimentally. Concerning the predictive power of the model, several deficiencies can be noted. Pion production is generally overestimated. Quasi-elastic peaks in ($p,n$) reactions are generally too narrow and sometimes underestimated. Finally, reaction cross sections are severely underpredicted below $\sim$100 MeV. Residue production cross sections are sometimes unsatisfactorily reproduced, especially for residues close to the target. For matter of convenience, we separate below the new features of the INCL4.6 model into those which are included in the intermediate INCL4.5 version and those which are posterior to this version.

\subsection{Main new features in the INCL4.5 model }
\label{features4.5}
\subsubsection{Introduction of known phenomenology}

\textit{A. Isospin and energy-dependent potential well for the nucleons}. The depth of the potential well felt by the nucleons is dependent upon the energy of the nucleons and is not the same for protons and neutrons. The energy dependence is taken from the phenomenology of the real part of the optical-model potential~\cite{HO94,JE76,JE91}. Roughly speaking, the potential depth decreases regularly with increasing energy, from ordinary values at the Fermi level to zero at roughly 200 MeV. The isospin dependence is such that  the neutron and proton  Fermi levels have the same energy. For more detail, see Ref.~\cite{AO04}. The influence of this modification is relatively small, except for special quantities, like the production cross sections for isotopes with an extra unit of charge compared to the target~\cite{AO04,AO09}.

\textit{B. Average potential for pions}. An average isospin-dependent potential well, of the Lane type~\cite{LA62}, is introduced for pions, as well as reflection and/or transmission at the border of this potential. The depth of the potential has been taken, as far as possible, from the phenomenology of the real part of the pion-nucleus optical potential (dispersive effects due to the strong imaginary part have to be removed). This depth amounts to 22 MeV for $\pi^+$'s and 38 MeV for $\pi^-$'s on  a $Pb$ target. The radius of the potential is taken as $R_0 +4a$, in rough accordance with phenomenology. This modification and its effect are presented in detail in Ref.~\cite{AO06}. In general, it reduces the pion production cross section, mitigating so the overestimate by INCL4.2, as illustrated below by Fig.~\ref{pion}.

\textit{C. Deflection of charged particles in the Coulomb field}. Once an impact parameter is selected for the incident nucleon, the cascade process is initiated with this nucleon located at the
intersection of the ``external'' Coulomb trajectory (corresponding asymptotically to the specific impact parameter) with the ``working sphere'' (see above). The same procedure is used to connect the direction of an outgoing particle at the nuclear periphery and its asymptotic direction.

These three modifications can be considered as mandatory. They do not introduce
any fitting parameter. Values of the parameters have been fixed once for all, 
largely inspired from known phenomenology.

\subsubsection{Emission of clusters}
\label{clmodule}
An improvement of the INCL4.2 model, concerning this feature, had been already proposed in Ref.~\cite{BO04}. The implementation of this feature in INCL4.5 is somehow different, although the basic idea is the same: an outgoing nucleon crossing the nuclear periphery is supposed to be  able to carry along other
nucleons to form a cluster, provided the involved nucleons are lying sufficiently near each other in phase space. We first describe the present implementation in INCL4.5 and then comment upon the difference with the work of Ref.~\cite{BO04}.

\begin{figure}[h]
\includegraphics[width=8cm,height=3cm]{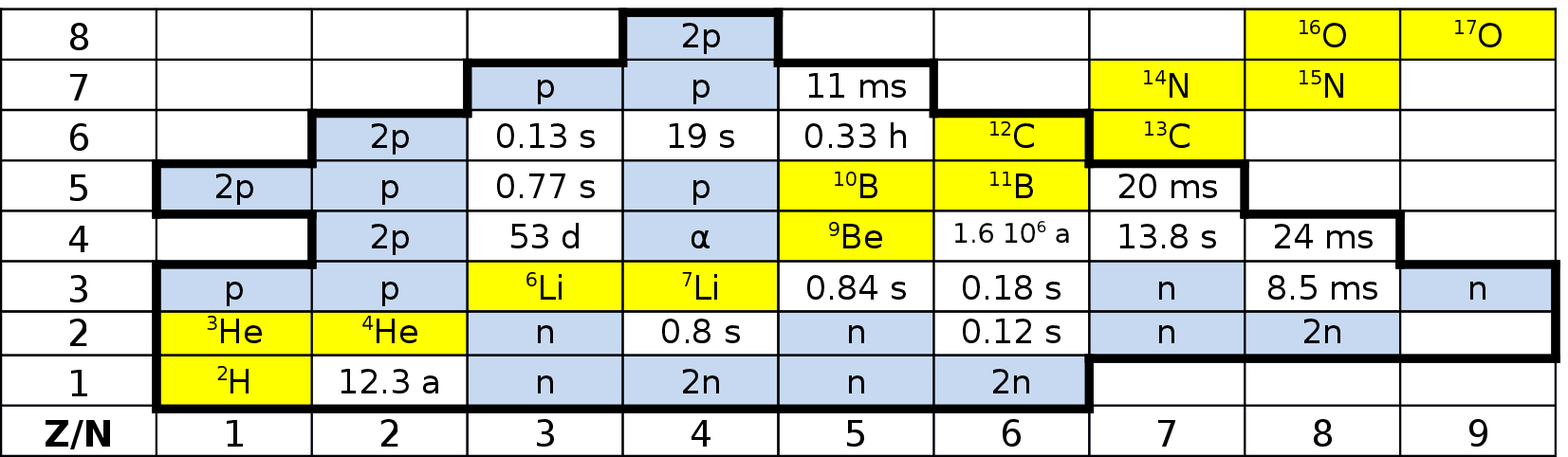}
\renewcommand{\baselinestretch}{0.5}
\caption{\small  In this graph, the considered clusters for $A_{cl}^{max}$=12 are identified by their charge $Z$ (vertical ordering) and neutron  $N$ (horizontal ordering) numbersand are displayed inside the perimeter delineated by the heavy line. Absolutely stable clusters are denoted by the ordinary symbols inside yeloow cells. Cells displaying a time correspond to  clusters with a lifetime larger than 1 ms, that are  considered as detectable clusters. Other cells correspond to clusters with a lifetime smaller than 1 ms. They are forced to decay by the indicated  mode.}
\renewcommand{\baselinestretch}{0.5}
\label{lcl}
\end{figure}

The features of the model for cluster production can be described as follows: 
\newcounter{marker}
\begin{list}{\arabic{marker}.}{\usecounter{marker} \setlength{\parsep}{0.3ex}}
\item An outgoing nucleon arriving at the surface of the ``working sphere'', whether or not it has made collisions earlier, is selected as a possible leading nucleon for cluster emission, provided its energy is larger than the threshold energy, otherwise it is reflected. 
\item Potential clusters are then constructed. The leading nucleon is drawn  on its (straight) line of motion back to a radial distance
\begin{equation} 
D=R_0+h,
\label{D} 
\end{equation} 
$R_0$ being the half density radius, and clusters are built by searching nucleons which are sufficiently close in phase space ($\Delta$'s are excluded)\footnote{If the line of motion of the nucleon does not cross the sphere of radius $R_0+h$, the nucleon is moved back to the minimum distance of approach of the center of the nucleus.}. Clusters of increasing sizes are built successively. All potential clusters up to a maximum size $A_{cl}^{max}$ are considered. The criterion of sufficient proximity is expressed with the help of Jacobian coordinates:
\begin{equation}
r_{i,[i-1]} p_{i,[i-1]} \leq h_0 (A_{cl}),\ \  for\ \ \  i=2,3,...,A_{cl}
\label{ph} 
\end{equation}
where $r_{i,[i-1]}$ and $p_{i,[i-1]}$ are the relative coordinates of $i-$th nucleon with respect to the subgroup constituted of the first $[i-1]$ nucleons ($i=1$ corresponding to the leading nucleon) and where $A_{cl}$ is the mass number of the cluster. The value of $h_0 (A_{cl})$ is discussed below. The test on Jacobian coordinates  is preferred to the usual test on the relative coordinates $r_{ij}$, $p_{ij}$ for any pair ($i,j$) of particles, because it disfavours exotic shapes (such as spaghetti) of the clusters.  Considered clusters up to $A_{cl}^{max}=$12 are listed in Fig. \ref{lcl}. For the moment, due to extremely fast increase of the combinatorics with the mass of the cluster and due to limitations in computing time, clusters up to $A_{cl}^{max}=$8 are considered. Extending $A_{cl}^{max}$ beyond this value may change the yields. We have checked that problems of convergence of the procedure manifest themselves by a slight excess of the yield for  clusters with a mass number equal to $A_{cl}^{max}$, as indicated in Ref.~\cite{CU11a}, so that in practice numerically stable results are obtained for clusters of mass up to 7, and that the predicted yield for production of mass 8 can be considered as an upper bound.   

\item The less ``virtual'' cluster is selected. Let $\sqrt s$ be the c.m. energy of the composing nucleons, built on the 4-momentum
of a cluster, defined as the sum of the 4-momenta of the nucleons inside the cluster. Let  us consider the quantity
\begin{equation}
\nu=(\sqrt{s}-\sum m_i )/ A_{cl}- B_{cl} /A_{cl}
\label{nu} 
\end{equation}
where $B_{cl}$ is the (nominal) binding energy of the cluster. The
cluster with the minimum value of $\nu$ is selected. This quantity can be viewed as the excitation energy per nucleon of the cluster diminished by twice the binding energy per nucleon. The introduction of this quantity is largely phenomenological and is solely justified by the relative success of the model.
\item The selected cluster is emitted provided three conditions are satisfied. First, it should have sufficient energy to escape, i.e. $T_{cl}=\sum (T_i-V_i )-B_{cl} >$0 , where the $T_i$ 's are the kinetic
energies of the nucleons and where the $V_i$ 's are the depths of their potential wells. Second, the cluster has also to succeed the test for penetration through the Coulomb barrier. Third, the cluster cannot be emitted too tangentially. If $\theta$ is the angle between the direction of the
cluster (defined as the direction of its total 3-momentum)
and the radial outward direction passing by the center of mass of the potential cluster, it is required that 
\begin{equation}
cos \  \theta > 0.7
\label{angle} 
\end{equation}
The idea behind this condition is that when a nascent cluster spends too much time in the nuclear
surface, it likely gets dissolved.  These choices are admittedly made to improve the results at low energy, though some supporting arguments can be produced. 
\item If these tests are successful, the cluster is emitted with the kinetic energy $T_{cl}$ in the 
direction of the total momentum of its components. If they are not, the leading nucleon is emitted alone provided it succeeds the test for penetration of the Coulomb barrier. If not, the leading nucleon is simply reflected.
\item At the end of the cascade process, short-lived
clusters with a lifetime less than 1 ms (e.g. $^5Li$) are forced to decay, isotropically in their c.m. frame. Clusters with a lifetime larger than 1 ms  are considered as detectable as such, prior to decay. Details are summarized in Fig.~\ref{lcl}.
\end{list}

In our previous work \cite{BO04}, which was primarily devoted to production of clusters (up to $A_{cl}$=4)  at high energy (i.e. from 500 to 2500 MeV), two parameters, $h$  (Eq.~\ref{D}) and a single parameter $h_0$ for all clusters in Eq.~\ref{ph}, with the selection of the heaviest possible cluster, were sufficient. In order to have a more or less satisfactory description at low energy, we have been forced to consider different values of the parameter $h_0$ for light clusters ($A_{cl} \leq 4$) and a simple law ($h_0 = h_1 (A_{cl}/5)^{1/3}$) for heavier clusters. In INCL4.5, the parameters have been determined once for all, by fitting data on a few illustrative cases. The values of the parameters are given in Table~\ref{paramclus}.

\begin{table}[b]
\caption{\it  Values of the parameters of the cluster formation model. The parameters are described in the text: $h$ is given in fm, $h_0$ and $h_1$ are given in fm MeV/c. }
\begin{tabular}{c|c|c|c|c}
\hline
  $h$ & $h_0(2)$ & $h_0(3)$ & $h_0(4)$ & $h_1$ ($ A_{cl} >4$) \\
\hline
  1.0 & 424 & 300  & 300 & 359\\

\end{tabular}
\label{paramclus}
\end{table}

\subsubsection{Modifications concerning the Pauli blocking}
\label{mcPb}
A strict Pauli blocking is applied to the first collision: the nucleons should lie outside the Fermi sea after the first collision. In INCL4.2, the Pauli blocking is applied stochastically, according to the product of the final state blocking factors. Conjugated with the fact that constructing the target with nucleons at random generates events with a nonuniform Fermi sea (even if it is uniform on the average over events), this
procedure has the drawback of allowing sometimes collisions which otherwise (i.e. in a perfectly filled Fermi sea) would be strictly forbidden. On the other hand, it allows to account for surface effects and for effects of the depletion of
the Fermi sea as the cascade process evolves. It is found in Refs.~\cite{CU03,HE05} that a good compromise is achieved when a strict Pauli blocking is adopted for the first collision and when the usual procedure is kept for the subsequent ones.

\subsubsection{Modifications for soft collisions}

\textit{A. Soft collisions}. In INCL4.2,  soft collisions (with c. m. energy $\sqrt{s} <\sqrt{s_0}=$ 1925 MeV) are neglected. Historically, this choice was made to avoid inconveniences linked with the raising NN cross sections at low energy. More profoundly, the underlying argument states that soft collisions (with low momentum transfer) do not change significantly the energy-momentum flow in the system and that their effect is likely to be more reasonably accounted for by the nuclear mean field. Furthermore, changing the boundary between soft and hard collisions has no sharp effect. The reason is that, in spallation reactions, soft collisions occur mainly when the colliding nucleons are lying close to the Fermi energy and thus that these collisions are  largely Pauli-blocked, as explained in Ref.~\cite{HE05}. This argument breaks down when a low energy incident nucleon makes a collision in the nuclear periphery, where Pauli blocking is not very efficient. We thus decided to lower $\sqrt{s_0}$, still  trying to keep the results roughly equivalent and to save computation time. The new value is now $\sqrt{s_0}$=1910 MeV.

\textit{B. Special treatment on the first collision}. For the first collision, we even lowered the value of $\sqrt{s_0}$ to the minimum, i.e. twice the nucleon mass, in apparent contradiction with the above arguments above. However, at low energy, only a few (1-3, 
on the average) collisions occur. Neglecting a
soft collision, allowed however by Pauli blocking, especially the first one, may amount to neglecting the event. This may have dramatic effects on the total reaction cross section, since the latter involves all kinds of events, be them hard or soft. It should be stressed that this procedure does not change the results at high energy (say, above $\sim$200 MeV). Indeed, in tis case, the first collision is always a hard one. Neglecting a subsequent soft collision is most of the time harmless. See Ref.~\cite{BO08} for more details. In addition, a special
procedure, named 'local E', is applied to the first collision. In INCL4.2, the momenta of the target nucleons are too large in the nuclear surface, near the turning points, since they experience constant square well potentials, instead of smoothly varying potentials (this is ``the price to pay'' to keep the simplicity of straight line motion in our code, see Ref.~\cite{BO02} for details).  In the 'local E' procedure, when two nucleons are selected for the first
collision, their momenta are ``corrected'', by replacing their values by those assumed  by the nucleons in a smoothly varying potential, at the same positions. When testing a pair of nucleons for collision, the NN cross section is calculated with the corrected momenta. After the collision, if it occurs, momenta are ``corrected back'' to the INCL4.2 prescription. Once again, this procedure has no effect on the first collision at high incident energy, nor when this first collision occurs in the bulk of the target nucleus. These two modifications are instrumental to give the predictions of the total reaction cross section in agreement with experimental data at low energy.

\subsubsection{Modification of the status of the participants}
\label{back}
If after a binary collision or after a $\Delta$-decay,
a nucleon (obviously a participant) has an energy
smaller than the Fermi energy plus a small quantity $\xi$, put arbitrarily equal to 18 MeV as a first attempt, it is considered thereafter as a spectator (it can be ``re-promoted'' as a participant if it gets a sufficiently large energy transfer in a subsequent collision with a participant). This procedure, which is inspired from the Isabel code~\cite{YA79,YA81,YA08}, may be motivated by various considerations: in a nucleus, the Fermi sea is not sharply defined, nucleons cannot be localized with precision when their energy is low and  correlations may render the difference between a nucleon above the Fermi level and a spectator rather
fuzzy. In fact, there is no compelling argument in favour of this procedure. It is included here  because it gives slightly better results in some cases, in particular for production of  clusters at low incident energy. Unfortunately, it leads, in our case, to unphysical results in neutron spectra at very low incident energy, as it is clearly illustrated in Fig.~\ref{n63} below. That is why this feature has been changed in the INCL4.6 version, see below. For convenience, we will refer to this procedure as the ``back to spectator'' recipe.

\subsection{Additional  features included  in the INCL4.6 model }
\label{features4.6}
\subsubsection{Further modification of the status of the participants}
As just stated above, the modification described in Section~\ref{back} produces unpleasant results in neutron and proton energy spectra, which will be illustrated when discussing comparison with experimental data.
In INCL4.6, participant neutrons are considered as spectators whence their energy gets  below the neutron emission threshold, corresponding roughly to $\xi$ defined above equal to 7 MeV. The same procedure is adopted for participant protons when their energy falls below their emission threshold plus two thirds of their Coulomb barrier. The whole procedure is certainly more acceptable, although it is not really justified on consistency arguments, but it is validated a posteriori by the results, as shown below (see Figs.~\ref{n63} and \ref{proton}).

\subsubsection{Use of experimental thresholds for nucleon emission}
In our standard model, a participant nucleon (of type $i=n,p$) can be emitted when it hits the surface of the potential with a  kinetic energy   $E$  larger than the potential depth, or equivalently when the difference  of  energy $E$ with the Fermi energy, $E-E_F^i$, is larger than the  model separation energy $S_i$. If it is emitted, the nucleon acquires an asymptotic kinetic energy $E_{\infty}$ given by  
\begin{equation}
E_{\infty}=E-E_F^i-S_i
\label{Eass1} 
\end{equation}
In our standard model, the values of $S_i$ are fixed for a given target nucleus and equal to the  differences between the  potential depths and the Fermi energies, respectively\footnote{The definition of $S_i$ is slightly more involved for energy-dependent potentials. We simplify the presentation on this minor point.}. In INCL4.6, the energy $E-E_F^i$ is compared to the  physical separation energy $S_i^{phys}$, taken from mass tables, for the emission  from the actual nucleus, i.e. the target nucleus  left over when the candidate particle is hitting the surface. Eq. \ref{Eass1} is replaced by   
\begin{equation}
E_{\infty}=E-E_F^i-S_i^{phys}
\label{Eass2} 
\end{equation}
 
The effect is expectedly small, except at low incident energy, where the precise value of the threshold energy for a reaction does matter, or when the evolution of the target in the course of the cascade reaches the border of available phase space.

\subsubsection{Modified value for $R_{max}$}
Using a value given by Eq.~\ref{Rmax} for the radius for the ``working sphere'' of $R_{max}=R_0+8a$ is safe at high energy, since this value allows a sampling of impact parameters concerning  all reaction events: it encompasses the outskirts of the nuclear density by a value which is greater than  $r_{int}=\sqrt{\sigma_{NN}^{tot}/\pi}$, that we denote for simplicity as the ``range of interaction'', $\sigma_{NN}^{tot}$ being the NN total cross section at the incident energy per nucleon\footnote{In INCL, collisions are decided on a minimum of relative distance basis, which is equivalent to introducing an effective interaction with a range equal to $r_{int}$.}. This is no longer true at low energy (from a few MeV to a few tens of MeV), where the ``range of  interaction''  may become very large, because of the raising NN cross sections. We thus decided to use 
\begin{equation}
R_{max}=R_0+8a+r_{int}=R_1 +8a.
\label{R1} 
\end{equation}
 
Accordingly, we have increased the maximum time of the cascade, which now corresponds to the time of passage of the incident particle through the ``working sphere'' along a diameter, when this time exceeds the usual stopping time, given by Eq.~\ref{tstop}. 

\subsubsection{Treatment of cluster-induced reactions}
\label{cir}
In INCL4.2, an incident cluster (up to an alpha particle) is considered as a collection of independent nucleons with internal Fermi motion superimposed to the motion of the incident cluster as a whole  (see Ref. \cite{BO02}), adjusted in such a way that the sum of the total energies of the constituting nucleons is equal to the nominal energy of the physical cluster. In other words, the cluster is replaced by independent on-shell nucleons with the correct nominal total energy, but with an incorrect (smaller than nominal) total momentum. This approximation is justified at high energy, where the missing momentum is relatively small, and  where this model gives more or less satisfactory results~\cite{BO02,BO10,KA10}. However, it is not really appropriate for reactions at low incident energy. In this domain, several features should be taken properly. First the total energy-momentum content of the projectile should be preserved. Second, the collective
motion of the projectile has to be  respected at low energy and Fermi motion has to be  progressively restored when the available energy allows it. Liberating Fermi motion at a too early stage (like in INCL4.2) may be harmful at low energy. Indeed a nucleon with a backward-oriented Fermi motion velocity larger than  the collective velocity the projectile may fly at once in the backward direction, which is indeed unphysical. Third, emission of projectile spectators
with the influence of their Fermi motion should be possible and  dominated by
geometrical properties. This property, which is well established at high energy~\cite{GO74,FE73,HU81}, is expected to survive, to some extent, at low energy, at least for large impact parameters. Finally, compound nucleus formation should  dominate at low incident energy. To fulfill these requirements, we  have implemented an ad hoc model, made of the following ingredients.

\textit{a. Initialisation of the incident cluster.} Nucleon momenta $ \vec{p'}_i$ and positions  $ \vec{r'}_i$ inside the cluster are generated as before~\cite{BO02} (note, however, that a special method is applied to ensure $\sum \vec{p'}_i=0$ and $\sum \vec{r'}_i=0$).  At the beginning of the event, the cluster center of mass is positioned on the classical  Coulomb trajectory in such a way that one of the nucleons is touching a sphere of radius $R_{Coul}$. The latter represents the Coulomb barrier. The value of $R_{Coul}$ is taken from the phenomenology of the Coulomb barrier  heights and has been tabulated as function of the target mass for $p$, $d$, $t$, $^3He$ and $^4He$ projectiles. It is given in the Appendix. Of course, for very large impact parameters and/or low incident energy, the Coulomb trajectory may miss the sphere of radius $R_{Coul}$. The incident cluster is then positioned at the minimum distance of approach.

For the energy-momentum content, the following procedure is adopted. Let $e'_i$ and $\vec{p'}_i$, the energy and the momentum of the nucleons generated (as in INCL4.2) in the frame of the incident cluster (one has $e'_i=\sqrt{\vec{p'}_i^2 +m_i^2}$, $\sum \vec{p'}_i=0$). The nucleons are put off-shell  by changing their energy $e'_i$ into $e_i=e'_i-v$, defined by: 
\begin{equation}
\sum e_i=\sum ( e'_i-v)=M_{inc},
\label{off1} 
\end{equation}
where $M_{inc}$ is the exact projectile  mass. When Lorentz-boosted with the incident velocity $\vec{\beta}_{inc}=\vec{P}_{inc} / M_{inc}$, $\vec{P}_{inc}$ being the nominal momentum of the projectile,  the nucleons acquire 4-momenta $(E_i, \vec{p}_i)$ such that 
\begin{equation}
\sum E_i=W_{inc}, \sum \vec{p}_i=\vec{P}_{inc}, 
\label{off2} 
\end{equation}
where $W_{inc}$ is the exact total energy  of the projectile. This rather crude procedure aims at preserving the energy-momentum content of the cluster when replacing it by nucleons. The latter are then off-shell, which  naturally accounts for their binding. Naively, the quantity $v$ can be viewed as the potential necessary to bind the nucleons to the right  binding energy of the cluster.  At large incident cluster energy, the 4-momenta of the nucleons inside the projectile, are  basically the same in the two schemes.
 
\textit{b. Generation of the geometrical spectators.} When positioned as indicated above, the nucleons of the projectile are separated into geometrical  participants and geometrical spectators, which are those nucleons whose direction of motion intercepts or not the working sphere, respectively~\footnote{This ``geometrical spectator'' term is used to distinguish these spectators, which are defined here by a geometrical criterion, from the usual definition of the spectators that are those nucleons, which, in a true reaction event, are intercepting the target, but avoid, by chance,  collisions.}. For performing this separation, the direction of motion of the nucleons is assumed to be parallel to the collective velocity vector of the cluster. Spectator nucleons are put on shell with momentum $\vec{p}_i$, defined above,  and energy $E_i^{spec}=\sqrt{\vec{p}_i^2 +m_i^2}$,  and  are frozen further on. In order to preserve the correct energy balance, the energy of the participants are decreased correlatively and equally.

\textit{c. Treatment of the geometrical participants.} The following critera are used. If a participant has an energy lower than the bottom of the target potential well, the event is discarded. This happens very rarely when, due to fluctuations, the generated Fermi motion energy is very small, but occurs close to thresholds for emissions of particles. 

If all the geometrical participant nucleons have a total energy above the bottom of the target potential and if one of these nucleons has a  trajectory cutting the sphere of radius defined in Eq.~\ref{R1}and if one of these nucleons at least has an energy {\it below} the Fermi energy, a compound nucleus is formed with all the geometrical participants and the target, with an energy which is equal to the available energy. The so constructed compound nucleus is ready for de-excitation (no further cascade is performed). Of course, its energy should be larger than the nominal ground state energy, otherwise the event is discarded. This choice is inspired by the fact  that a compound nucleus is expectedly formed at low energy when the inhibiting effects of the Coulomb barrier are overcome.
 
Finally, when all the geometrical participants have an energy above the Fermi energy, the usual cascade is applied. However, before colliding for the first time, geometrical participants are propagated with the  velocity of the incident cluster as a whole. Right before their first collision, they are given back their Fermi motion and are put on shell. We take Fermi motion into account for calculating the cross section and the kinematics of the collision.

\textit{d. Coulomb polarisation of incident deuterons.} When an incident deuteron is selected, it is positioned initially with one of its nucleons touching the sphere of radius $R_{Coul}$. For heavy targets, if this nucleon is a proton, it is interchanged with the neutron. This procedure mocks up the polarisation of deuteron by the Coulomb field of the target. It is often advocated that this effect is necessary to describe properly deuteron-induced reactions \cite{AO00}.
    
The model described above may appear as a rather ad hoc procedure. It has been inspired by our willingness of coping with several aspects of the low-energy dynamics. But it can compete with other models, like fusion models. It manages Coulomb effects and Fermi blocking effects at low energy, but in addition, it leads automatically and dynamically to departures from compound nucleus formation, as the incident energy increases. Of course, we are well aware of the fact that a valid justification of our procedure can mainly come from a successful confrontation  with experimental data.

We have presented here the initializations of the nucleon-induced and cluster-induced events separately. We want to stress  that these preparations are not disjoint however. Actually, the initialisation of a cluster-induced event reduces to the one of a nucleon-induced event when the cluster is reduced to a single nucleon without Fermi motion.

\section{Comparison with experimental data for nucleon-induced reactions}
\subsection{Introduction }
We will not make a thorough comparison with experimental data, nor with the predictions of  competing models, such as Bertini~\cite{BE63,BE69} Isabel~\cite{YA79,YA81}, CEM03~\cite{MA83,MA00}, BUU~\cite{BUU}, IQMD\cite{IQMD}, JQMD~\cite{JQMD} and the BIC cascade of Geant4~\cite{Geant4} (see Ref.~\cite{FI08}, for a recent description of most of these models), since such a comparison can be found in the IAEA Intercomparison~\cite{IAEA} for INCL4.5. We will restrict ourselves to a comparison with data in cases where the influence of the modifications brought by INCL4.5 and INCL4.6 are important or when the physical significance of the modifications brought by INCL4.6 in comparison with INCL4.2 can be tested. For nucleon-induced reactions, the differences between INCL4.5 and INCL4.6 predictions are rather of minor importance, except for low-energy neutron spectra, which is illustrated in Section~\ref{ns-spec}. All INCL4.5 and INCL4.6 predictions have been obtained with the adjunction of the ABLA07 de-excitation model (actually the  ABLA07V5 version), described in Ref.~\cite{KE08}. 

\subsection{Total reaction cross sections}
\label{react}
\begin{figure}[h]
\includegraphics[height=10cm]{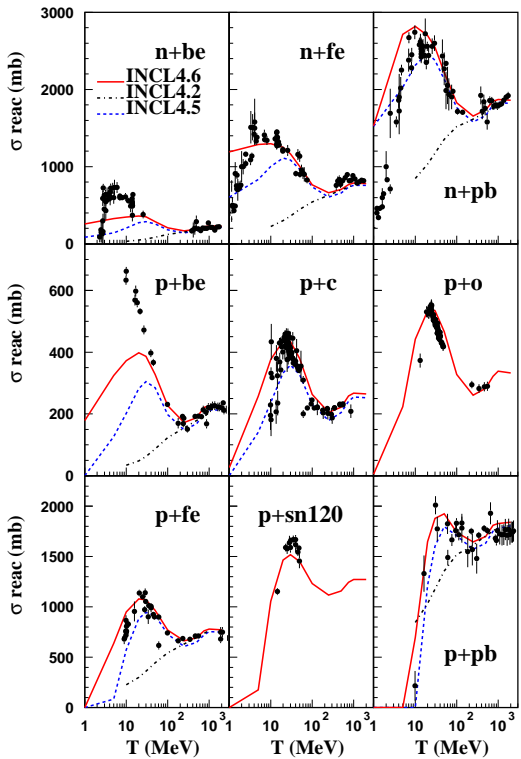}
\renewcommand{\baselinestretch}{0.5}
\caption{\small Comparison of INCL4.2 (black), INCL4.5 (blue) and INCL4.6 (red) predictions with experimental data for the total reaction cross section as a function of the incident kinetic energy $T$, for neutron-induced reactions on $^9Be$, $^{56}Fe$  and $^{208}Pb$  (first row) and proton-induced reactions on $^7Be$, $^{12}C$, $^{16}O$, $^{56}Fe$, $^{120}Sn$  and $^{208}Pb$ (2d and 3rd rows). Data are from Refs.~\cite{PR97,BA93,CA96}. }
\renewcommand{\baselinestretch}{0.5}
\label{sigR}
\end{figure}

This observable is entirely determined by the cascade stage and even by the first collision in this stage. Total reaction cross sections calculated with INCL4.2, INCL4.5 and INCL4.6 are displayed in Fig.~\ref{sigR}. Whereas the predictions are roughly the same for the three models above 200 MeV, there is a dramatic improvement brought by INCL4.5 and especially INCL4.6 in the 10 MeV-100 MeV range. We recall that the difference between INCL4.5 and INCL4.6 predictions is mainly coming from a too small radius of the working sphere $R_{max}$ in INCL4.5, which, especially at low energy, underestimates the number of interacting events for large impact parameters. It is remarkable that INCL4.6 is able to account for the bump of the cross sections appearing at a few tens of MeV and for its main properties, except for the $Be$ target. The systematic parametrization of the data given in Ref.~\cite{CA96} seems to indicate that reaction cross sections are not sensitive to structure effects and are smoothly varying with incident energy and target mass number, except for light nuclei at low incident energy, as it is illustrated  by Fig.~\ref{sigR}.

\begin{figure}[h]
\vspace{1cm}
\includegraphics[width=8cm]{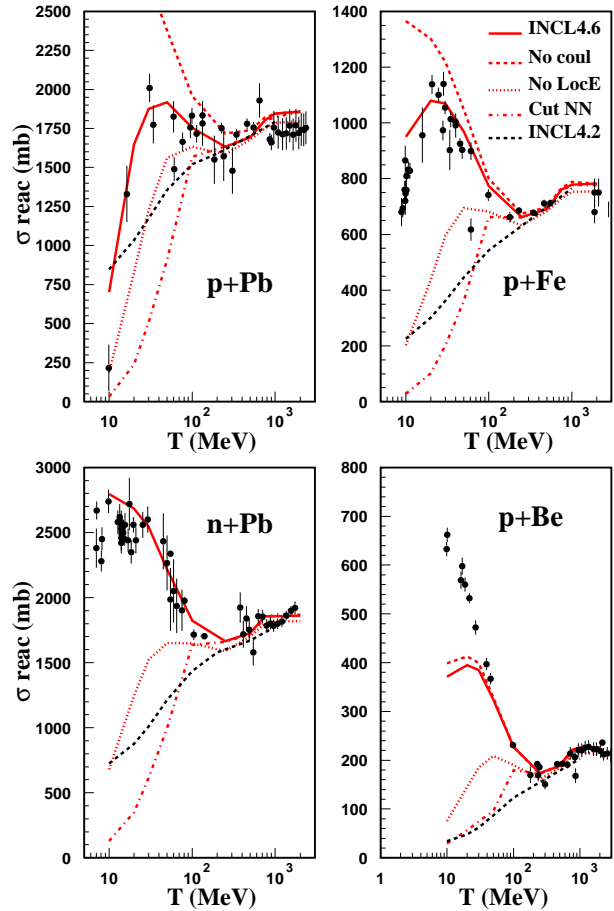}
\renewcommand{\baselinestretch}{0.5}
\caption{\small Comparison of INCL4.2 (black dotted lines) and INCL4.5 (full red lines) predictions with experimental data for the total reaction cross section as a function of the incident kinetic energy $T$, in various systems. Dashed red lines  correspond to INCL4.6 predictions when Coulomb deflection is neglected. Dotted (dot-and-dashed) lines correspond to INCL4.6 predictions when so-called 'local E' correction (soft collision suppression) are neglected. See text for detail. Data are from Refs.~\cite{PR97,BA93,CA96}. }
\renewcommand{\baselinestretch}{0.5}
\label{sigRd}
\end{figure}

Above 200 MeV, where there is no real difference between the predictions of the three models, the variation of the total reaction cross section with the incident energy  closely follows the variation of the NN total cross section, a feature that comes naturally in Glauber models~\cite{TR01} for total reaction cross section.  It is interesting to look at the origin of the differences between the INCL4.2 and INCL4.5 (or 4.6) predictions below 200 MeV. This is indicated by Fig.~\ref{sigRd}, where we have compared INCL4.2 predictions with INCL4.6 predictions when one of the additional features of this latter model is removed: Coulomb deflection, removal of soft collisions and the so-called 'local E' correction (see above). This Figure clearly shows that, in proton-induced reactions, the rise of the reaction cross section when the incident energy goes down from 200 MeV to ~30-40 MeV,  is due to the  removal of  the cut on soft collisions for the first collision, which in fact takes full account of the strong increase of the raising NN cross section with decreasing incident energy in this range, when the 'local E' correction, defined above, is applied. This correction alone  is already giving an important increase of the  cross section, as indicated by the dotted curves in Fig.~\ref{sigRd}. Of course, at very low energy, the reaction cross section for incident protons is decreased by the Coulomb repulsion, which leads to a global sharp decrease below the Coulomb barrier. In fact, it appears from Fig. \ref{sigRd} that it is important to take account of all the features which determine the probability of the first interaction. These effects are rather well identified in macroscopic models for the reaction cross sections~\cite{CA96,TR01}. Similar considerations can be done for neutron-induced reactions. However, in this case, the fall-off of the reaction cross section at low energy is basically due, in our model,  to the Pauli blocking of collisions at very low incident energy.

\subsection{Emission of light particles}
\label{elp}
\subsubsection{Neutron multiplicity in the cascade stage}
Neutron multiplicity is of crucial importance for nuclear energy applications. For our purpose here, it is convenient, following Ref.~\cite{LE02}, to consider the multiplicity of neutrons with a kinetic energy larger than 20 MeV, since this quantity is sensitive to the cascade stage only.
Results are given in Table \ref{multn}. One can see that our predictions with the latest version INCL4.6 are globally satisfactory, although the agreement with experimental data is somehow marginal at 1600 MeV. The decrease of the $E>$20 MeV neutron multiplicity from INCL4.2 to INCL4.6 is due to the ``back to spectator'' procedure. It can also be seen from Table \ref{multn} that the predicted neutron multiplicities in the 2-20 MeV range are  rather well described. In this case, of course, the merit is to be shared by ABLA07 and INCL4.6, since the latter determines the excitation energy of the remnant, prior to evaporation.

\begin{table}[t]
\caption{\it Neutron multiplicities (second column), obtained by integration of the experimental double differential cross-sections of Ref. \cite{LE02} in proton-induced reactions on $^{208}Pb$ nuclei at three values of the incident kinetic energy $T$, compared with the predictions of various INCL-ABLA couplings.}
\vspace{4mm}
\renewcommand{\baselinestretch}{1.5}
\begin{tabular}{|c|c|c|c|c|}
\hline
Neutron energy & Exp & INCL4.2 & INCL4.5  & INCL4.6 \\
 & & ABLA & ABLA07  & ABLA07 \\
\hline
\multicolumn{5}{|c|}{ \hspace{1cm}  $T$ = 800 MeV} \\
\hline
2-20 MeV & 6.5 $\pm$ 0.7 & 6.8 & 6.33 & 6.73\\
20 MeV-E$_{max}$ & 1.9 $\pm$ 0.2 & 2.5 &1.84  & 1.91  \\

\hline
\multicolumn{5}{|c|}{ \hspace{1cm} $T$ = 1200 MeV} \\
\hline

2-20 MeV & 8.3 $\pm$ 0.8 &8.1  & 7.8 & 8.27\\
20 MeV-E$_{max}$ & 2.7 $\pm$ 0.3 & 3.1 &2.39     & 2.48\\
\hline
\multicolumn{5}{|c|}{ \hspace{1cm} $T$ = 1600 MeV} \\
\hline
2-20 MeV & 10.1 $\pm$ 1.0 & 8.8  & 8.6  & 9.16\\
20 MeV-E$_{max}$ & 3.4 $\pm$ 0.5 & 3.7 & 2.79  &  2.90\\
\hline
\end{tabular}
\renewcommand{\baselinestretch}{0.5}
\label{multn}
\end{table}

\subsubsection{Neutron energy spectra}
\label{ns-spec}
\begin{figure}[h]

\includegraphics[width=8cm, height=10cm]{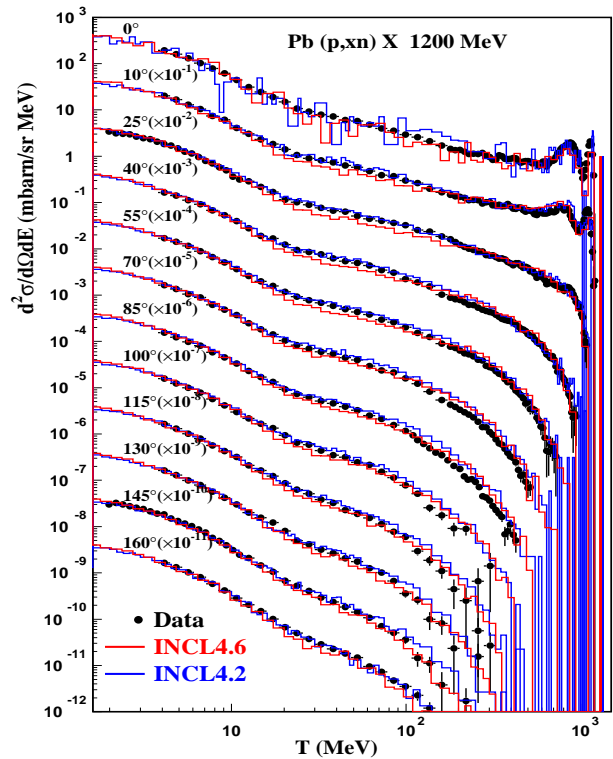}

\caption{\small Comparison of INCL4.2 (blue) and INCL4.6 (red) predictions with experimental data for neutron double differential cross sections  in $p$+$^{208}Pb$ collisions at 1200 MeV, as functions of the neutron kinetic energy $T$ and at different angles, indicated on the figure. For the sake of clarity, the  spectra at the various angles have been multiplied by the indicated constants. Data are from Ref.~\cite{LE99}. }
\label{neutron}
\end{figure}

These quantities are also of importance for applications. We will not comment very much on this point since extensive results can be found in the IAEA Intercomparison and since globally, all our models are giving rather satisfactory results for neutrons with a kinetic energy larger than ~20 MeV, i.e. for those neutrons which are produced in the cascade stage.  As an illustrative example, we show in Fig.~\ref{neutron} the predictions of INCL4.2 and INCL4.6 for the neutron spectra in $p$+$^{208}Pb$ collisions at 1200 MeV. 

The INCL4.5 (not shown) and INCL4.6 models give essentially the same results, for all targets and all incident energies, except below $\sim$100 MeV (see below). The predictions of these models are somehow less good than those  of INCL4.2. The difference particularly concerns neutrons with a kinetic energy between 20 to 50 MeV at angles between 10 and 50 degrees, where INCL4.5 and INCL4.6 underestimate the cross sections. This deficiency partly comes from the implementation of the production of composite particles by coalescence (inexistent in INCL4.2), which in a sense ''eats up'' neutrons and protons in this energy range, and partly from the use of energy-dependent nucleon average potentials. Note that this deficiency is reduced at 800 MeV and seems to be slightly increased for lighter targets ($Fe, Zr$). It can also be seen from Fig.~\ref{neutron} that  INCL4.6 is slightly better than INCL4.2 for the high energy part of the spectra.

\begin{figure}[h]
\begin{minipage}[t]{4cm}
\includegraphics[width=4cm]{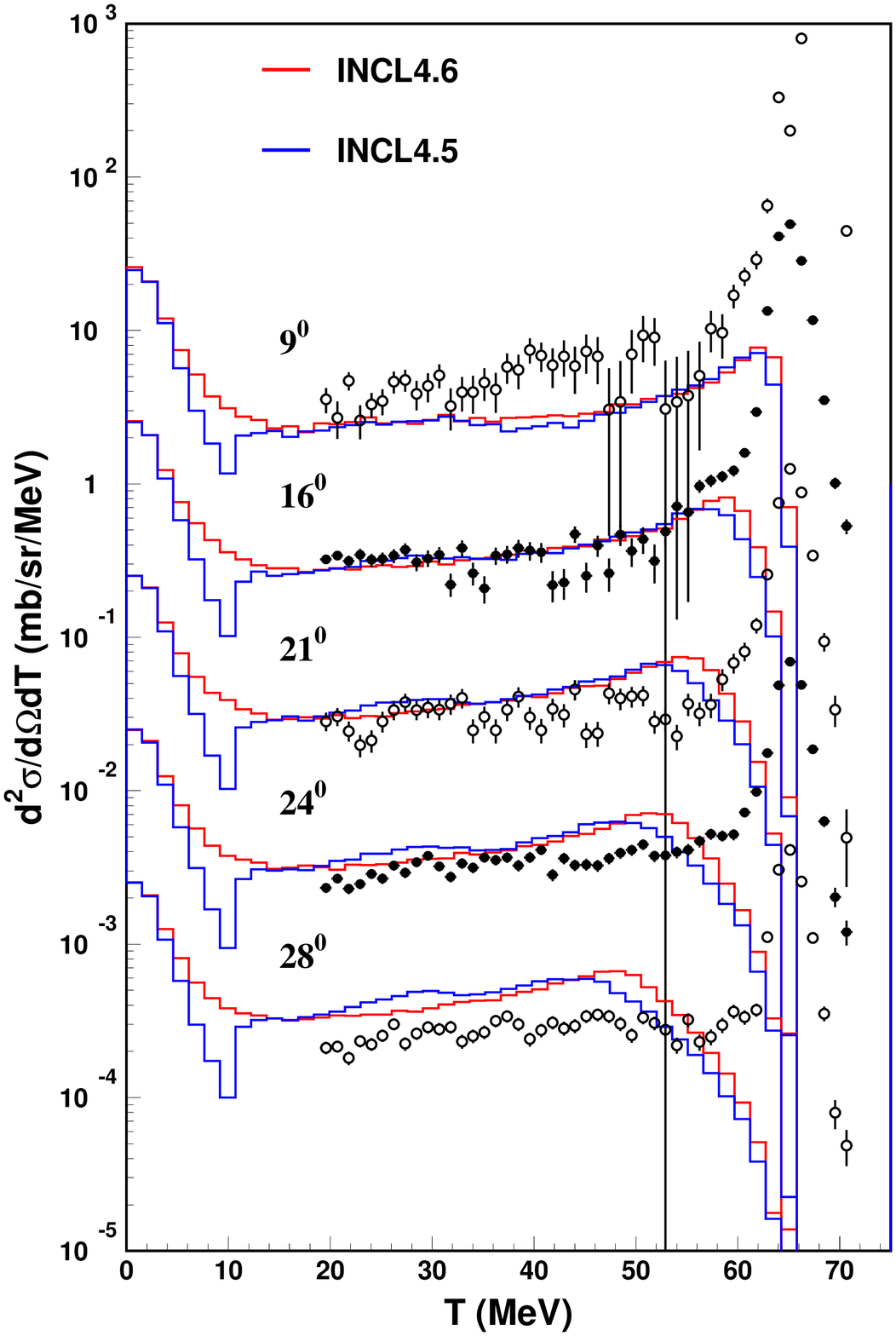}
\renewcommand{\baselinestretch}{0.5}
\end{minipage}
\begin{minipage}[t]{4cm}
\includegraphics[width=4cm]{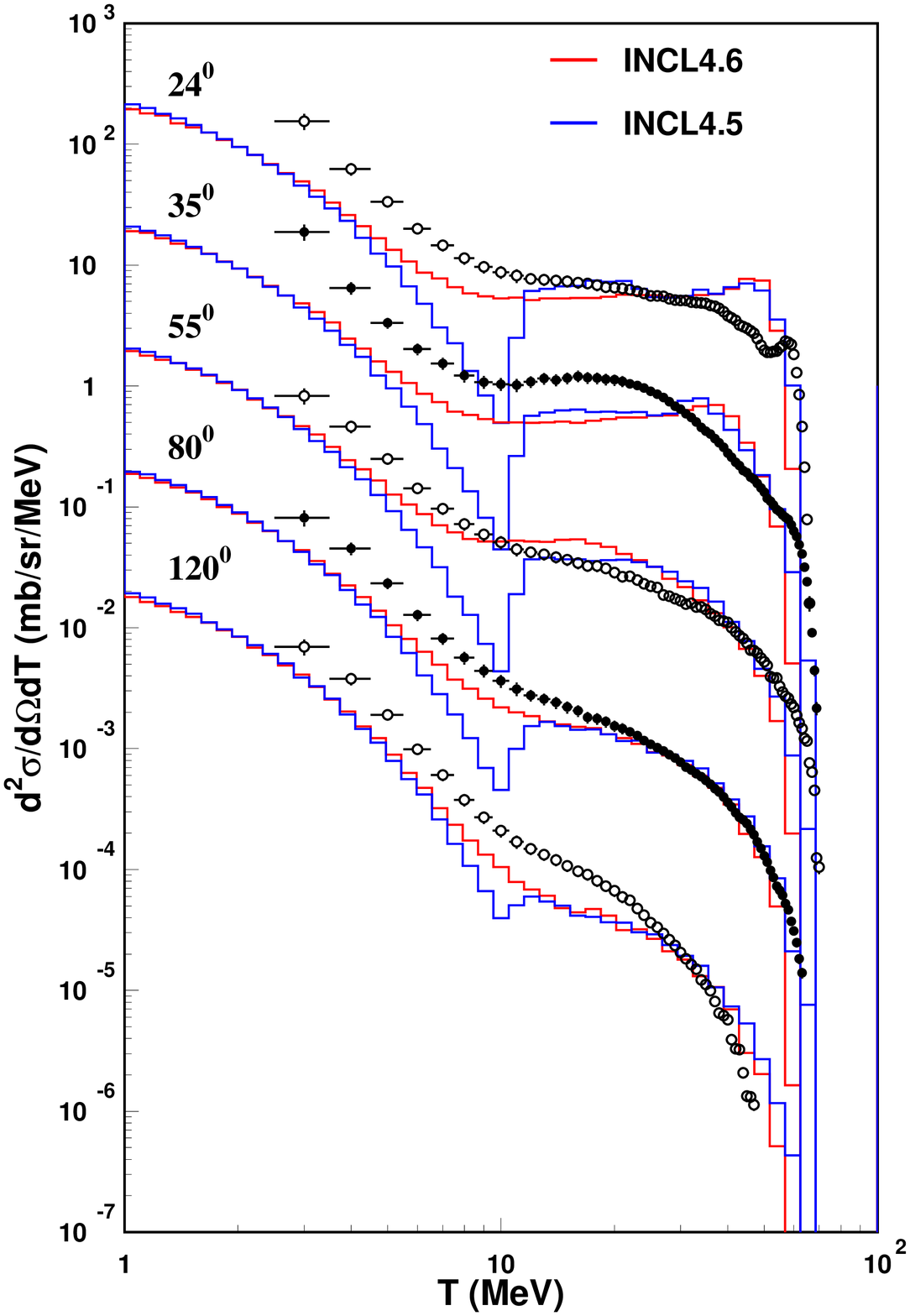}
\renewcommand{\baselinestretch}{0.5}
\end{minipage}
\hfill
\caption{\small Comparison of INCL4.5  (blue) and INCL4.6 (red) predictions for the neutron double differential cross sections  in $n$+$^{56}Fe$ collisions at 65 MeV (left panel, data from Ref.~\cite{HJ96}) and in $p$+$^{208}Pb$ collisions at 63 MeV (right panel, data from Ref.~\cite{GU05}).  }
\renewcommand{\baselinestretch}{0.5}
\label{n63}
\end{figure}

We show in Fig.~\ref{n63} the kind of results that we obtain around 60 MeV incident kinetic energy. The striking feature is the ``dip'' in the neutron spectra around 10 MeV produced by the INCL4.5 model. This originates from the ``back to spectator'' recipe, explained in  Section \ref{back}. This unphysical feature disappears in INCL4.6, since the quantity $\xi$ is then reduced to ~7 MeV for neutrons and the same quantity plus two thirds of the Coulomb barrier height for protons. Dips are so removed from the nucleon spectra. Dips of this kind are not visible at high energy (say above 100 MeV, see for instance Fig.~\ref{neutron}) because they are hidden by the overwhelming evaporation contribution. The theoretical shapes of the spectra above the evaporation peaks in Fig.~\ref{n63} are rather satisfactory, although the degree of agreement is substantially smaller than at high energy (see Fig.~\ref{neutron}). As a matter of fact, the shapes are almost perfectly reproduced at 800 MeV and are progressively departing from experiment when the incident energy is decreasing.

\begin{figure}[h]

\begin{minipage}[t]{4cm}
\includegraphics[width=4cm]{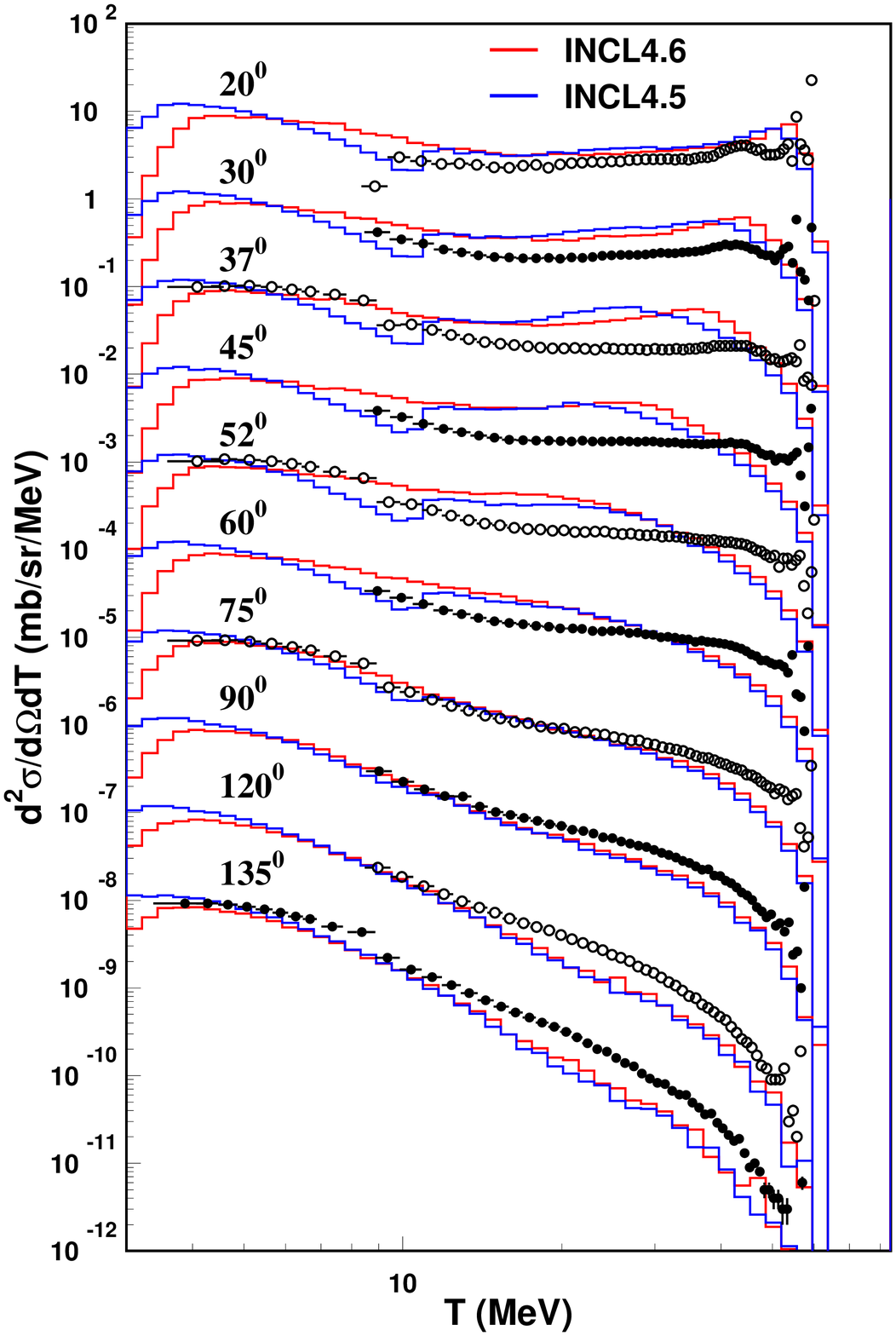}
\renewcommand{\baselinestretch}{0.5}
\end{minipage}
\hfill
\begin{minipage}[t]{4cm}
\includegraphics[width=4cm]{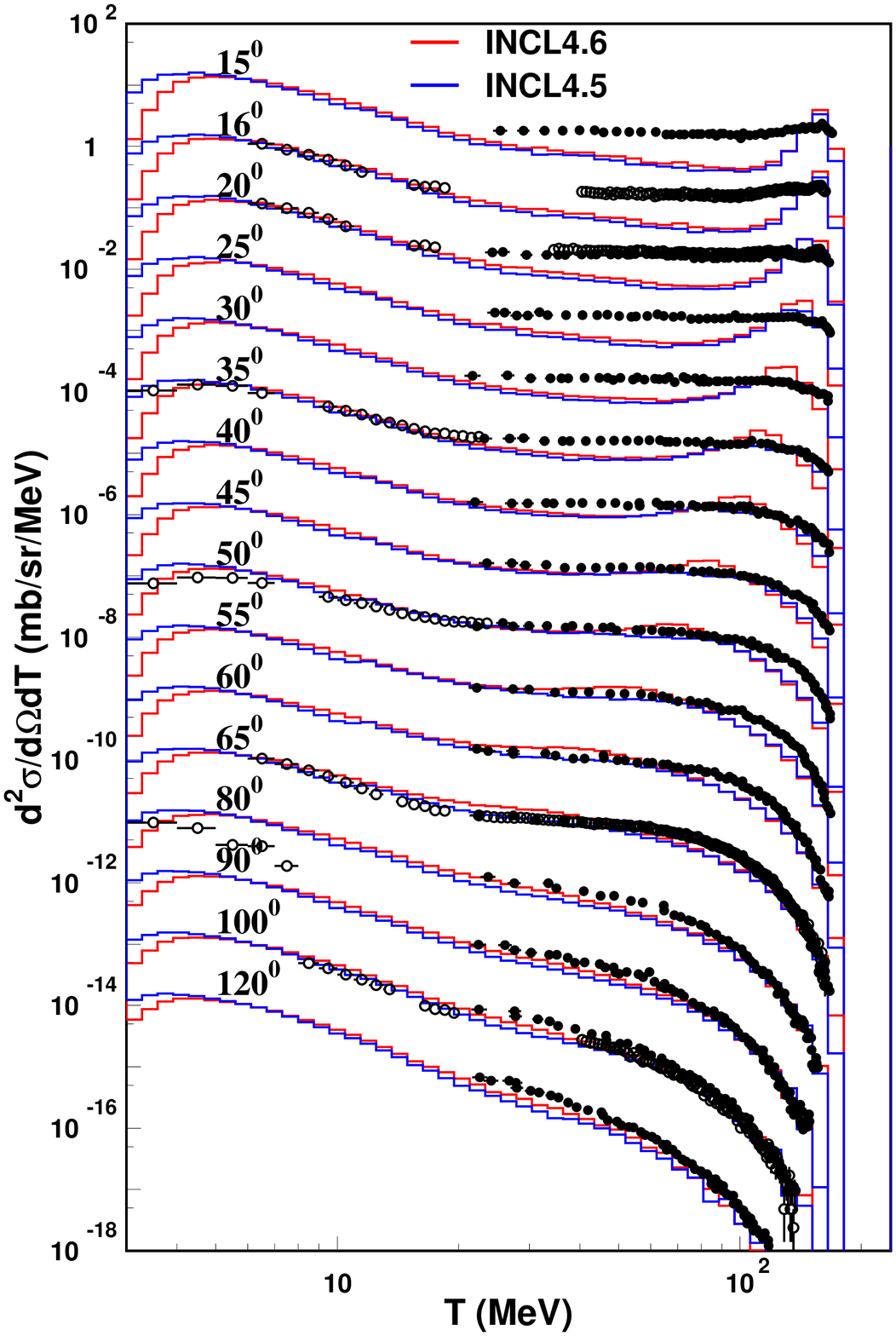}
\renewcommand{\baselinestretch}{0.5}
\end{minipage}
\newline
\begin{minipage}[t]{4cm}
\includegraphics[width=4cm]{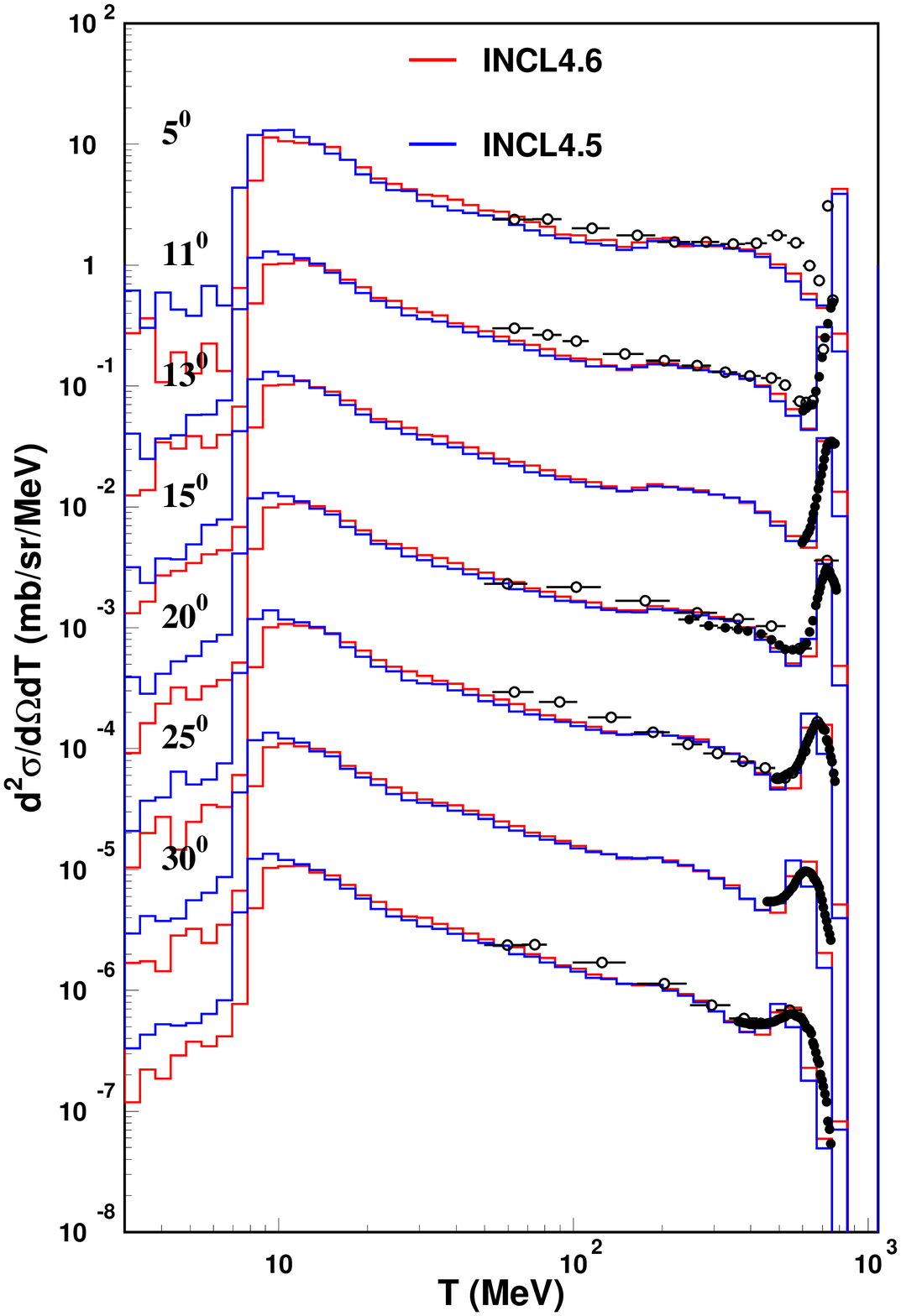}
\renewcommand{\baselinestretch}{0.5}
\end{minipage}
\hfill
\begin{minipage}[t]{4cm}
\includegraphics[width=4cm]{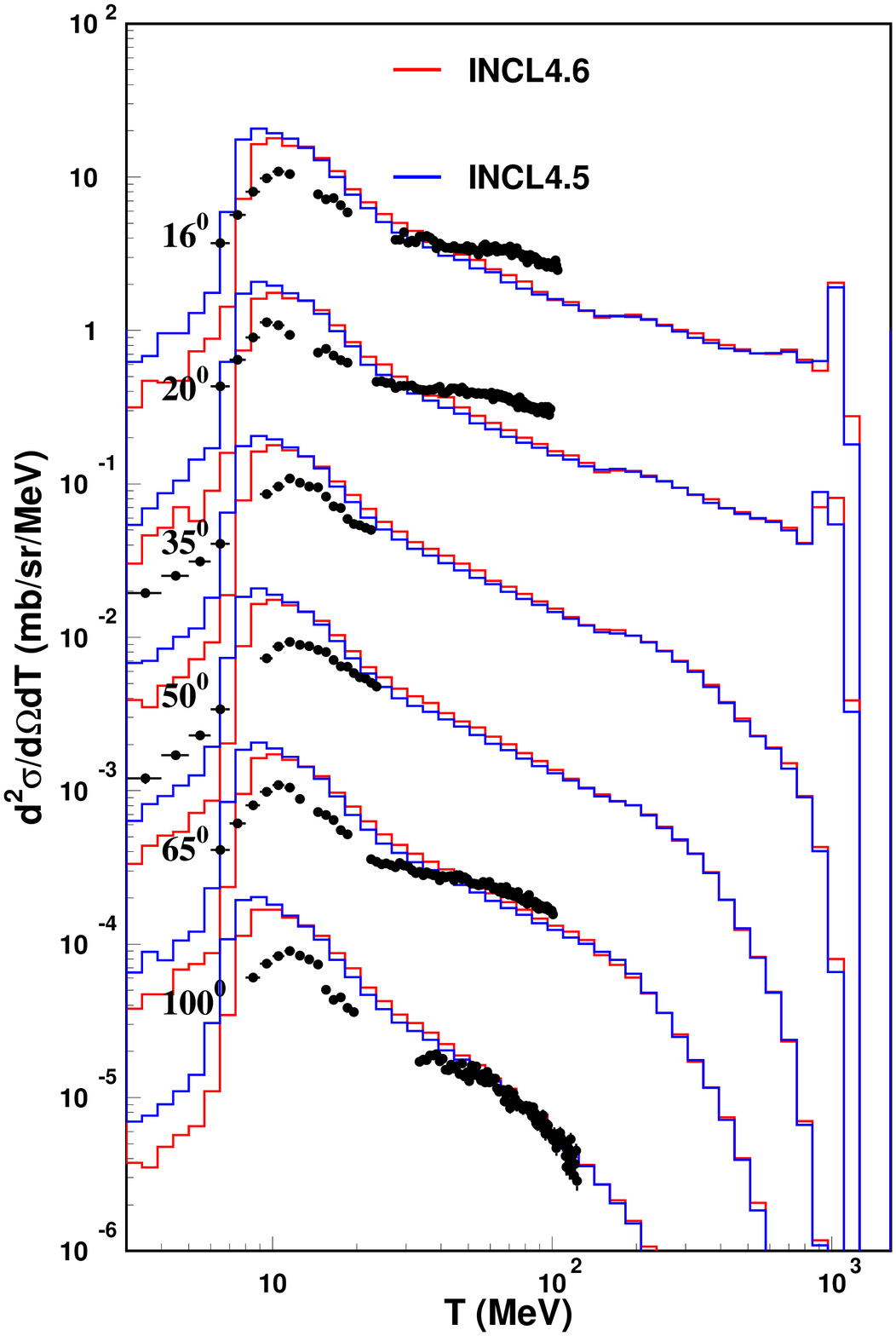}
\renewcommand{\baselinestretch}{0.5}
\end{minipage}

\caption{\small Upper row: comparison of INCL4.5  (blue) and  INCL4.6 (red) predictions with the experimental data (symbols) for the proton double differential cross sections  in $p$+$^{56}Fe$ collisions at 62 MeV (left panel, data from Refs.~\cite{BE73,EXFOR}) and in $p$+$^{58}Ni$ collisions at 175 MeV (right panel, data from Refs.~\cite{EXFOR,BU09}). Lower row: same for the proton double differential cross sections  in $p$+$^{208}Pb$ collisions at 800 MeV (data of Refs.~\cite{CH80,Mc84}, left panel) and  in $p$+$^{197}Au$ collisions at 1200 MeV (data of Ref.~\cite{BU08}, right panel).}
\renewcommand{\baselinestretch}{0.5}
\label{proton}
\end{figure}

\subsubsection{Proton energy spectra}

We spot some of our results in Fig.~\ref{proton}. Clearly, the kind of discontinuity observed in low energy proton spectra at 62 MeV incident kinetic energy and due  to the use of the ``back-to-spectators'' trick in INCL4.5, has disappeared in INCL4.6, as shown in the upper left panel. 

The INCL4.6 results are  globally somehow satisfactory. However,  systematic deficiencies can be detected. Cross sections are overpredicted for proton energy in the 10 to $\sim$ 50 MeV range at lowest angles (upper left panel), although results for heavier targets (not shown) are much better. This seems to come from an overestimate of the quasi-elastic component. Spectra at small angles and in the 40-100 MeV energy range are underpredicted at 175 MeV proton incident energy (upper right panel) and to a lesser extent at incident energy larger than 800 MeV (lower right panel). This defect seems to disappear around 800 MeV (lower left panel), where the agreement with experiment is remarkable. The origin of this discrepancy is not clear since it occurs in a region of the spectrum  neighbouring the quasi-elastic   and quasi-inelastic peaks. The latter  are believed to come from with a single NN collision, and  thus from events with a low number of collisions. It is thus surprising that our model, be it INCL4.5 or INCL4.6, is able to describe reasonably well the quasi-elastic  and quasi-inelastic peaks (at least above ~250 MeV) and not this few collision component, except around 800 MeV.

Let us also mention that our calculations overestimate the proton yield in the evaporation region at high incident energy (see lower right panel of Fig.~\ref{proton}). This may point to a deficient competition between the various particle emissions in the different versions of the ABLA model (or to an incorrect excitation energy of the target in INCL4.6). Let us however remind that the spectra of evaporated neutrons, which provides the dominant channel, in this energy range, are rather described by the ABLA model, as indicated by Fig.~\ref{neutron}.

We finally comment on the comparison with INCL4.2 (not shown here, see Ref.~\cite{BO02}). The predictions of the latter concerning proton spectra  are slightly better at high energy, with a less pronounced deficiency at small angles and intermediate proton energy. At low incident energy, say below 150 MeV, the shape of the spectra are slightly better (see Ref.~\cite{CU03}), but the normalizations are noticeably lower than those of experimental data, since the total reaction cross sections are badly underestimated, as noticed in Section \ref{react}.        

\subsubsection{Light cluster  energy spectra}

\begin{table}[t]
\caption{\it Average multiplicity of the clusters produced in the cascade and in the evaporation stages, for  two systems, as  given by the INCL4.6-ABLA07 model.}
\vspace{4mm}
\renewcommand{\baselinestretch}{1.5}
\begin{tabular}{|c|c|c|c|c|}
\hline
cluster & \multicolumn{2}{|c|}{ $p$ (1.2 GeV) + $^{197}Au$} & \multicolumn{2}{|c|}{ $p$ (63 MeV) + $^{208}Pb$} \\
\hline
 & casc  & evap & casc  & evap \\
\hline
$d$ &   0.505 & 0.166 & 0.0335 & 0.00035\\
$t$ &  0.124 & 0.081  & 0.0102 & 0.00011  \\
$^3He$ & 0.046 & 0.0067 & 0.00147  & 0.  \\
$^4He$ & 0.167 & 0.352 & 0.0141  & 0.0136  \\
\hline
\end{tabular}
\renewcommand{\baselinestretch}{0.5}
\label{multcl}
\end{table}
This is an important feature of INCL4.6, since INCL4.2 can not accommodate emission of  clusters during the cascade stage. We have made extensive calculations with our new  INCL4.6 model for double differential cross sections of $d$, $t$, $^3He$ and $\alpha$ production and compared with representative experimental data of Refs~\cite{BU08,FR90,CO96,BU07,LET02,BE73,BU09,GU05,HE06} covering a large incident energy range (62 MeV-2.5 GeV) and a target mass range spanning from $Al$ to $Bi$. Extensive results can be found in IAEA Intercomparison report \cite{IAEA}. See also Ref.~\cite{LE10} which contains results concerning excitation functions for production of $He$ and $H$ isotopes, using basically INCL4.5+ABLA. We just display here some typical cases. 

\begin{figure}[h]
\begin{minipage}[t]{4cm}
\includegraphics[width=4cm]{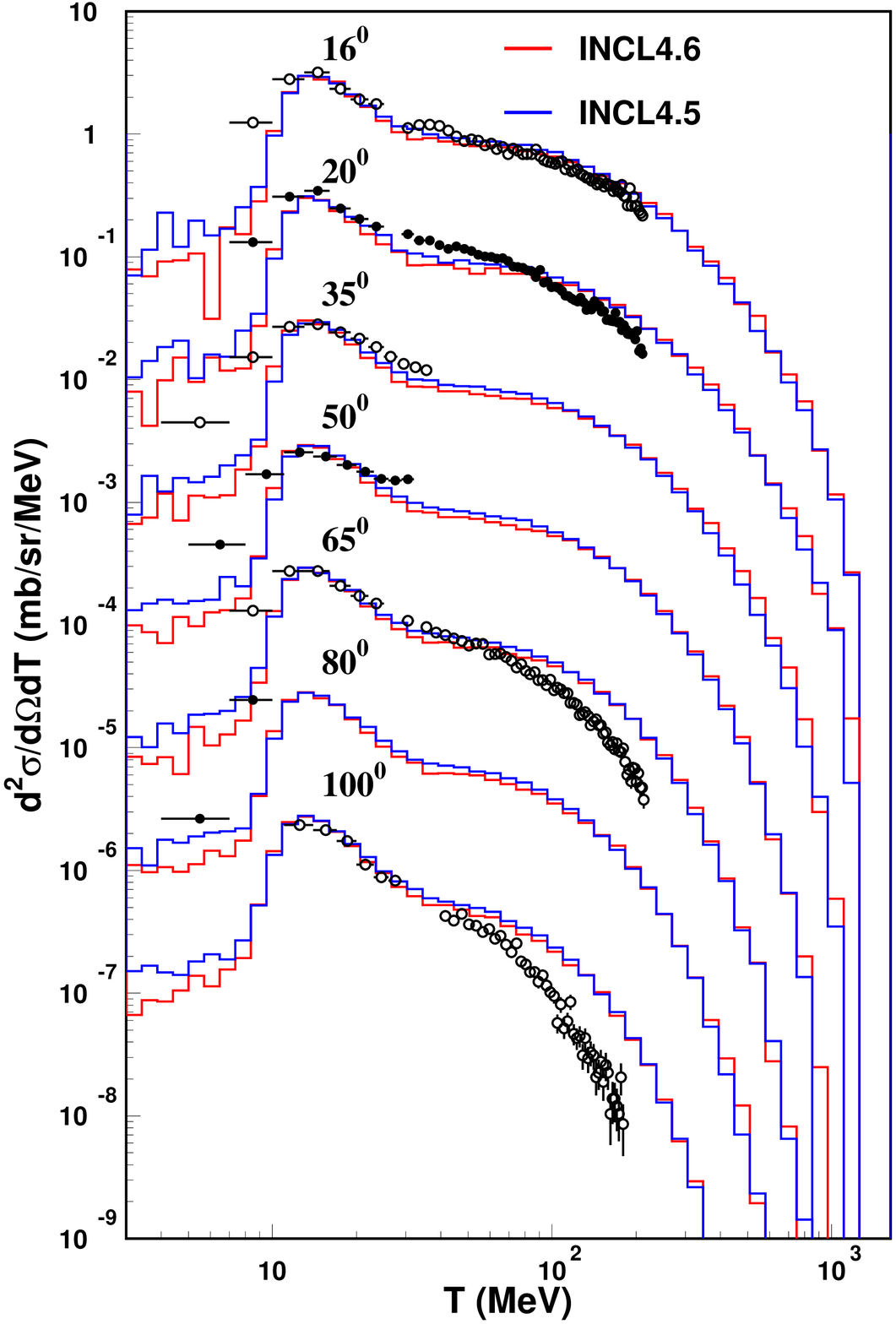}
\renewcommand{\baselinestretch}{0.5}
\end{minipage}
\begin{minipage}[t]{4cm}
\includegraphics[width=4cm]{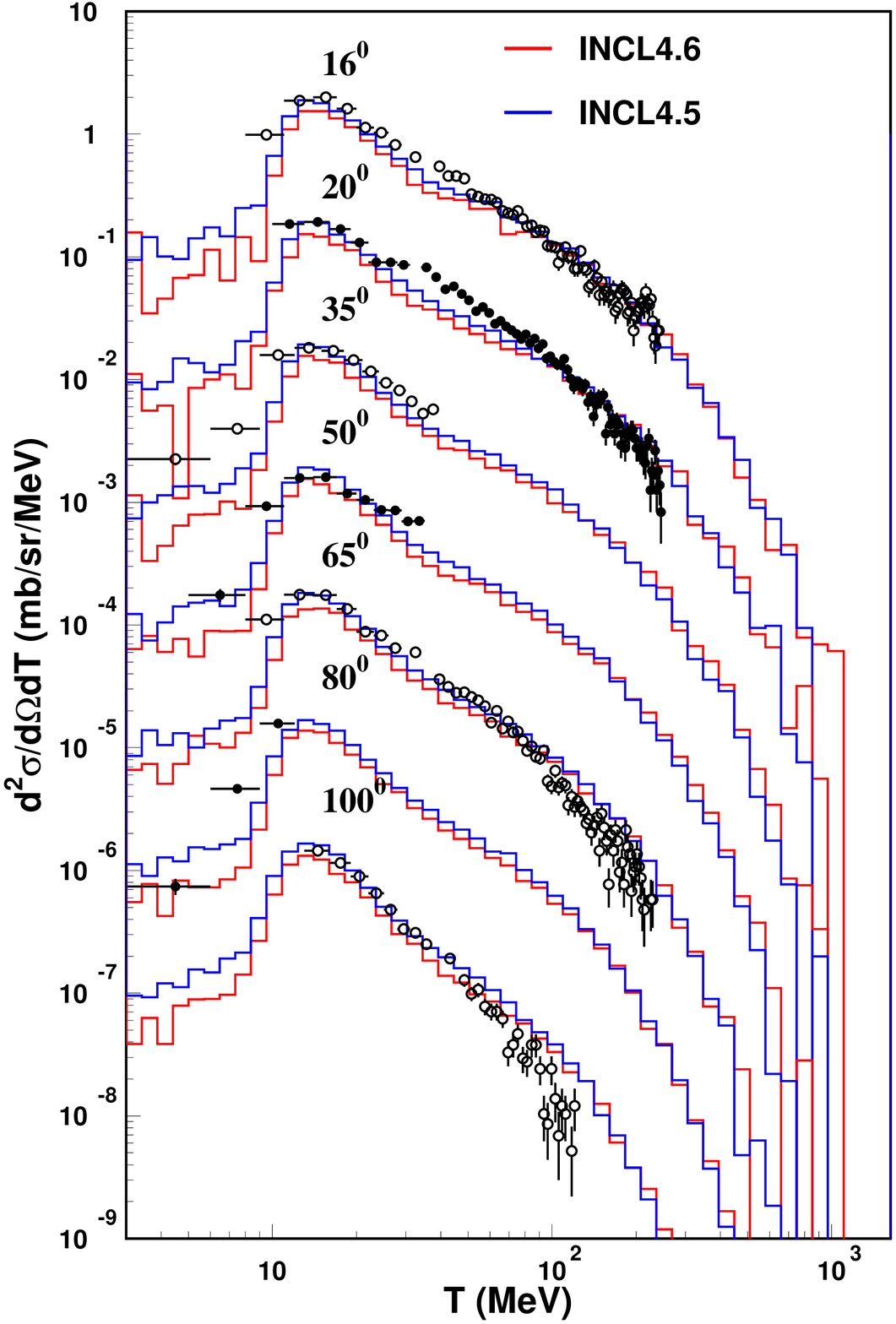}
\renewcommand{\baselinestretch}{0.5}
\end{minipage}
\newline
\begin{minipage}[t]{4cm}
\includegraphics[width=4cm]{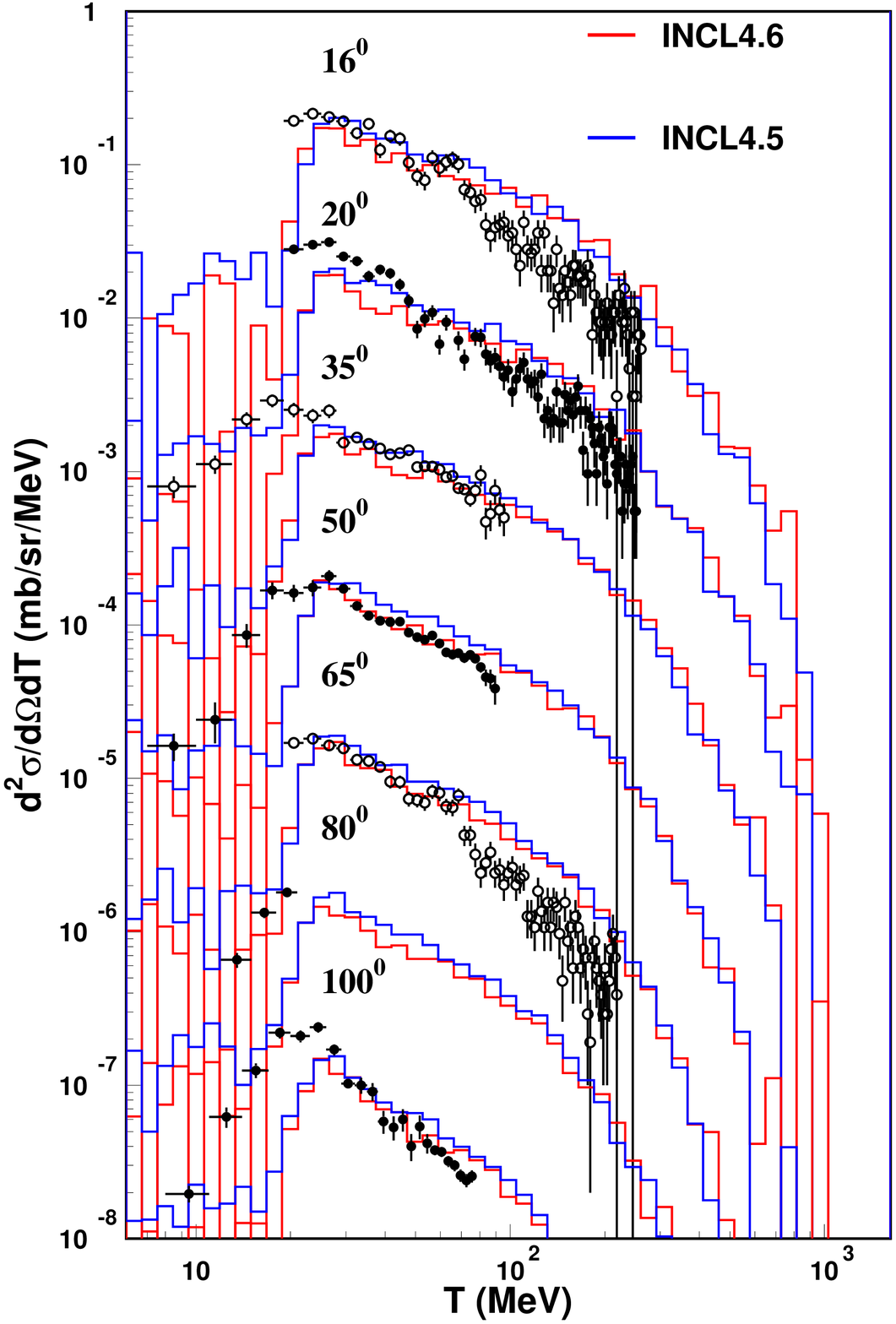}
\renewcommand{\baselinestretch}{0.5}
\end{minipage}
\begin{minipage}[t]{4cm}
\includegraphics[width=4cm]{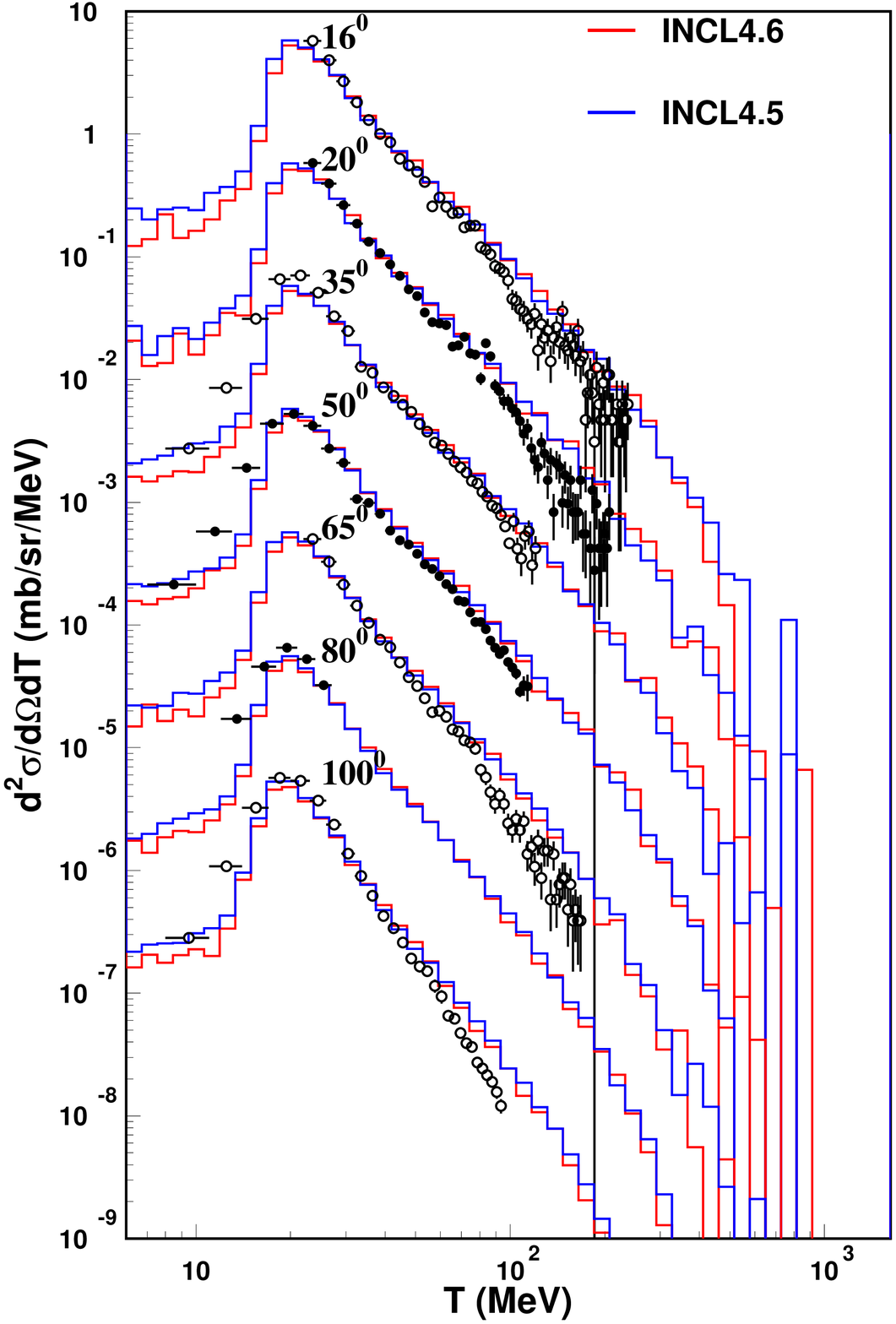}
\renewcommand{\baselinestretch}{0.5}
\end{minipage}
\hfill
\caption{\small Comparison of INCL4.5  (blue) and INCL4.6 (red) predictions for the double differential cross sections for deuteron (upper left), triton (upper right), $^{3}He$ (lower left) and $^{4}He$ (lower right) production in $p$+$^{197}Au$ collisions at 1200 MeV. Data from Ref.~\cite{BU08}.  }
\renewcommand{\baselinestretch}{0.5}
\label{cl1200}
\end{figure}

We first present results for  $p$+$^{197}Au$ collisions at 1200 MeV in Fig.~\ref{cl1200}. Globally, the agreement is fairly good, except for high energy deuterons at large angles. We remind that at this incident energy~\cite{BO04}, alpha particles are predominantly produced in the evaporation stage, $^3He$'s are dominantly produced in the cascade stage, deuterons and tritons are produced in both  stages, but definitely more in the cascade, as indicated in Table~\ref{multcl} (see also Ref.~\cite{LE10} for a discussion of these dominances and of their energy dependences). The impressive agreement between calculations and data for $\alpha$ production is  due to both INCL4.6 and ABLA07 models (evaporation yield is determined by the excitation energy left after the cascade). On the other hand the good description of the high energy tails of all particle spectra is solely due to our cascade model. There is no practical difference between the predictions of INCL4.6 and those of INCL4.5. A comparison with INCL4.2 itself is impossible as cluster production is not included in this model. However, a similar cluster production model (up to alpha particles) was implemented in the intermediate version~\cite{BO04}, refered later as INCL4.2 with clusters. The predictions of the latter are really satisfactory in the energy range (600-2500 MeV), in which it has been tested~\cite{BO04}, except for high energy $\alpha$ production, whose yield is underestimated by this intermediate version. Compared to INCL4.2 with clusters, INCL4.6 is significantly better on this latter point but the most
important improvement regards production at low incident energy and
production of heavier clusters (see Section \ref{ehc} below).

\begin{figure}[h]
\begin{minipage}[t]{4cm}
\includegraphics[width=4cm]{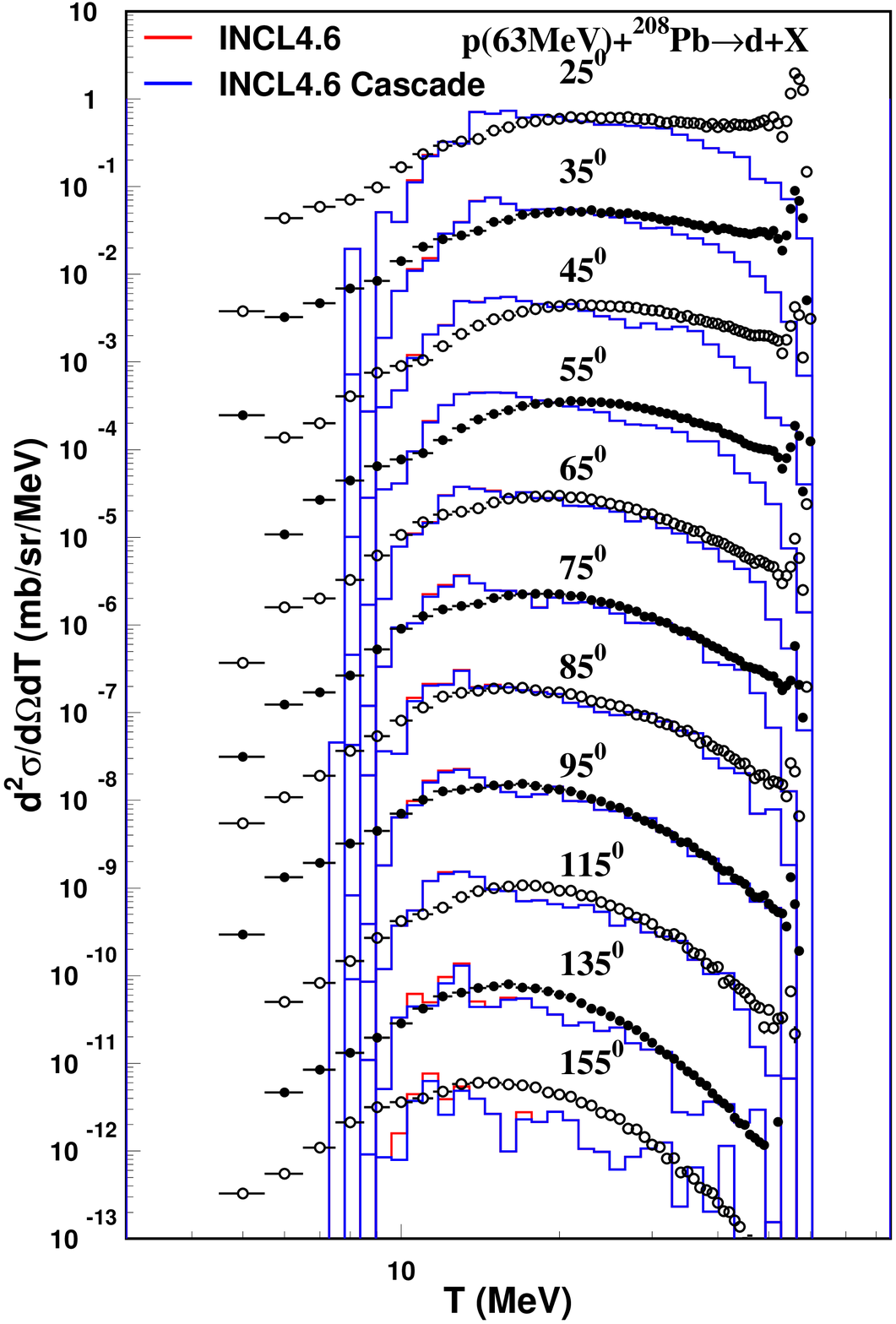}
\renewcommand{\baselinestretch}{0.5}
\end{minipage}
\hfill
\begin{minipage}[t]{4cm}
\includegraphics[width=4cm]{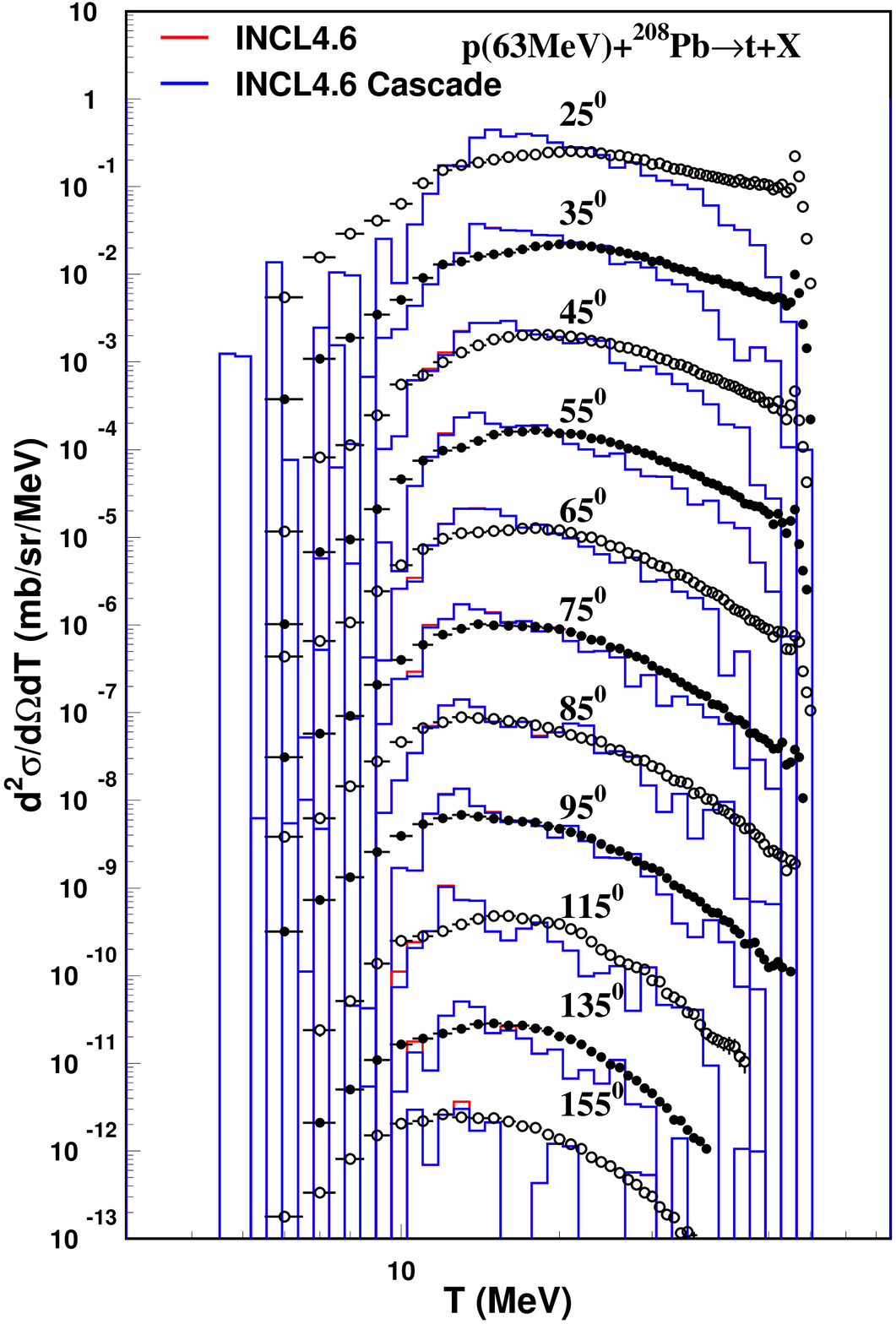}
\renewcommand{\baselinestretch}{0.5}
\end{minipage}
\newline
\begin{minipage}[t]{4cm}
\includegraphics[width=4cm]{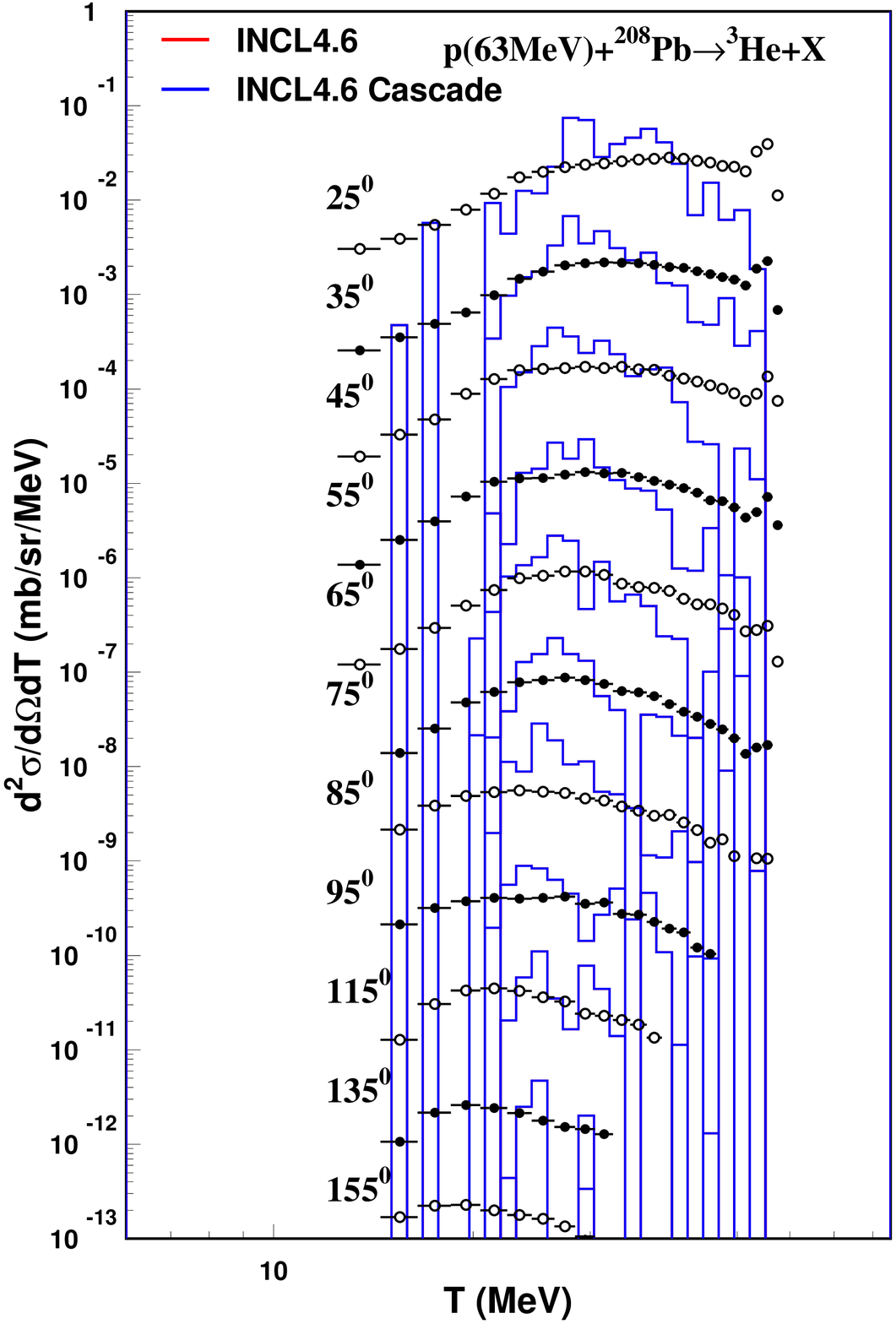}
\renewcommand{\baselinestretch}{0.5}
\end{minipage}
\hfill
\begin{minipage}[t]{4cm}
\includegraphics[width=4cm]{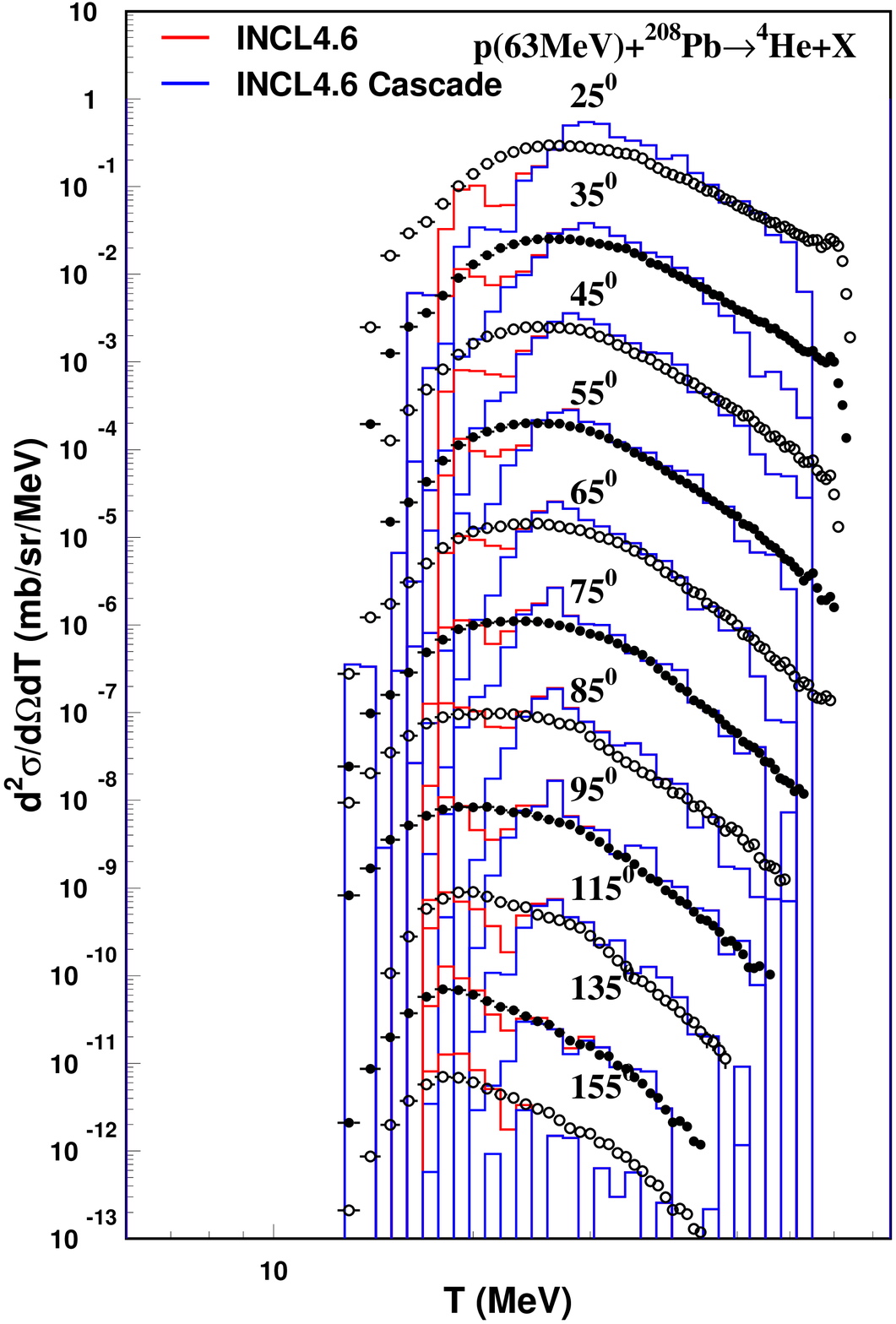}
\renewcommand{\baselinestretch}{0.5}
\end{minipage}

\caption{\small Comparison of INCL4.6 predictions (red) to the experimental data (symbols) for the double differential cross sections for deuteron (upper left), triton (upper right), $^{3}He$ (lower left) and $^{4}He$ (lower right) production in $p$+$^{208}Pb$  collisions at 63 MeV. The blue curves  give the cascade contribution (only the blue curve is visible when the blue and red results are identical).  Data from Ref.~\cite{GU05}.  }
\renewcommand{\baselinestretch}{0.5}
\label{cl63}
\end{figure}

Results for a slightly different system ($p$+$^{208}Pb$) at low energy are displayed in Fig.~\ref{cl63}. At such an incident energy, the average excitation energy of the remnant is very small, evaporation of charged particles is strongly hindered and  lcp's, even $\alpha$ particles, are predominantly produced in the cascade stage, as illustrated in Table \ref{multcl}. So the comparison with the data of Ref.~\cite{GU05} offers a severe test of INCL4.6 and especially of the cluster emission module. It can be said that, on the average, the model captures the magnitude and the shape of the spectra fairly well, although the agreement is not as good as at high energy. Of course, partial deficiencies can be seen. For $d$, $t$ and $^{3}He$ emission, cross sections are overestimated above  the expected value of the Coulomb barrier ($\sim$13 MeV for unit charge particles), and underestimated at energies close to the incident energy and small angles. The importance of this last deficiency can hardly be assessed, since this part of the particle spectra is partly coming from coherent processes, which are outside the scope of the cascade simulations. Finally, for  $\alpha$ emission, the height of the Coulomb barrier used in the cascade ($\sim$26 MeV) is definitely too large. It is larger than the Coulomb barrier height used in the ABLA07 version, around 15 MeV. The effect of this barrier is clearly seen in the $\alpha$ energy  spectra shown in Fig.~\ref{cl63} because the evaporation of $\alpha$ particles, though not important, is not negligible. These considerations explain the presence of a dip in the spectra around 23 MeV. Such a dip is not present or not visible in the spectra of the other light clusters.

We want to mention that the kind of agreement achieved in Fig.~\ref{cl63} crucially relies on criterion (Eq.~\ref{angle}). If this condition is removed, the yield of low energy clusters is badly overestimated.

We cannot multiply Figures here, due to lack of space, but we can say that more or less the same kind of agreement with experimental data is also reached for target nuclei as light as $^{27}Al$.

\subsubsection{Heavier cluster  energy spectra}
\label{ehc}
\begin{figure}[h]
\begin{minipage}[t]{4cm}
\includegraphics[width=4cm]{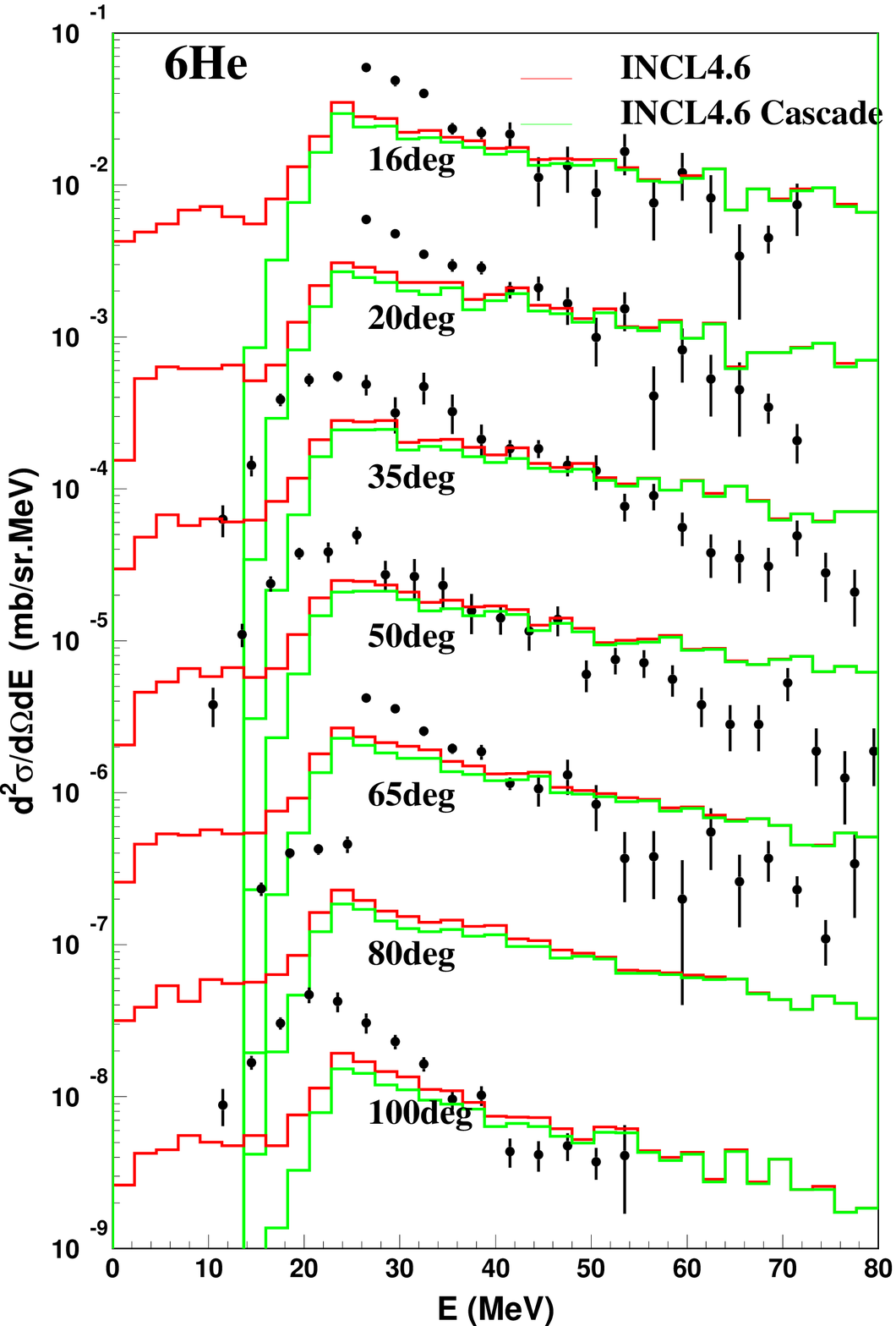}
\renewcommand{\baselinestretch}{0.5}
\end{minipage}
\begin{minipage}[t]{4cm}
\includegraphics[width=4cm]{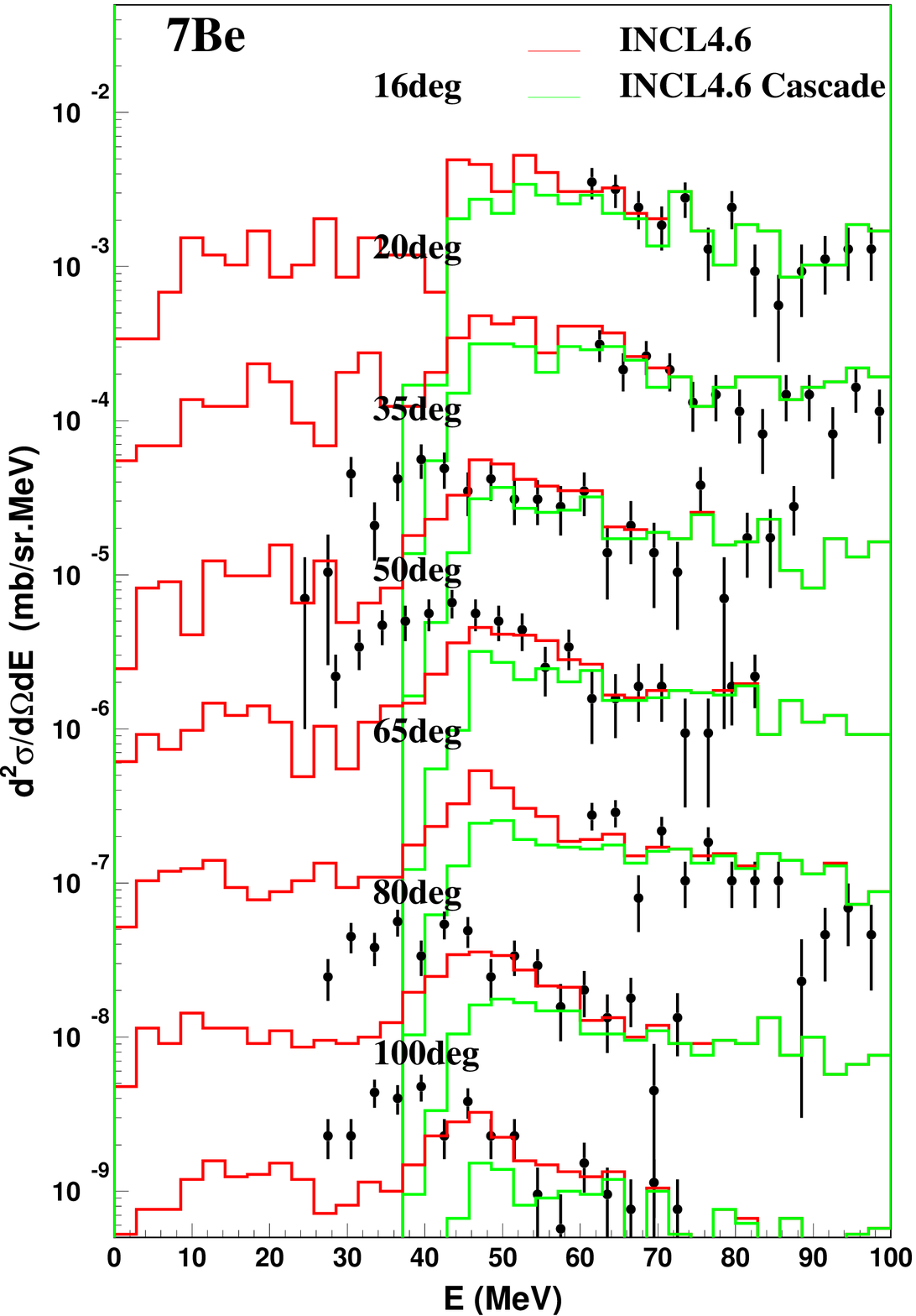}
\renewcommand{\baselinestretch}{0.5}
\end{minipage}

\caption{\small Double differential cross section for $^6He$ (left panel) and    $^7Be$  (right panel) cluster production  in $p$+$^{197}Au$ collisions at 1200 MeV. The predictions of INCL4.6 + ABLA07 (red histograms) are compared with the experimental data of Ref.~\cite{HE06} (black points) at different angles, indicated in the panels. The green histograms indicate the cascade component. }
\renewcommand{\baselinestretch}{0.5}
\label{hc}
\end{figure}

\begin{figure}[h]
\begin{minipage}[t]{4cm}
\includegraphics[width=4cm]{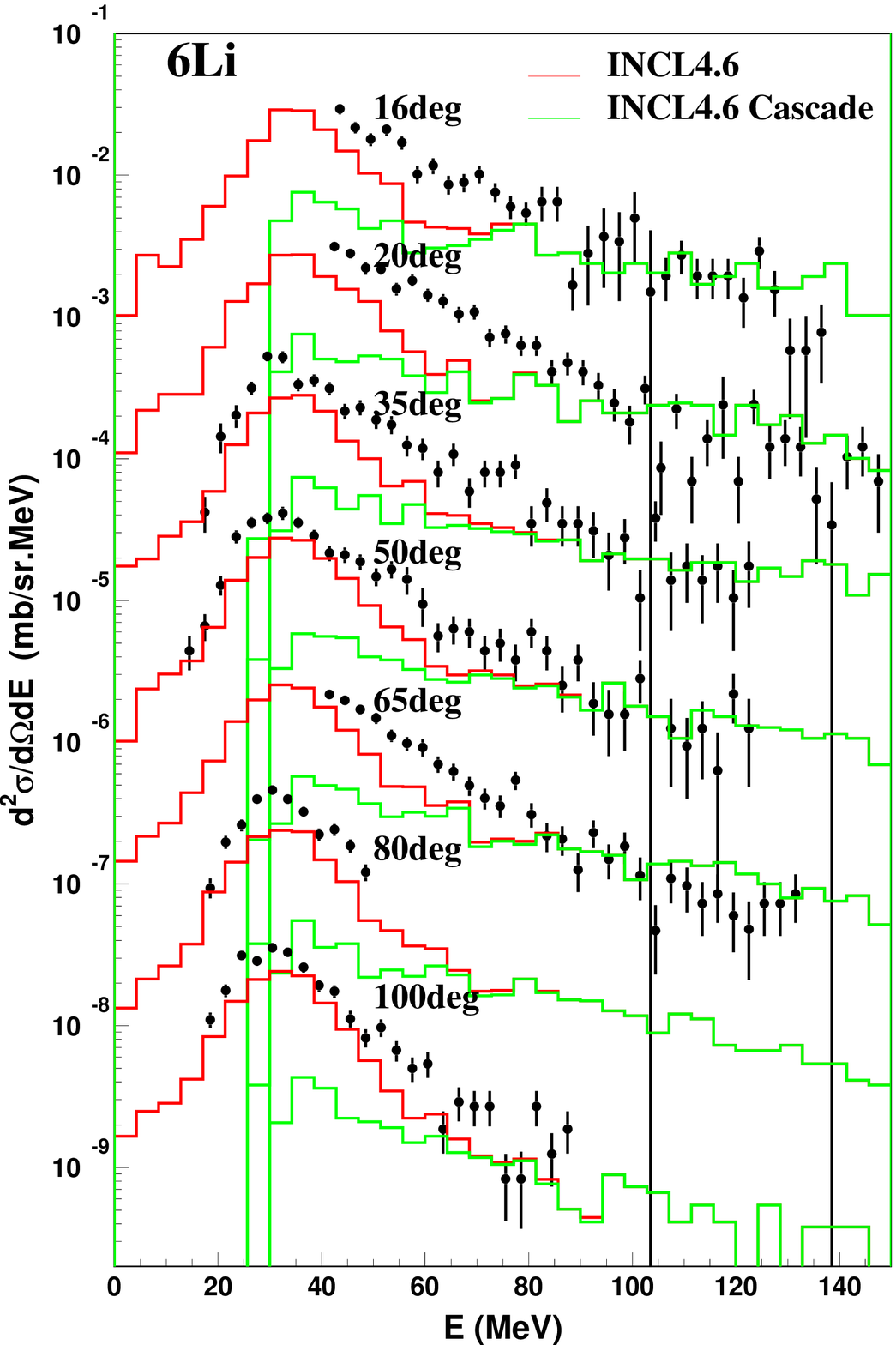}
\renewcommand{\baselinestretch}{0.5}
\end{minipage}
\begin{minipage}[t]{4cm}
\includegraphics[width=4cm]{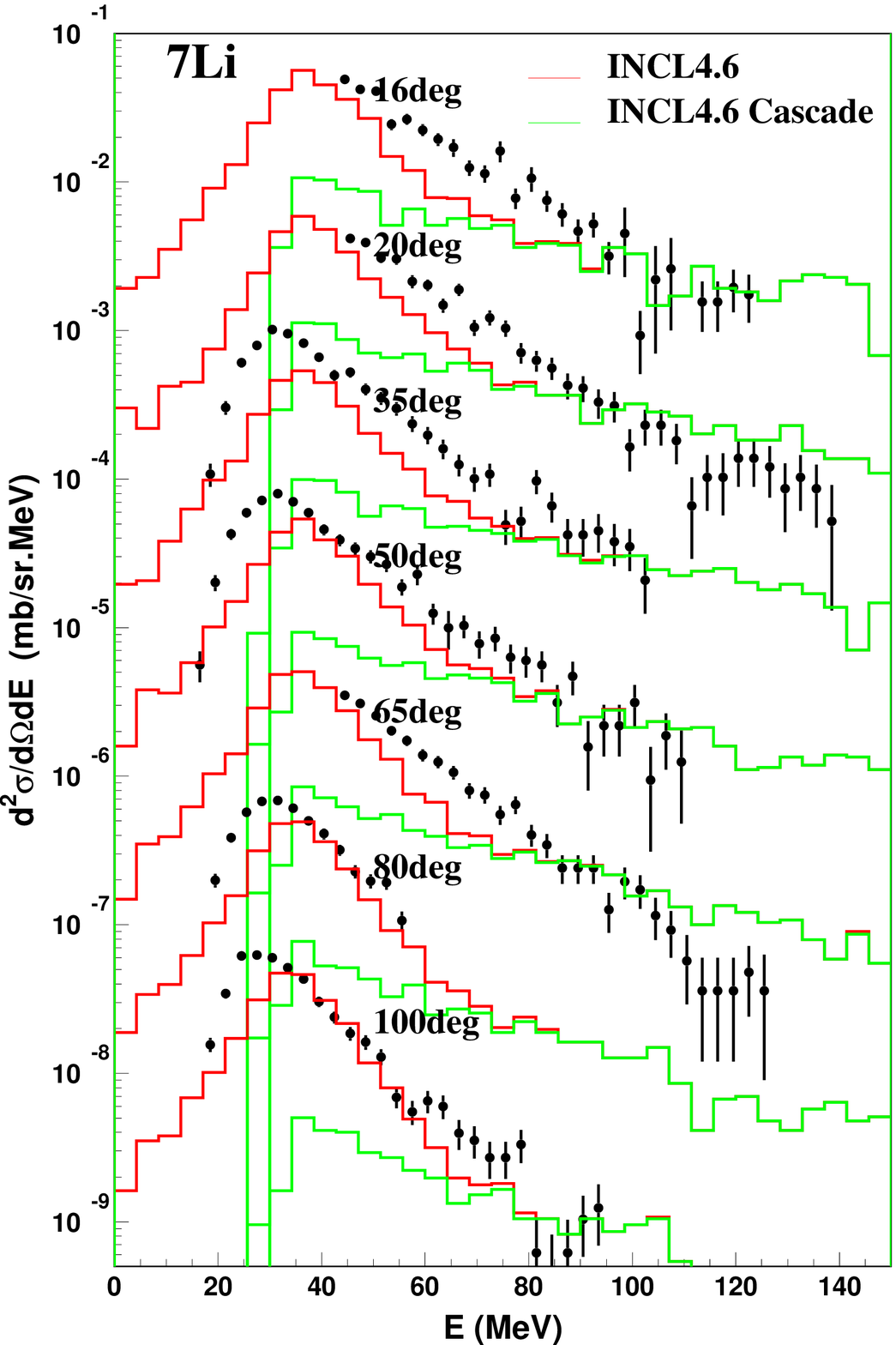}
\renewcommand{\baselinestretch}{0.5}
\end{minipage}

\caption{\small Same as Fig.~\ref{hc} for $^6Li$ and $^7Li$ production cross sections. }
\renewcommand{\baselinestretch}{0.5}
\label{hc2}
\end{figure}

This paragraph refers to emission of clusters heavier than alpha particles. The separation between light clusters ($A_{cl} \leq 4$) and heavier clusters may appear  arbitrary. However, if the production of light clusters by a sort of coalescence mechanism is more and more accepted, the one of heavier clusters by the same mechanism may appear doubtful. Here we want to point out some illustrative results obtained with the INCL4.6 + ABLA07 model, including the dynamical coalescence model described in Section~\ref{clmodule}. We show in Fig.~\ref{hc} the double differential production cross sections of $^6He$ and $^7Be$ clusters in $p$+$^{197}Au$  collisions at 1.2 GeV. One can see that our model reproduces rather well the magnitude of the cross sections and the shape of the spectra. Notice that this is not an obvious result since, for instance,  the  $^6He$ cross section is almost an order of magnitude larger than the $^7Be$ one. Fig.~\ref{hc} also shows that these ions  are  predominantly produced in the cascade, at least for kinetic energies larger than $\sim$20 MeV for $^6He$  and $\sim$30 MeV for  $^7Be$. The situation is  different for the production of $^6Li$ and $^7Li$, shown on Fig.~\ref{hc2}. These clusters are predominantly produced in the evaporation, but the large energy part of the spectra, say above 50 MeV, is explained by the production in the cascade. Notice that these results are less satisfactory than those displayed in Fig.~\ref{hc}, since our predictions seem systematically too low around 50-60 MeV. We have calculated  production cross sections for  other targets and other energies. Because of lack of space, results will not be given here and will be described in a future publication. Some partial results are contained in Ref.~\cite{CU11a}. Let us just mention here that our model yields similar results as those displayed in Fig.~\ref{hc}. However, at low incident energy ($\sim$200 MeV~\cite{MA06} or less), most of the energy spectra are too hard~\cite{DA11}. This matter is in progress. Nevertheless, our present results are rather unique in explaining the production of heavy clusters by a dynamic coalescence model. At least, they give credit to the plausibility of this mechanism, even if the details may not totally correspond to our coalescence model of Section~\ref{clmodule}.

\subsubsection{Pion  energy spectra}

\begin{figure}[h]
\begin{minipage}[t]{4cm}
\includegraphics[width=4cm]{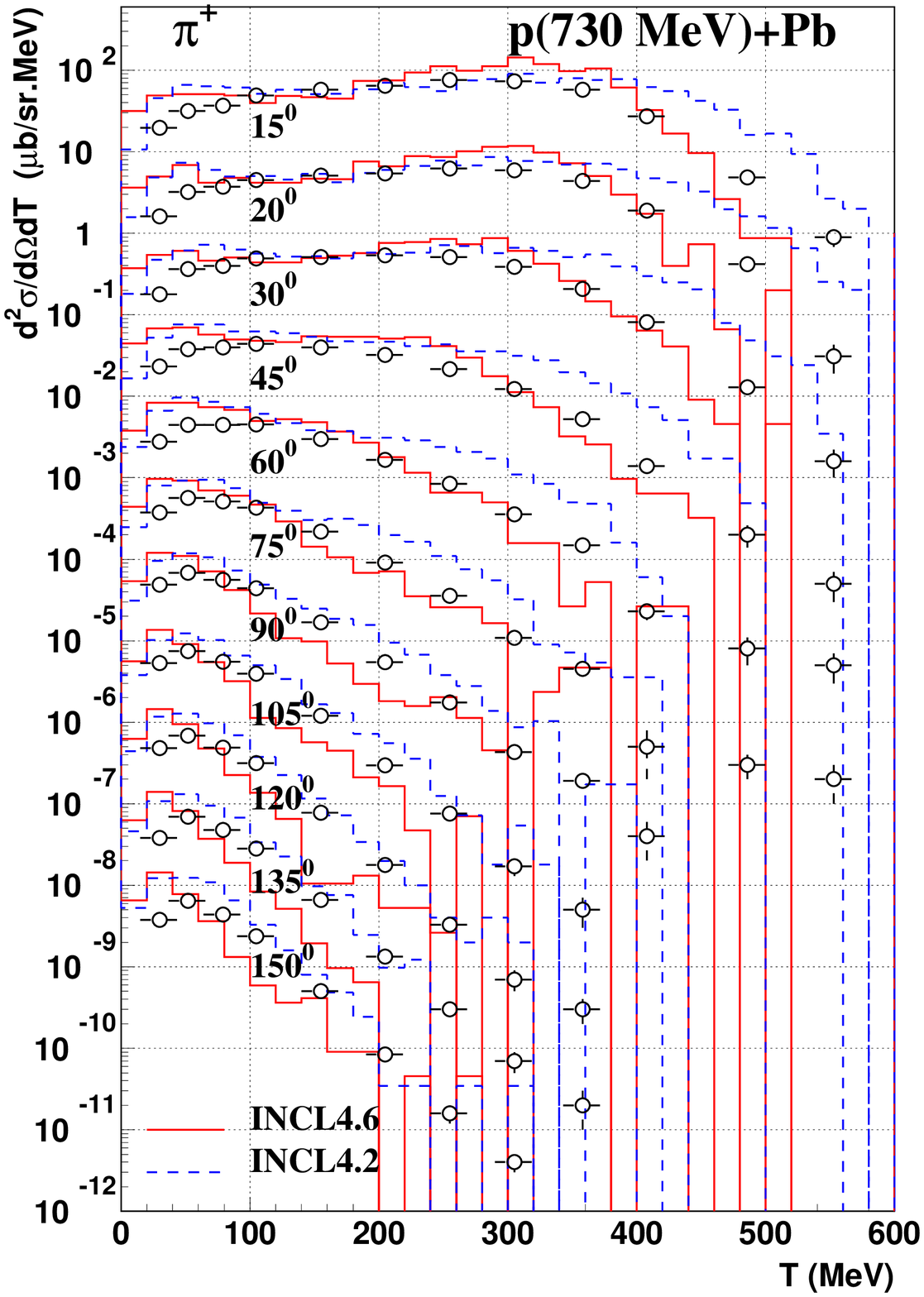}
\renewcommand{\baselinestretch}{0.5}
\end{minipage}
\hfill
\begin{minipage}[t]{4cm}
\includegraphics[width=4cm]{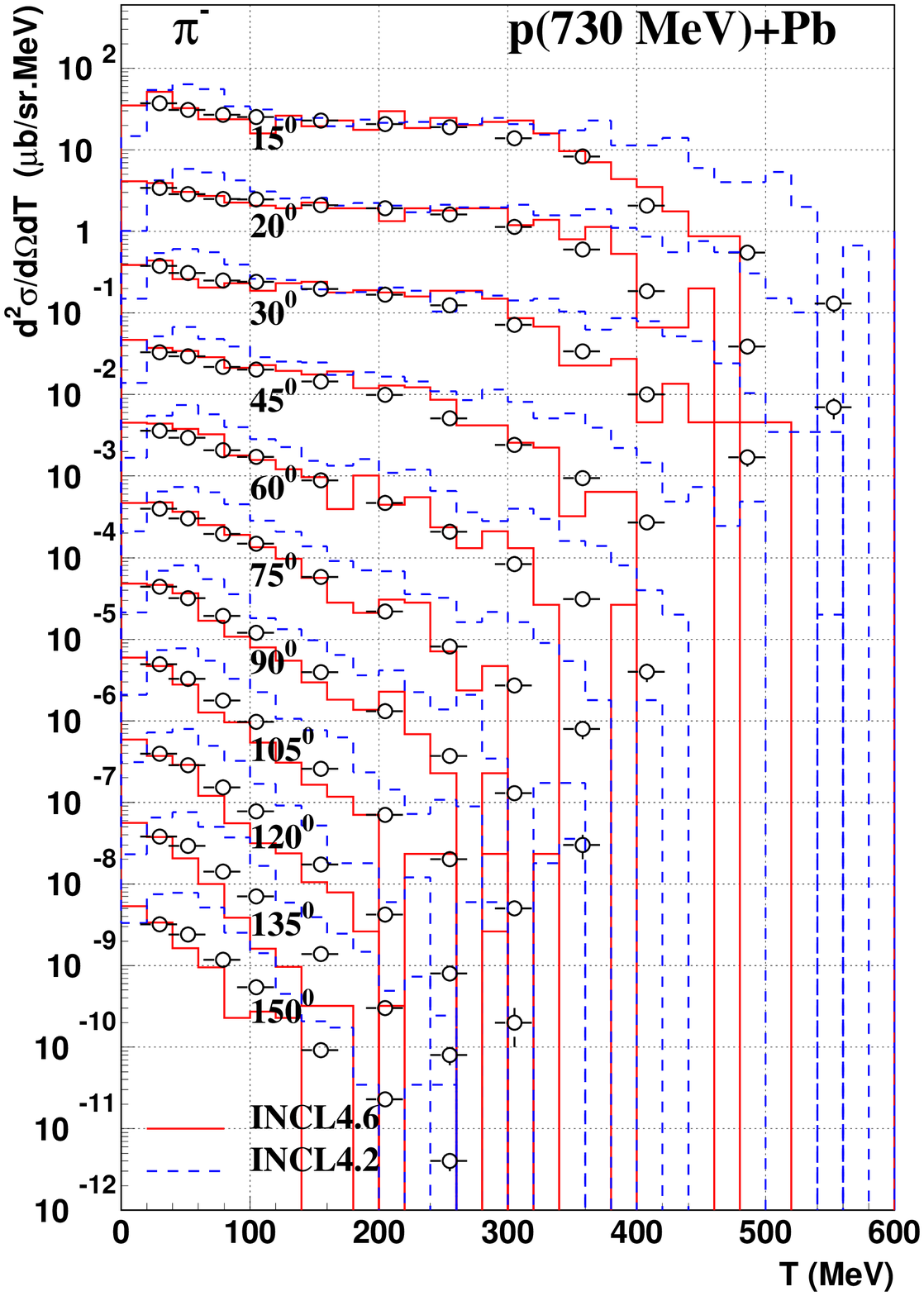}
\renewcommand{\baselinestretch}{0.5}
\end{minipage}

\caption{\small Comparison of INCL4.2 (dashed blue lines) and INCL4.6 (red lines) predictions with experimental data for inclusive $\pi^{+}$ (left panel) and $\pi^{-}$ (right panel) double differential cross sections  in $p$+$^{208}Pb$ collisions at 730 MeV. Data are from Ref.~\cite{CO72}. }
\renewcommand{\baselinestretch}{0.5}
\label{pion}
\end{figure}

We will not expand very much on this point, since it is of small direct importance for applications (note however that pion production involves $\Delta$ excitation, which influences sizably the energy flow at high incident energy). We present some results in Fig.~\ref{pion}. It can be seen that INCL4.6 is describing the data quite well, especially for negative pion production. These results present a real improvement compared to INCL4.2. The respective magnitudes of the $\pi^{+}$ and $\pi^{-}$  cross sections are well reproduced, indicating a correct isospin dependence of the elementary cross sections. The differences at low energy are coming rather from the introduction of the pion  potential well and the Coulomb barrier effects. Whereas the predictions are correct for $\pi^{-}$, an  overprediction of the $\pi^{+}$  cross sections appears slightly above the Coulomb barrier. This shortcoming was already present in INCL4.2, but is now significantly reduced.

\subsection{Residue production}
\subsubsection{Introduction}
Residue production properties are determined by both the cascade and the evaporation stages, whose respective influences cannot often be  easily  disentangled. Furthermore, the ABLA model  which have been used with INCL4.2 is not the same as the (new) ABLA07 model  which have been used with INCL4.6 (and INCL4.5).  This complicates the analysis. Therefore we will limit ourselves to a few significant results which may enlighten the merits and deficiencies of INCL4.6. More information can be found in Refs.~\cite{IAEA,CU10,IAEA2}.

\subsubsection{Residue mass and charge spectra}
\label{resmc}

\begin{figure}[h]

\includegraphics[width=8cm, height=10cm]{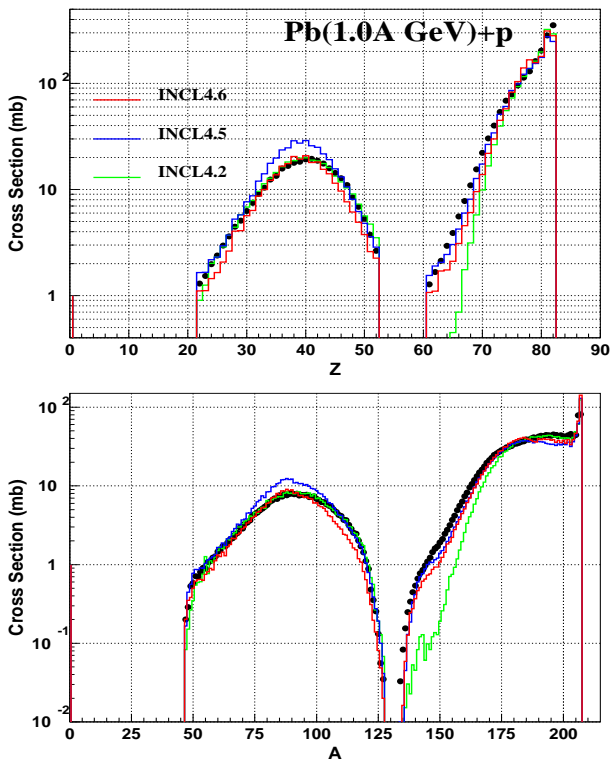}

\caption{\small Comparison of INCL4.2 (green lines), INCL4.5 (blue lines) and INCL4.6 (red lines) predictions with experimental data for residue mass spectrum  in $p$+$^{208}Pb$ collisions at 1 GeV. Data are from Ref.~\cite{EN01}. }
\label{massPb1}
\end{figure}

We present in Fig.~\ref{massPb1} the predictions of INCL4.2 and INCL4.6 for mass spectrum  with the experimental data concerning the illustrative case of $p$+$^{208}Pb$ at 1 GeV. The theoretical results have been improved on the low mass side of the so-called evaporation residue peak and have been deteriorated in the A=180-200 region and to a lesser extent in the fission peak. 

\begin{table}[t]
\caption{\it Average value of the mass number $A$, charge number $Z$, excitation energy $E^*$ (in MeV) and angular momentum $J$ (in $\hbar$ units) of the remnant in  $p$+$^{208}Pb$ collisions at 1 GeV, as calculated by different versions of INCL.}
\vspace{4mm}
\renewcommand{\baselinestretch}{1.5}
\begin{tabular}{|c|c|c|c|c|}
\hline
INCL version & ~$<A>$~ & ~$<Z>$~ & ~$<E^*>$~ & ~$<J>$~  \\
\hline
INCL4.2          & 202.73   & 80.51 & 135.9   &  16.81 \\
without clusters &   &  &   &  \\
\hline
INCL4.2          & 201.92   & 80.13 & 137.5   &  \\
with clusters    &   &  &   &  \\
\hline
INCL4.5          & 203.0    & 80.4  & 166.4   &  16.44 \\
\hline
INCL4.6          & 202.7    & 80.4  & 157.5   &  16.04 \\

\hline
\end{tabular}
\renewcommand{\baselinestretch}{0.5}
\label{remn}
\end{table}
We now try, for this typical case, to disentangle the changes brought by the modifications of the our cascade model from those introduced by the changes in the ABLA model. In the changes from INCL4.2 and INCL4.6, listed in Sections \ref{features4.5} and \ref{features4.6}, the ones that are the most relevant for the residue mass spectrum  are the introduction of cluster emission and the ``back-to-spectators'' trick. They, of course, influence the mass, charge, excitation energy and angular momentum of the remnants. To enlighten the discussion we give in Table~\ref{remn}, the average value of these quantities for the different versions of INCL. It can be seen that the introduction of cluster production slightly increases the average excitation energy and slightly decreases the average mass of the  remnant. The ``back-to-spectators'' trick produces, as expected, a strong increase of $<E^*>$ (even when this trick is ``corrected'' as in INCL4.6) and a slight increase of the average mass of the remnant. The increase of the excitation energy is likely at the origin of the better description of the mass spectrum at the low mass end of the so-called evaporation peak. It is presumably responsible for the underevaluation of the cross section in the A=185-200 region. However, the competition with fission may play a role as well in this region. The various changes introduced in the ABLA code~\cite{KE08,KE11} precludes a detailed determination of the respective influences of the model. Roughly speaking, one can say that the fission probability, for  excitation energy and fissility parameter typical of the system above, has decreased from ABLA to ABLA07. Further small changes in ABLA07 when coupled to INCL4.5 and INCL4.6 have increased the fission probability. Fig.~\ref{massPb1} seems to indicate that the fission probability is probably rather satisfactory and that the remaining lack of cross section in the A=185-200 region is due to other features.         

\begin{figure*}[t]

\begin{minipage}[t]{6cm}
\includegraphics[width=6cm, height=8cm]{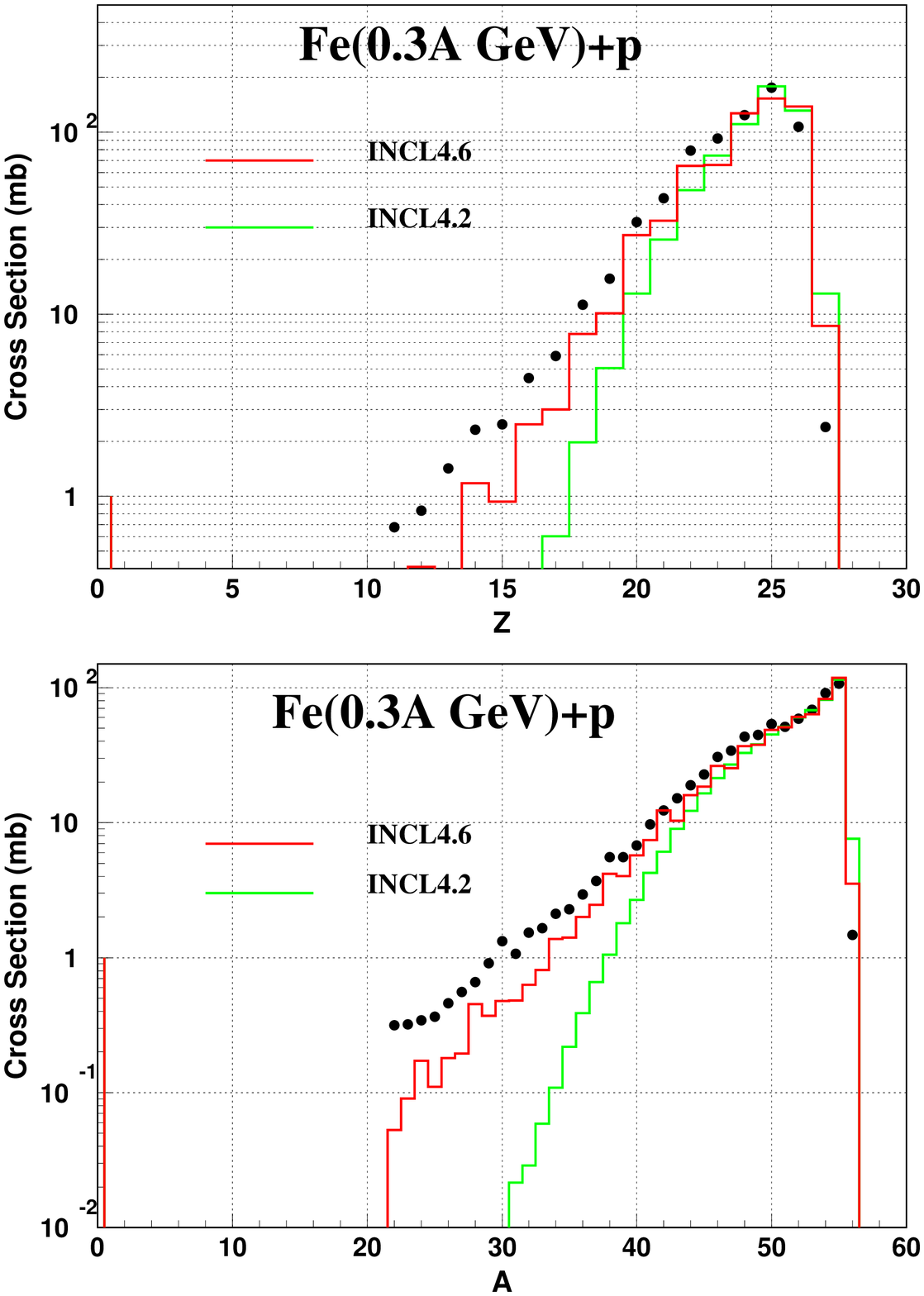}
\renewcommand{\baselinestretch}{0.5}
\end{minipage}
\hspace{1cm}
\begin{minipage}[t]{6cm}
\includegraphics[width=6cm, height=8cm]{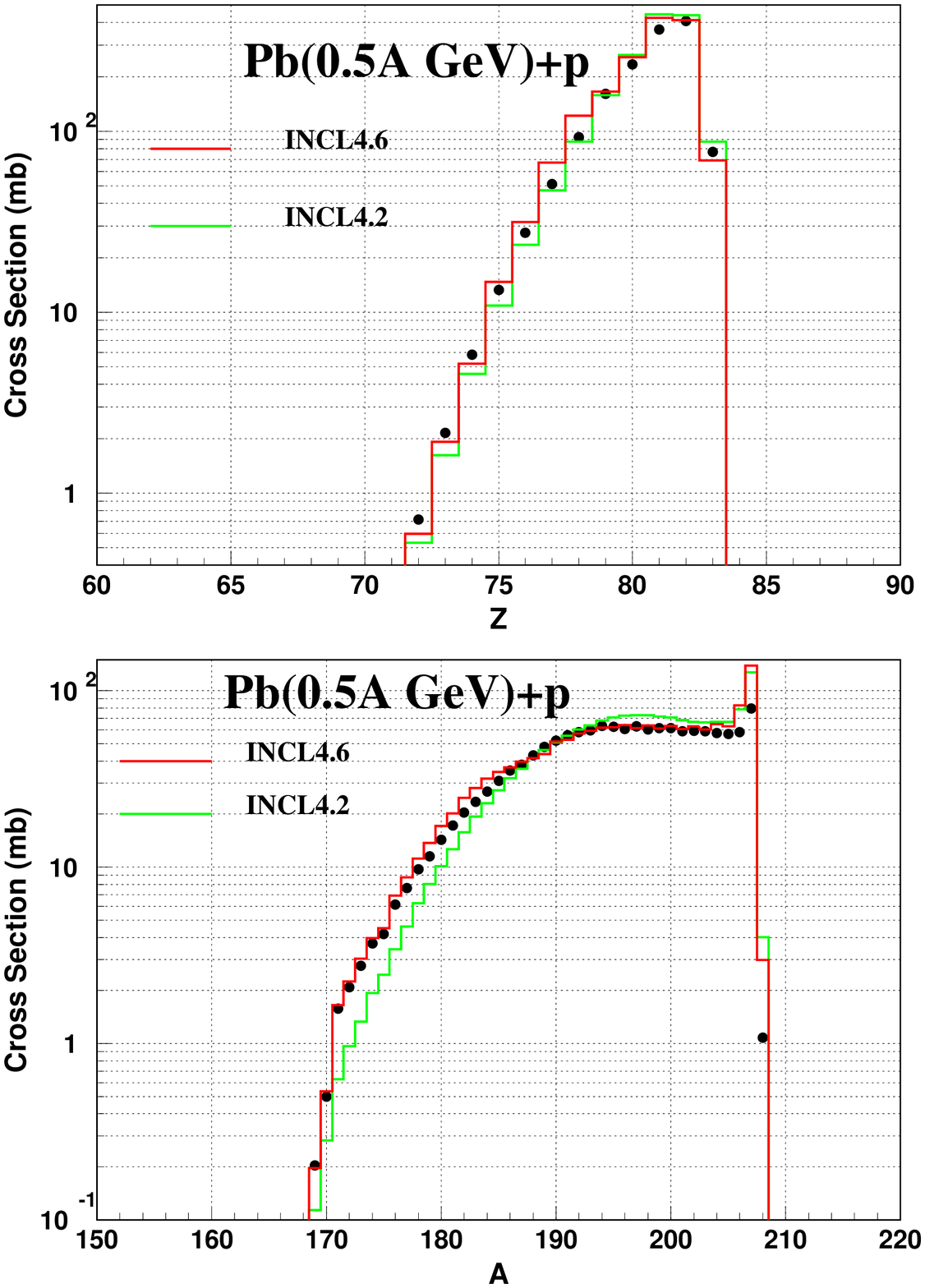}
\renewcommand{\baselinestretch}{0.5}
\end{minipage}
\\
\begin{minipage}[t]{6cm}
\includegraphics[width=6cm, height=8cm]{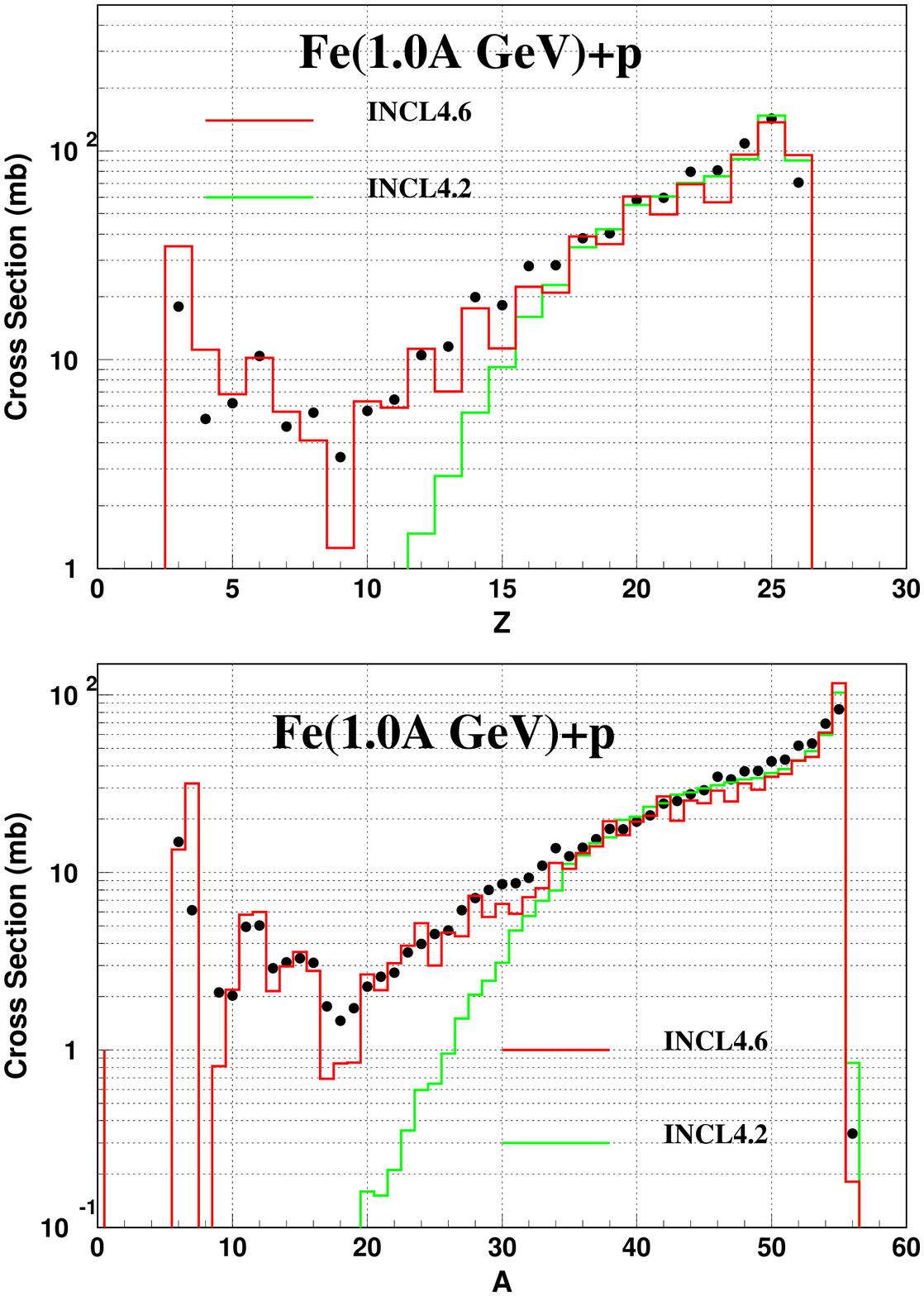}
\renewcommand{\baselinestretch}{0.5}
\end{minipage}
\hspace{1cm}
\begin{minipage}[t]{6cm}
\includegraphics[width=6cm, height=8cm]{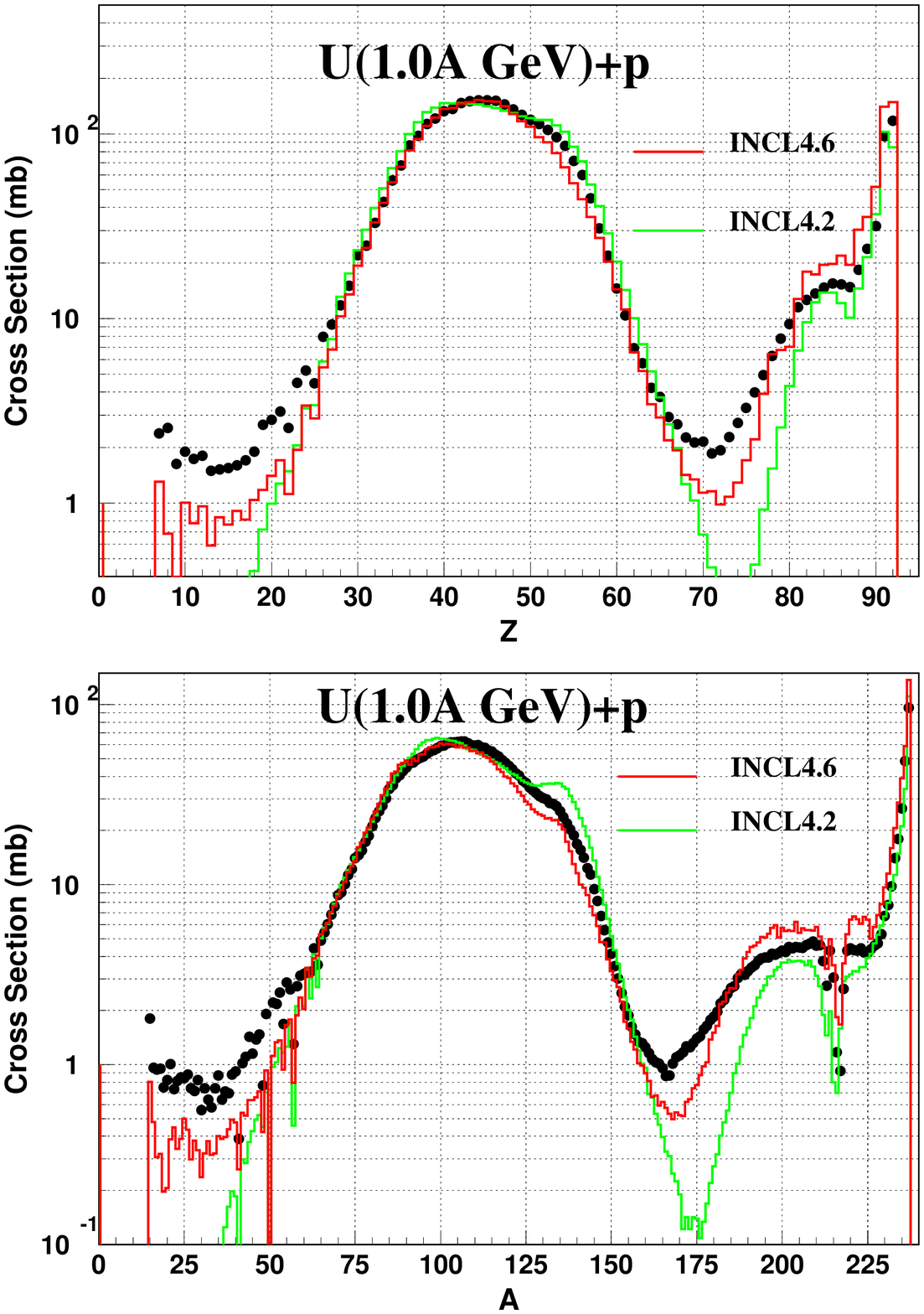}
     \renewcommand{\baselinestretch}{0.5}
\end{minipage}
\caption{\small Comparison of INCL4.2-ABLA (green lines) and INCL4.6-ABLA07 (red lines) predictions with experimental data for residue charge (1st and 3rd rows) and mass (2d and 4th rows) spectra  in $p$+$^{56}Fe$  at 300 MeV (upper left quarter), $p$+$^{208}Pb$ at 500 MeV (upper right quarter), $p$+$^{56}Fe$ at 1 GeV (lower left quarter) and  $p$+$^{238}U$ at 1 GeV (lower right quarter), respectively. Data are from Refs.~\cite{VI07,FE05,AU06,TA03,BE03,BE06,RI06}. }
\renewcommand{\baselinestretch}{0.5}
\label{masses}
\end{figure*}

For the sake of illustration, we give the charge and mass spectra for other systems in Fig.~\ref{masses}. One can see that the predictions of the INCL4.6-ABLA07 model are in a good global agreement with the experimental data. We comment on the remaining discrepancies. The so-called deep spallation products are understimated  in $p$+$^{56}Fe$  at 300 MeV. This is also the case for the A$\sim$15 residues in  the $p$+$^{56}Fe$  at 1 GeV  and for the  A$\sim$160-170 residues in the $p$+$^{238}U$ system at 1 GeV. This might indicate an underestimation of the frequency of high excitation events. But this may as well indicate an underevaluation of the emission of the large clusters in the de-excitation. At least this seems to be corroborated, for the $p$+$^{238}U$ system, by the underpredicted yield of the A$\sim$20-40  residues, since this kind of residues are expected to be produced by evaporation (beyond multifragmentation, which very likely is not operative in this system~\cite{MA11}). 

We have to underline the almost perfect description of the mass spectrum in $p$+$^{208}Pb$ collisions at 500 MeV. Only the A=207 yield is overestimated. 

Let us comment on the fission fragment distribution in the $p$+$^{238}U$ system. Although the magnitude and the width of the fission peak are well reproduced, some features are not satisfactory. On the one hand, the shape of the fission peak is not so well accounted for, probably signalling inappropriate shell effects. On the other hand, the plateau in the A=200-225 region is somehow overestimated, pointing probably to an imperfect evaporation-fission competition at large excitation energy\footnote{The dip around A=214 is due to very short-lived $\alpha$-emitters which escape detection.}.

Finally, let us notice that odd-even effects are too large in  $p$+$^{56}Fe$,  presumably due to a too large pairing used in ABLA07.

\begin{figure*}[t]

\begin{minipage}[t]{7.5cm}
\includegraphics[width=7.5cm, height=10cm]{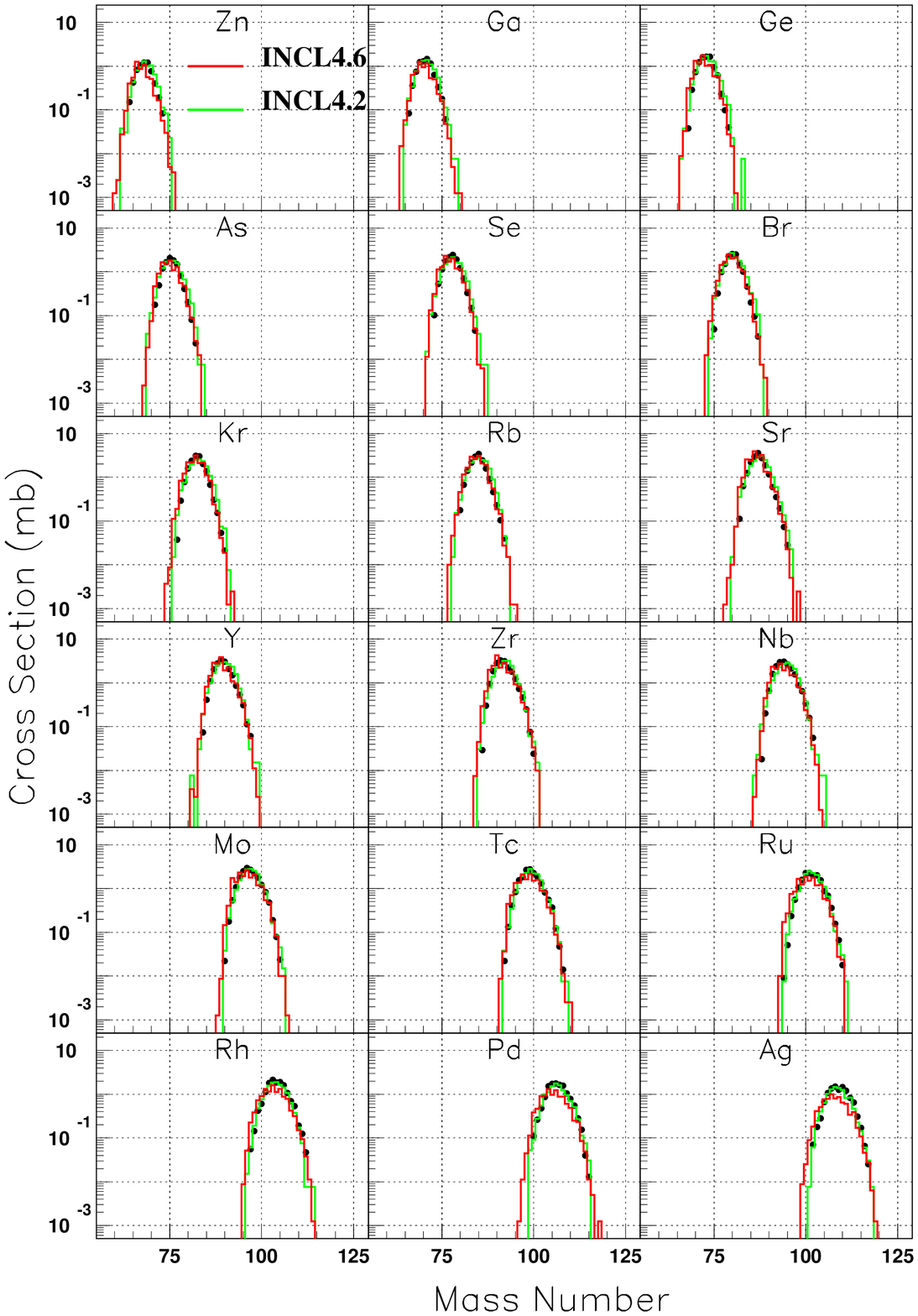}
\renewcommand{\baselinestretch}{0.5}
\end{minipage}
\hfill
\begin{minipage}[t]{7.5cm}
\includegraphics[width=7.5cm, height=10cm]{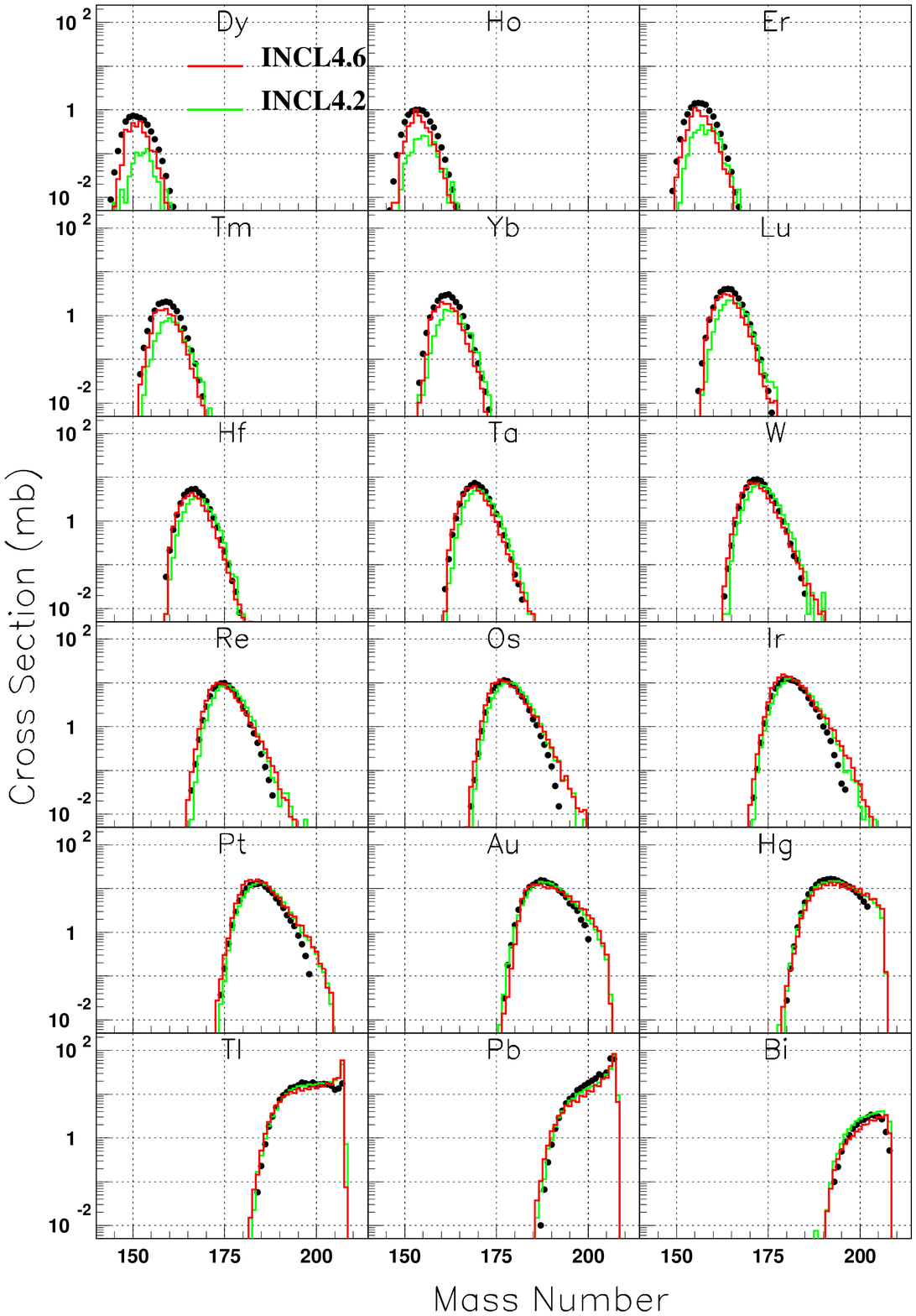}
\renewcommand{\baselinestretch}{0.5}
\end{minipage}
\hfill

\caption{\small Comparison of INCL4.2-ABLA (green lines) and INCL4.6-ABLA07 (red lines) predictions with experimental data, for so-called isotropic distributions in $p$+$^{208}Pb$ collisions at 1 GeV. The left panel shows typical isotopes produced as fission fragments and the right panel show typical evaporation residues. Data are from Ref.~\cite{EN01}. }
\renewcommand{\baselinestretch}{0.5}
\label{diso}
\end{figure*}

As for the isotopic distributions, we  restrict ourselves to show our results for the illustrative case of $p$+$^{208}Pb$ at 1 GeV in Fig.~\ref{diso}. The shapes of these distributions for fission products are very well reproduced by both  INCL4.2-ABLA and  INCL4.6-ABLA07 models, with a slightly better performance for the first model. The isotopic distributions for evaporation residues (right panel)  are well reproduced by the INCL4.2-ABLA model, except for the overall yield for the lightest ones. On the contrary, the INCL4.6-ABLA07  model does not suffer from this disease, but some deficiencies appear in the shapes of the distributions for the residues close to the target: the $Pb$ isotopic distribution is underestimated around A=200, the $Os$ to $Hg$ ($Z=Z_{T}-6$ to $Z=Z_{T}-2$) distributions  are overestimated on the neutron-rich side and the $^{207}Tl$ ($Z=Z_{T}-1, A=A_{T}-1$) production cross section is sizably overestimated (this was also the case in INCL4.2). These deficiencies seem to come from a too low excitation energy for those remnants that are located on the neutron-rich border of the so-called evaporation corridor (i. e. the region of the ($N,Z$) plane populated by the residues), close to the target. We have not been able to convincingly relate these deficiencies to the change of a specific parameter or to the removal of a specific hypothesis. Let us notice that the INCL4.6-ABLA07 model reproduces rather nicely the distribution for the $Bi$ isotopes ($Z=Z_{T}+1$) produced through $(p,xn)$ reactions.

Similar deficiencies occur in the other systems referenced in Fig.~\ref{masses} for isotopes close to the target, although these deficiencies have a smaller amplitudes in the $p$+$^{56}Fe$ system at 1~GeV and they are almost vanishing in the same system at 300 MeV.

\begin{figure}[h]

\begin{minipage}[t]{4cm}

\includegraphics[width=4cm, height=4cm]{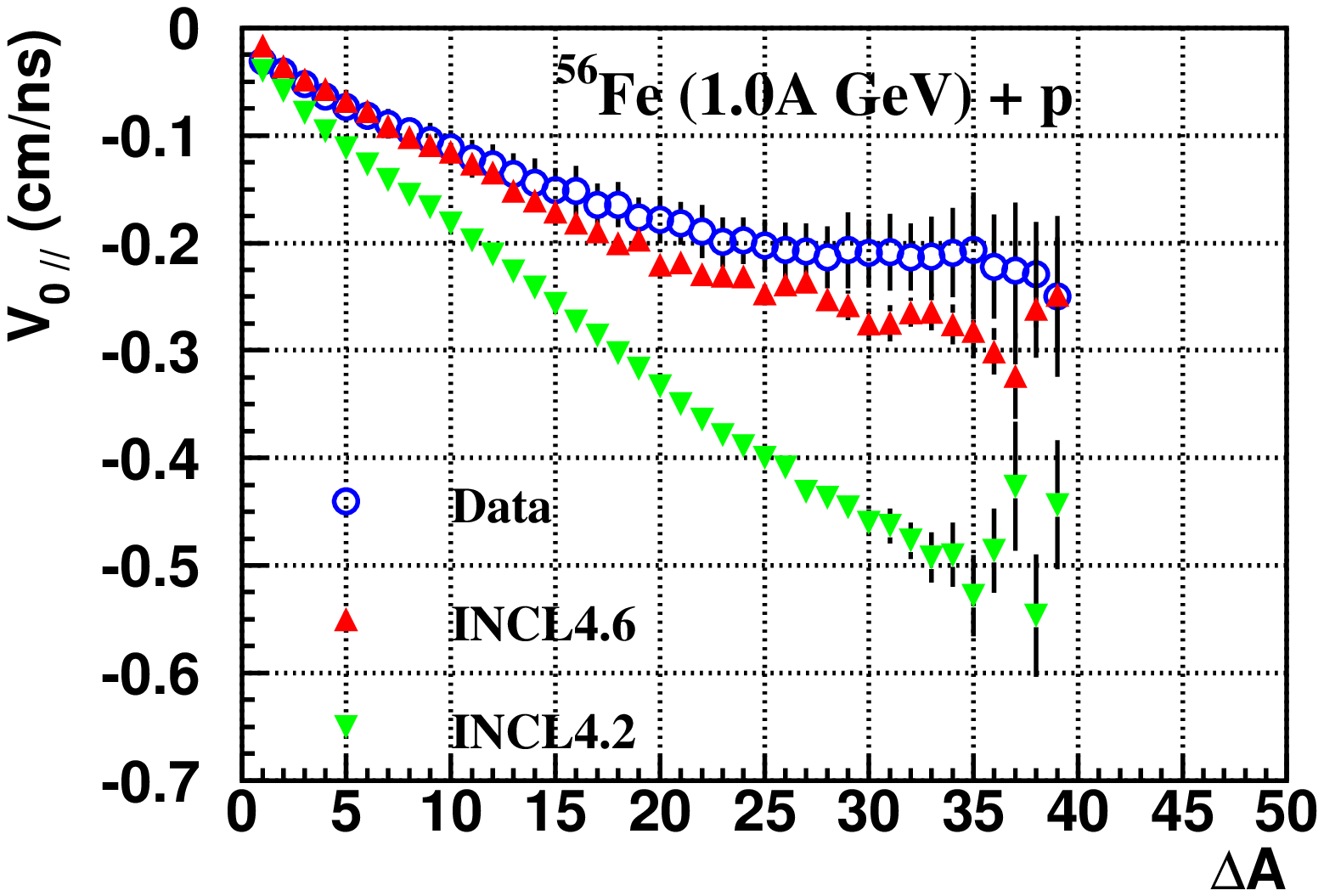}
\renewcommand{\baselinestretch}{0.5}
\end{minipage}
\hfill
\begin{minipage}[t]{4cm}
\includegraphics[width=4cm, height=4cm]{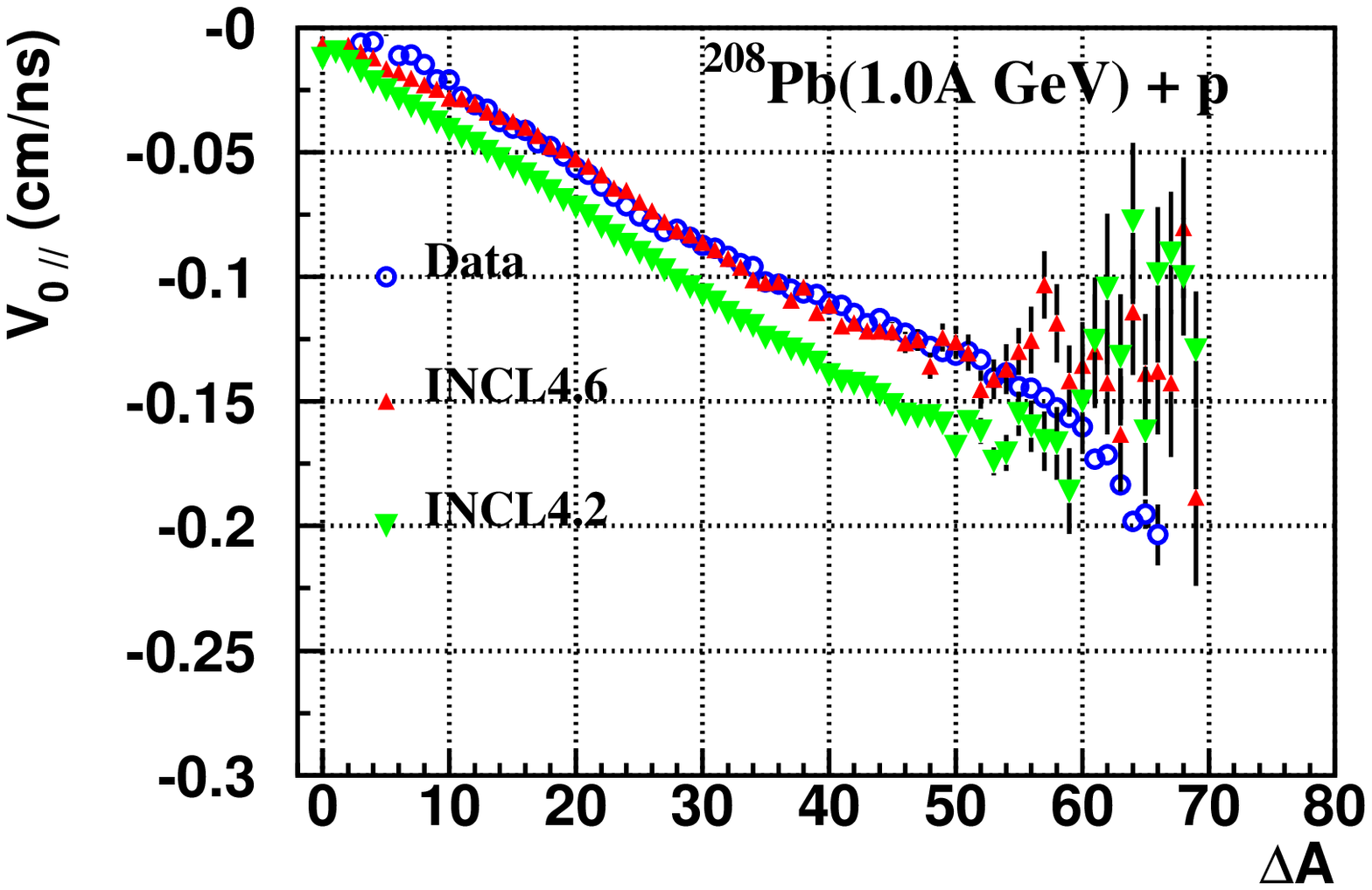}
\renewcommand{\baselinestretch}{0.5}
\end{minipage}
\\
\begin{minipage}[t]{4cm}
\includegraphics[width=4cm, height=4cm]{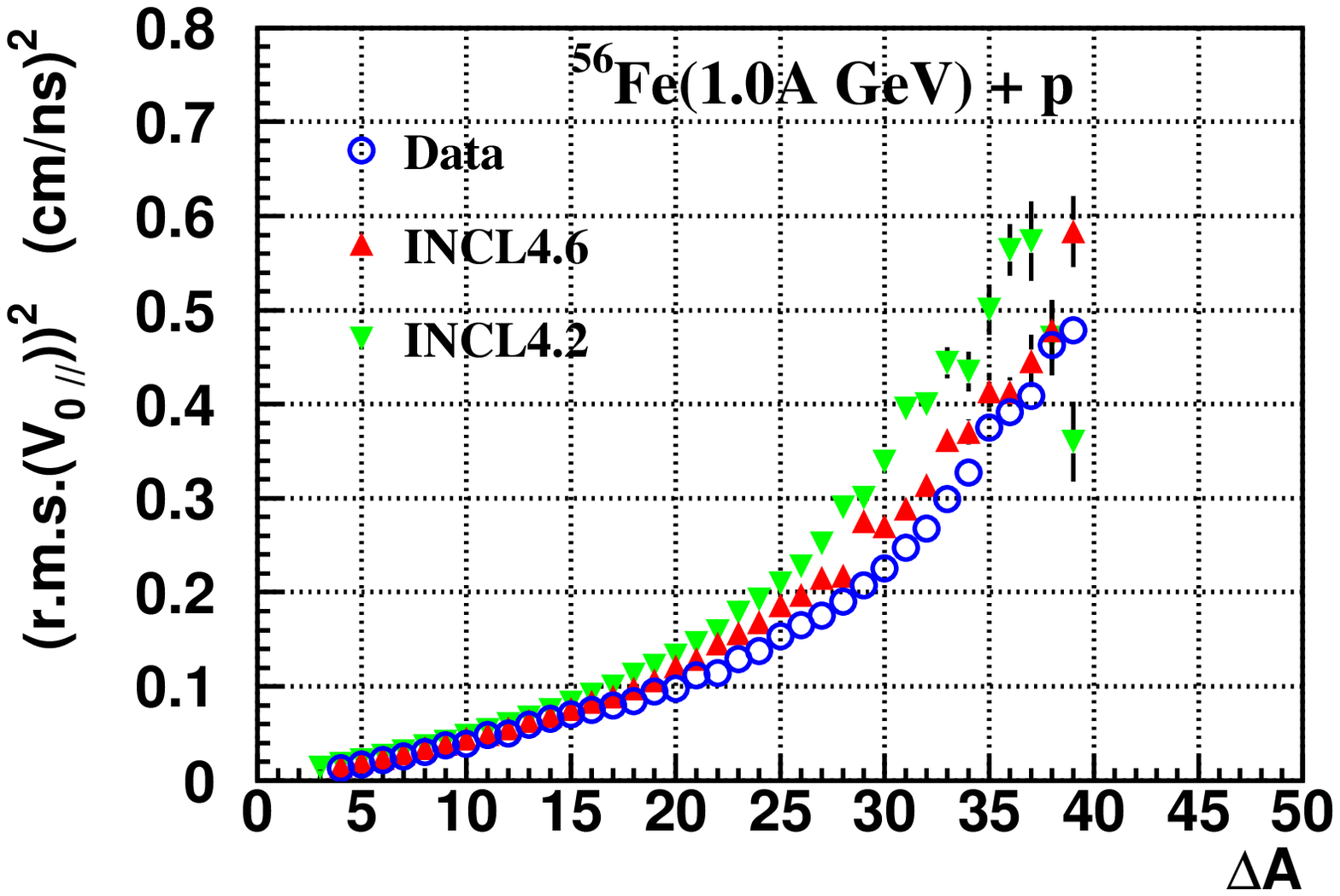}
\renewcommand{\baselinestretch}{0.5}
\end{minipage}
\hfill
\begin{minipage}[t]{4cm}
\includegraphics[width=4cm, height=4cm]{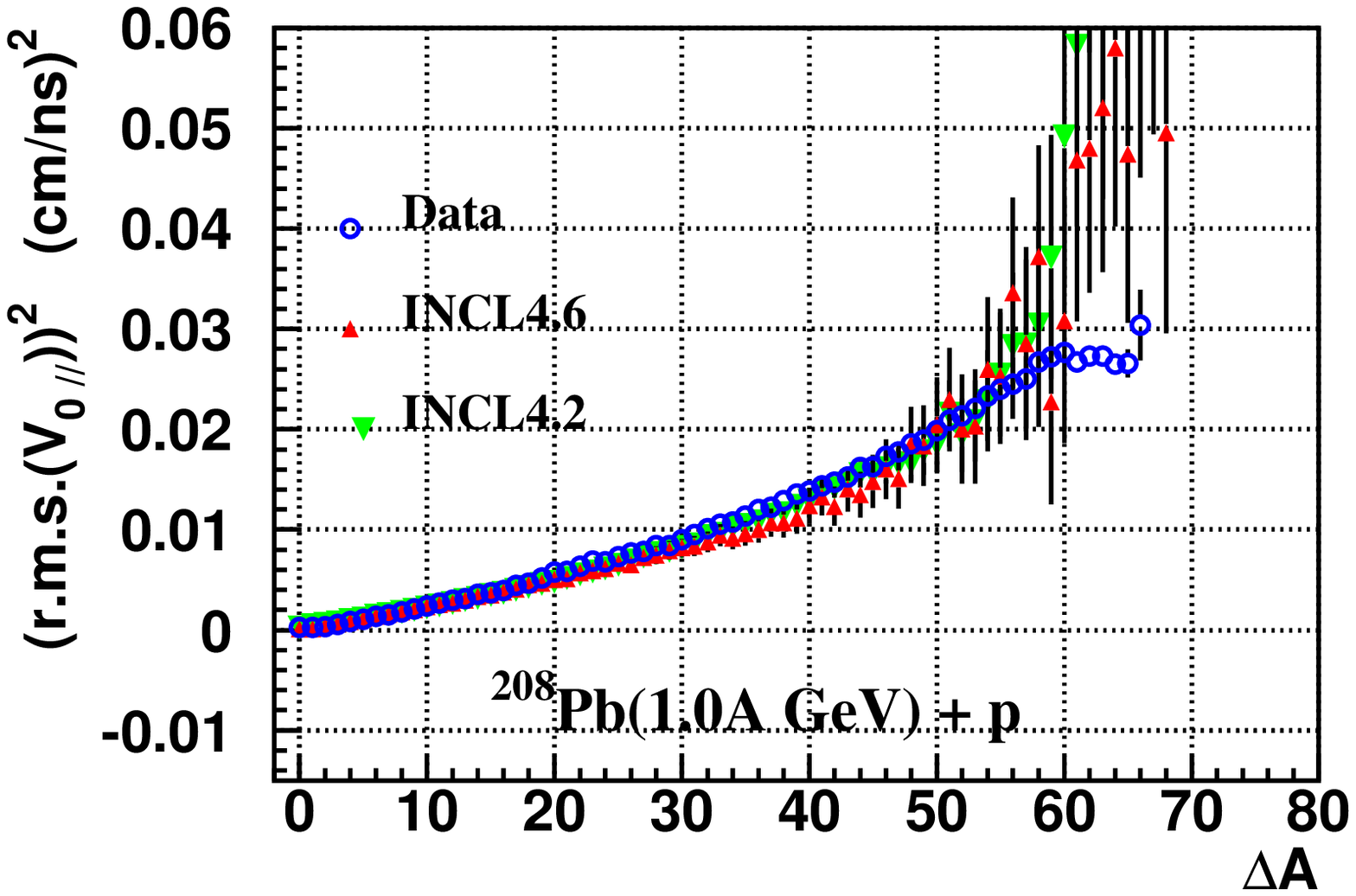}
\renewcommand{\baselinestretch}{0.5}
\end{minipage}


\caption{\small Recoil velocity of the residues. Comparison of INCL4.2-ABLA (green symbols) and INCL4.6-ABLA07 (red symbols) predictions with experimental data (blue circles), for the average velocity (upper row) and for the variance of the velocity distribution  (lower row), as functions of the mass loss $\Delta$A, equal to the target mass number minus the residue mass number. The left (right) panels refer to  $p$+$^{56}Fe$ ($p$+$^{208}Pb$) collisions at 1 GeV. Data are from Refs.~\cite{VI07,EN01}. }
\renewcommand{\baselinestretch}{0.5}
\label{recoils}
\end{figure}

\subsubsection{Residue recoil velocity}
\label{rec}
Longitudinal recoil velocity measurements have been performed in inverse kinematics experiments carried on at GSI. We present typical results in Fig.~\ref{recoils} along with our predictions, for proton-induced reactions on $^{56}Fe$ and $^{208}Pb$ at 1 GeV. The predictions of INCL4.6-ABLA07 for the average values of the longitudinal  recoil velocity of the residues agree quite well with the data, for mass losses $\Delta A= A_T-A$ up to $\sim$50 for $p$+$^{208}Pb$ and to $\sim$15 for $p$+$^{56}Fe$ (roughly one quarter of the target mass number for both systems). For  the former system, the apparent lack of agreement for $\Delta A \gtrsim$ 50 is presumably (or at least partly) due to large fluctuations in the calculation arising from very small cross sections (see Fig.~\ref{massPb1}). For  the other system, the production cross sections remain sizable for very large mass losses (relative to the target mass), and Fig.~\ref{recoils} indicates that our predictions start to  depart from  experiment when mass loss gets larger than 20.

The lower part of  Fig.~\ref{recoils} indicate that the predictions of INCL4.6-ABLA07 concerning the fluctuations are in remarkable agreement with the experimental values, even for larger intervals of mass loss (predictions of the previous INCL4.2-ABLA model are also remarkably close), for both systems. 

These results provide a severe test for the cascade model. The average velocities are largely determined by the cascade process and the subsequent evaporation, being basically isotropic (in the remnant frame), does not bring significant contribution. This is also true, to a lesser extent, for the r.m.s. fluctuations of the longitudinal velocity, except when evaporated particles are  more numerous than the cascade ejectiles, i. e. when $\Delta A \gtrsim$ 40 for $p$+$^{208}Pb$ and to $\sim$15 for $p$+$^{56}Fe$. 

Our INCL4.6 cascade model is thus very successful in describing these observables and in picking up the basic features of the generation of the recoil. It is turning less good in describing the average velocities for large mass loss in  $p$+$^{56}Fe$  at 1 GeV presumably because this system is characterized by a strong depletion of the target, for which  models like INC are becoming less suitable. 

For relatively small mass losses ($\lesssim$15  for $Fe$ and $\lesssim$30 for $Pb$), the contribution of evaporation to r.m.s. fluctuations remains small. In these conditions, which allow a test of the cascade alone, the average value of the  longitudinal recoil velocity and its variance are remarkably linear in  the mass loss variable, both in INCL4.6 and in the experiment. As discussed in Ref.~\cite{BO02}, this is an indication of the recoil resulting from a diffusion process, arguably  due to independent successive NN collisions. The quality of the predictions of our cascade model appears as a strong support of its basic premises. We  will not elaborate on this point here. We will instead discuss a little bit the difference between the results obtained by INCL4.6 and INCL4.2. Definitely, the predictions of the former are improving on those of the latter for the average velocity, whereas the  differences are rather minute for the r.m.s. fluctuation. For the average velocity, we checked that more than half of the difference  for the $Pb$ case, and almost  the whole difference for the $Fe$  case, seems to come from the introduction of the emission of clusters in the cascade. In some sense, this emission allows a larger mass loss (in the cascade) without changing very much the pattern of the collisions for energetic participants (which are supposed to contribute the most to the recoil), i.e. without changing the recoil velocity and the excitation energy. This leads for a given recoil velocity to a corresponding increased mass loss. In simple terms, this pushes the curve for the average  velocity in Fig.~\ref{recoils} toward the right of the Figure. We could not isolate clearly the cause of the remaining difference. Presumably it is due to the introduction of forbidding collisions below the Fermi level (item 3.B of Section~\ref{features4.5}) and  to the introduction of ``the back-to-spectators'' trick (item 1 of Section~\ref{features4.6}). Both do not affect very much the collisions involving energetic participants, but slightly increase the excitation energy. So, they increase the mass loss for a given recoil velocity, pushing also the recoil velocity curve of INCL4.2  toward the right.         

For completeness, we mention that we made the same calculation for $p$+$^{56}Fe$ collisions at 500 MeV (not shown here). The results are slightly better than at 1 GeV.

\section{Results for cluster-induced reactions}
\label{CIR}
\subsection{Introduction}

As we said in the Introduction, INCL4.2 can accommodate light clusters as projectiles, considering them  as  collections of on-shell nucleons with Fermi motion and a collective velocity adjusted in such a way that the total energy of the projectile nucleons (in the target frame) is equal to the nominal projectile incident energy. This method is justified at high incident energy (where the difference between the adjusted velocity and the nominal velocity is small anyway). Indeed, in Refs.~\cite{BO02,HE05}, good results were obtained for deuteron-induced collisions around 1 GeV and departures from the simple additivity of a proton and a neutron cascades were illustrated. Satisfactory results were also obtained for other clusters in Refs.~\cite{BO10,KA10}. At low energy, this simple method cannot be accurate. Even more, the adjusted velocity cannot be defined for incident kinetic energy lower than  the binding energy of the incident cluster. To cure this situation, we have introduced the method described in  Section~\ref{cir}. Our interest is  mainly motivated by applications, in particular  to accelerator-driven systems and spallation targets, as quoted in the Introduction. As an example, astatine isotopes can be produced in $Pb-Bi$ thick targets through, for instance, ($\alpha, xn$) reactions induced by  secondary $\alpha$-particles~\cite{TA08,MA06a,LE12}. Following this motivation, we will mainly devote our attention here to the residue production cross sections. 

\subsection{Total reaction cross sections}
\label{clindreac}
\vspace{1cm}
\begin{figure*}[t]

\begin{minipage}[t]{7cm}
\includegraphics[width=7cm, height=7cm, angle=270]{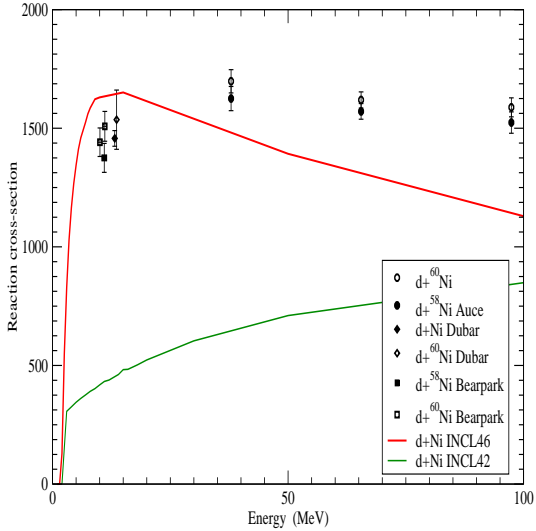}
\renewcommand{\baselinestretch}{0.5}
\end{minipage}
\hfill
\begin{minipage}[t]{7cm}
\includegraphics[width=7cm, height=7cm, angle=270]{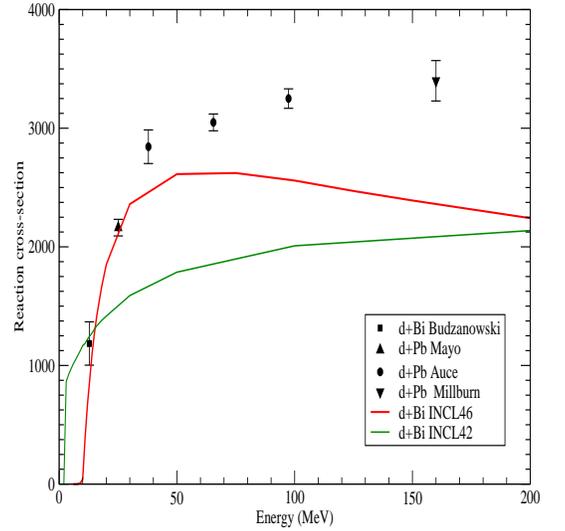}
\renewcommand{\baselinestretch}{0.5}
\end{minipage}
\vspace{2cm}
\newline
\begin{minipage}[t]{7cm}
\includegraphics[width=7cm, height=7cm, angle=270]{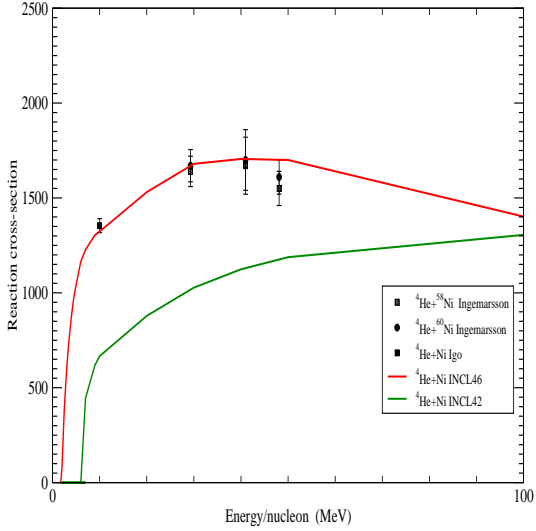}
\renewcommand{\baselinestretch}{0.5}
\end{minipage}
\hfill
\begin{minipage}[t]{7cm}
\includegraphics[width=7cm, height=7cm, angle=270]{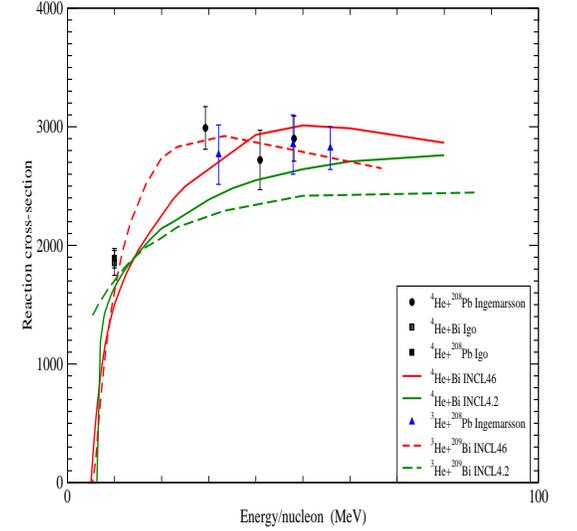}
\renewcommand{\baselinestretch}{0.5}
\end{minipage}

\caption{\small Total reaction cross sections (in $mb$) for deuterons (upper row) and for $\alpha$ or $^3He$ particles (lower row), on $Ni$ and $Fe$ targets (left panels) and on $Pb$ and $Bi$ targets (right panels), as functions of the incident kinetic energy (for deuterons) or of the incident kinetic energy per nucleon (for  $^3He$ and $^4He$). The predictions of the  INCL4.2 (INCL4.6) model are given by the green (red) curves. In the last panel, the dotted curves refer to $^3He$-induced reaction cross sections. Data are taken from  Refs.~\cite{BA74,HE05a,RI90}. }
\renewcommand{\baselinestretch}{0.5}
\label{adsigtot}
\end{figure*}

A set of total reaction cross sections is given in Fig.~\ref{adsigtot}. Since this quantity is slowly varying with target mass and even with projectile mass, for the same velocity, we mix data for neighbouring target nuclei. Although the experimental data are rather scarce, one can see that our model roughly picks up the basic properties of the energy dependence of the reaction cross sections, namely a rapid rise at low energy followed by a slowly varying plateau.  The magnitude of the plateau value is particularly well reproduced for incident  $\alpha$-particles. For the $d$-induced reactions, this plateau value is underestimated. This may be explained by the contribution of the Coulomb dissociation of the deuteron, which is not taken into account. Evaluations of this contribution performed in Refs.~\cite{HE05,RI99,TO98,OK94,AV12,SI11} point toward a sizable cross section, of the order of 100-300 mb for the $d-Pb$ system. This contribution is expectedly smaller for $d-Fe$ and somehow negligible for $\alpha$-induced reactions, because of the large binding energy and the compactness of this projectile. One may notice the spectacular improvement obtained by INCL4.6 especially for deuteron-nucleus reaction cross sections. Of course, this is consistent with the better reproduction of the  nucleon-nucleus cross section by INCL4.6 in comparison with INCL4.2, as disussed in Section \ref{react}, due itself to a better treatment of the first collision at low incident energy. The predictions of INCL4.2 are closer to those of INCL4.6 for $\alpha$-particles, compared to deuterons, because for peripheral collisions, the probability of interaction is increasing with the number of nucleons of the projectile. Finally, the threshold behaviour is different in INCL4.2 and INCL4.6: in the former, it is mainly dictated by the value of binding energy per nucleon of the projectile, wheras in the latter, it is  mainly dominated by the introduction of the Coulomb deflection and of the fusion (compound nucleus) process.
 
Let us add a comment on the plateau values. First of all, the maximum of the cross section is losely correlated  with the maximum which occurs in the proton-nucleus cross sections for the same target at low energy (see  Fig.~\ref{sigR}). The value of the reaction cross section at the maximum for cluster-induced reactions is larger than the maximum value of the proton-nucleus cross section and reflects the size of the cluster, which allows interactions for larger impact parameters. Let us finally indicate that the slow decrease of the cross sections in Fig.~\ref{adsigtot} is presumably related to the decrease of the NN cross section in this energy range.

\subsection{Residue production cross sections}
\label{rpX}
\begin{figure}[h]

\begin{minipage}[t]{4cm}
\includegraphics[width=4cm]{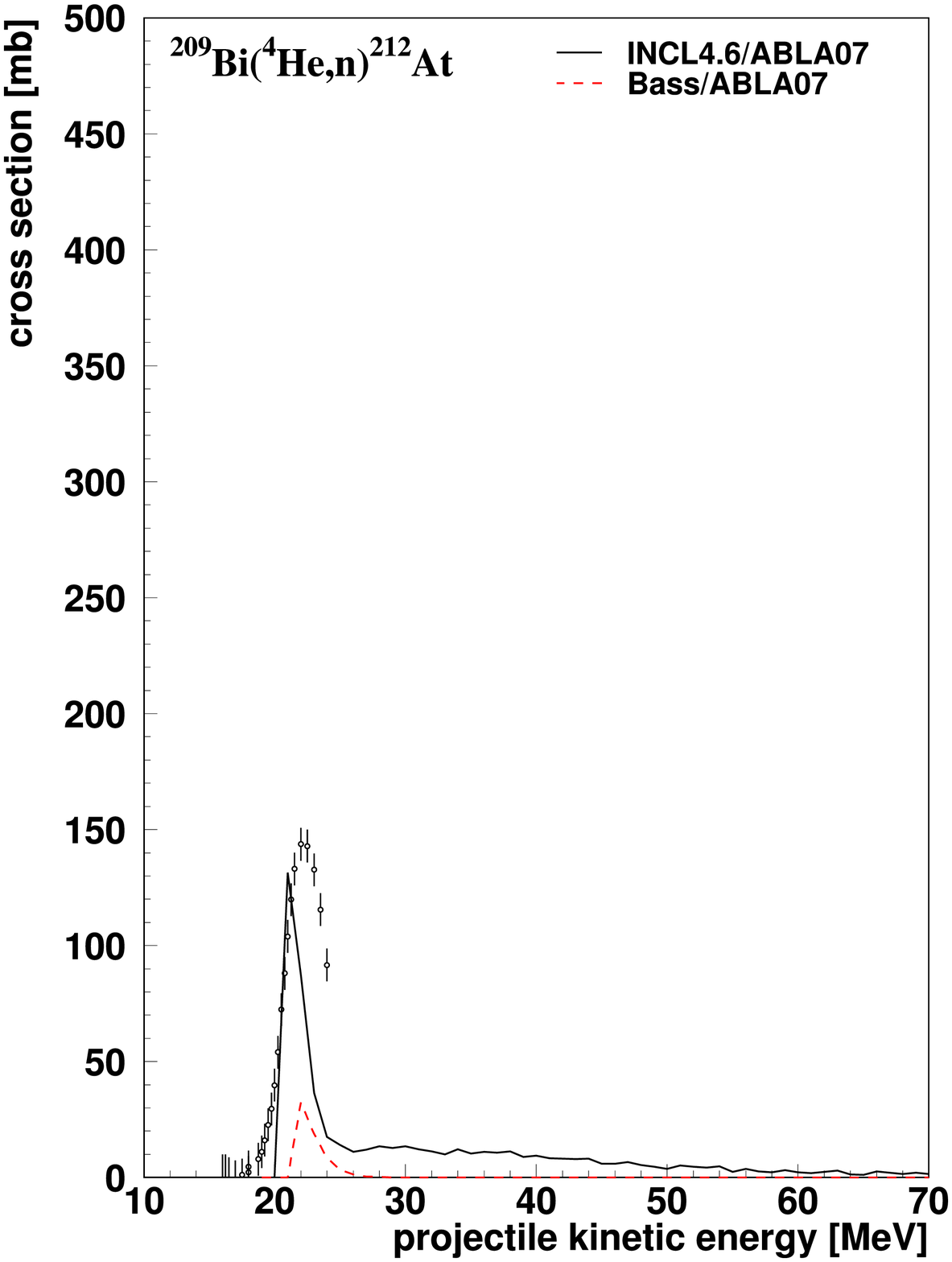}
\end{minipage}
\hfill
\begin{minipage}[t]{4cm}
\includegraphics[width=4cm]{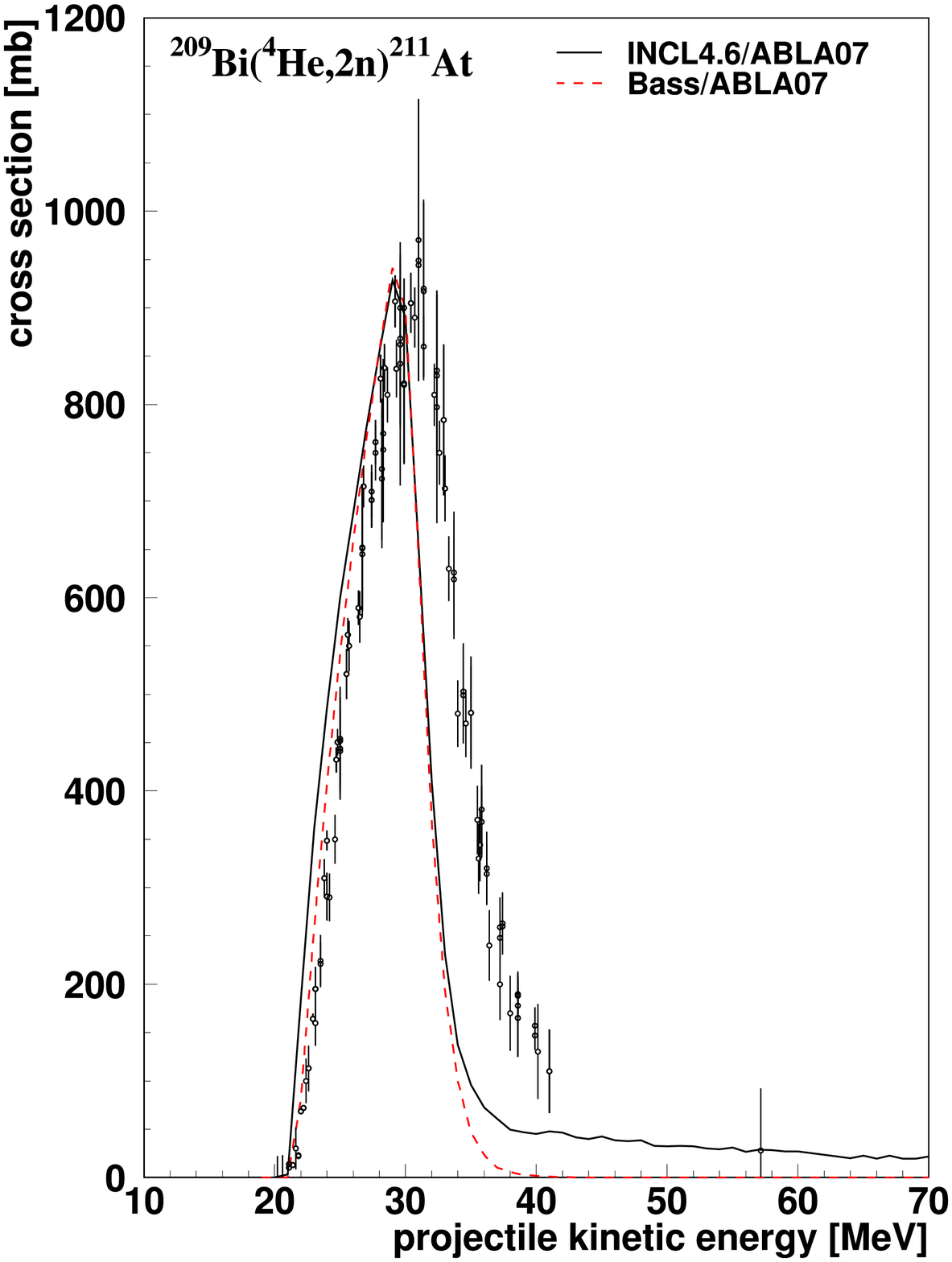}
\end{minipage}
\newline
\begin{minipage}[t]{4cm}
\includegraphics[width=4cm]{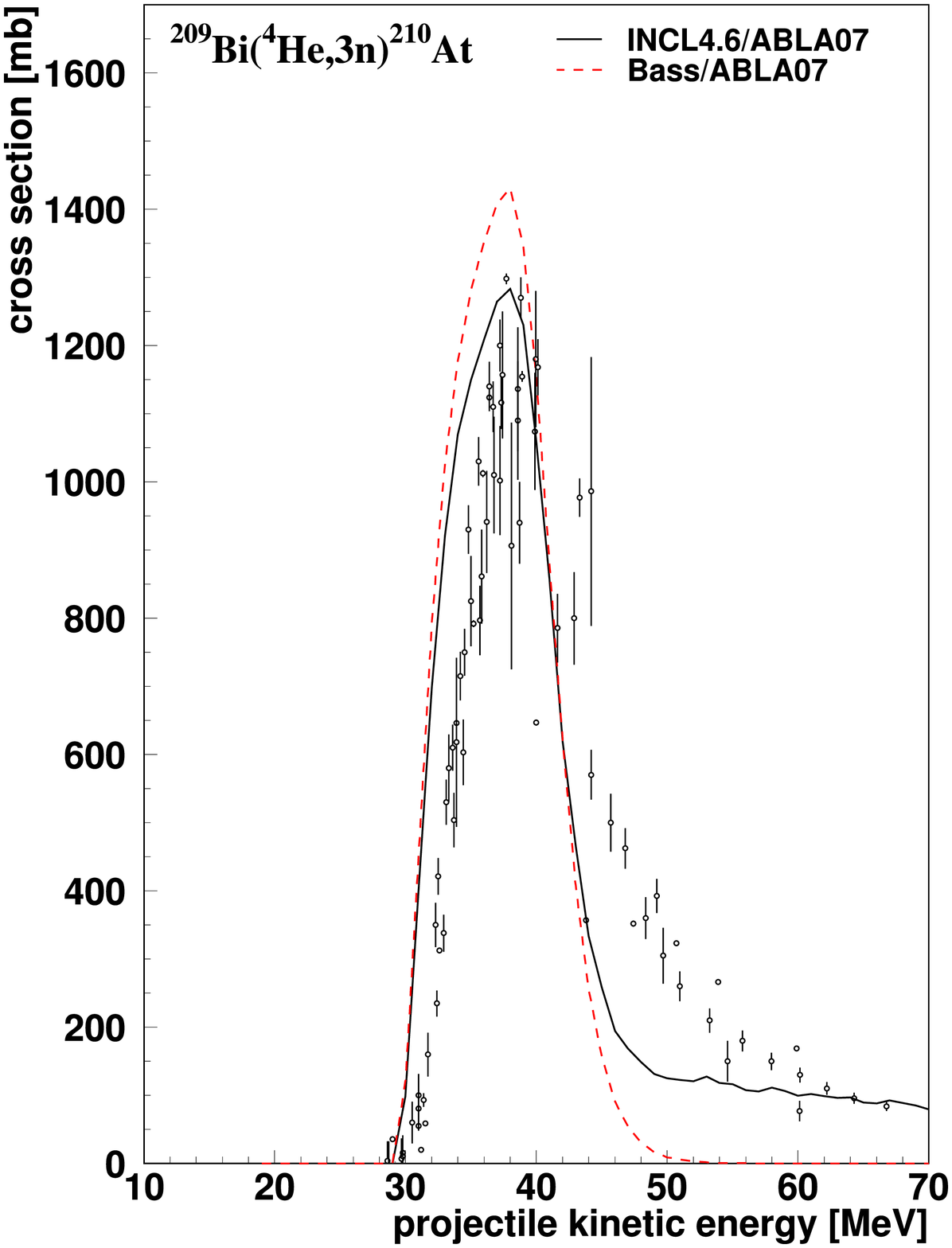}
\end{minipage}
\hfill
\begin{minipage}[t]{4cm}
\includegraphics[width=4cm]{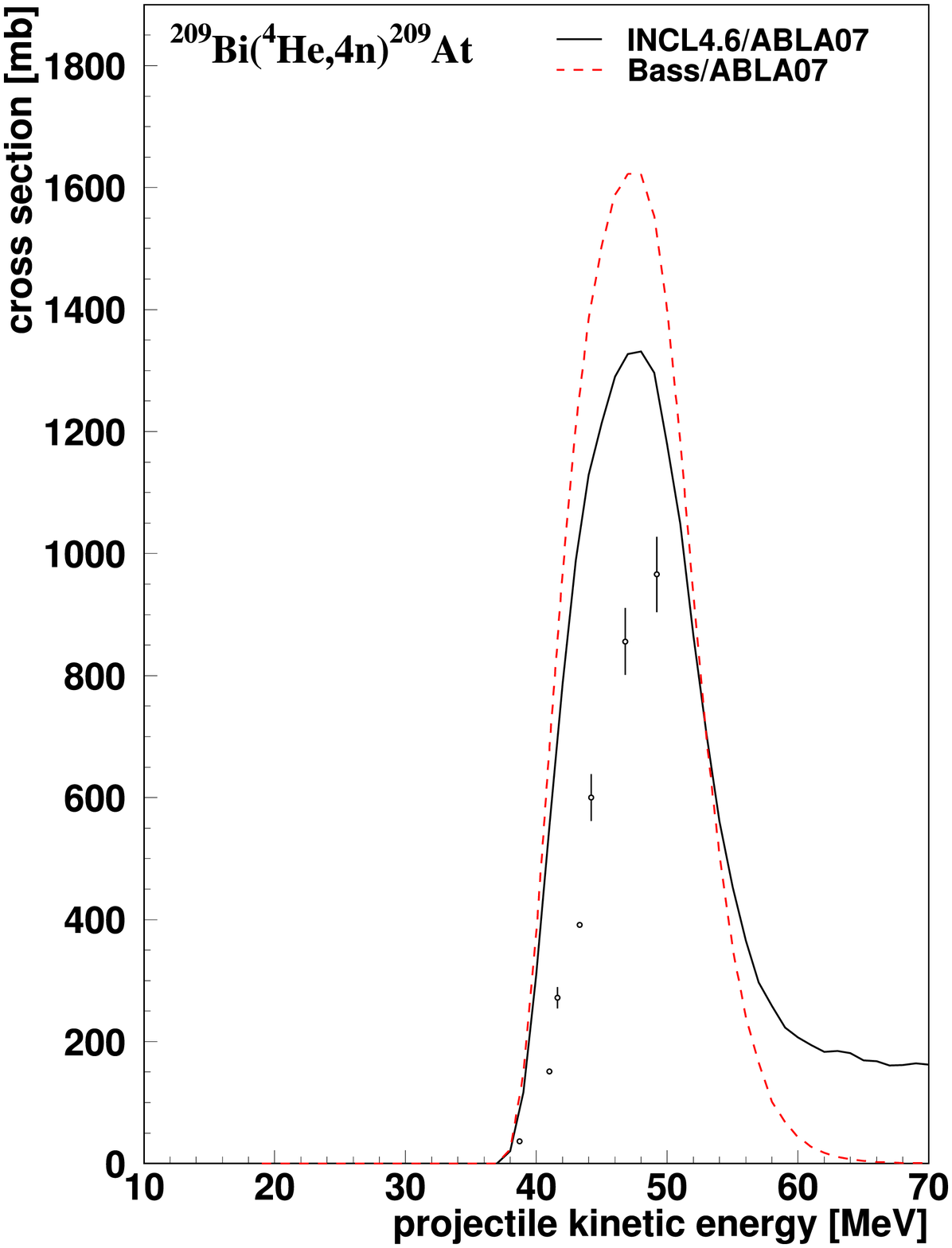}
\end{minipage}

\hfill

\caption{\small $^{209}Bi (\alpha, xn) $ cross sections  for $x$=1, 2, 3 and 4, respectively, as functions of the $\alpha$ incident kinetic energy. The full black curves correspond to the predictions of the INCL4.6+ABLA07 model. The dashed red lines represent the results using the Bass fusion model, followed by the ABLA07 model. Data are taken from  Refs.~\cite{BA74,HE05a,RI90}. }
\label{Biaxn}
\end{figure}

\begin{figure*}[t]

\begin{minipage}[t]{5cm}
\includegraphics[width=5cm, height=7cm]{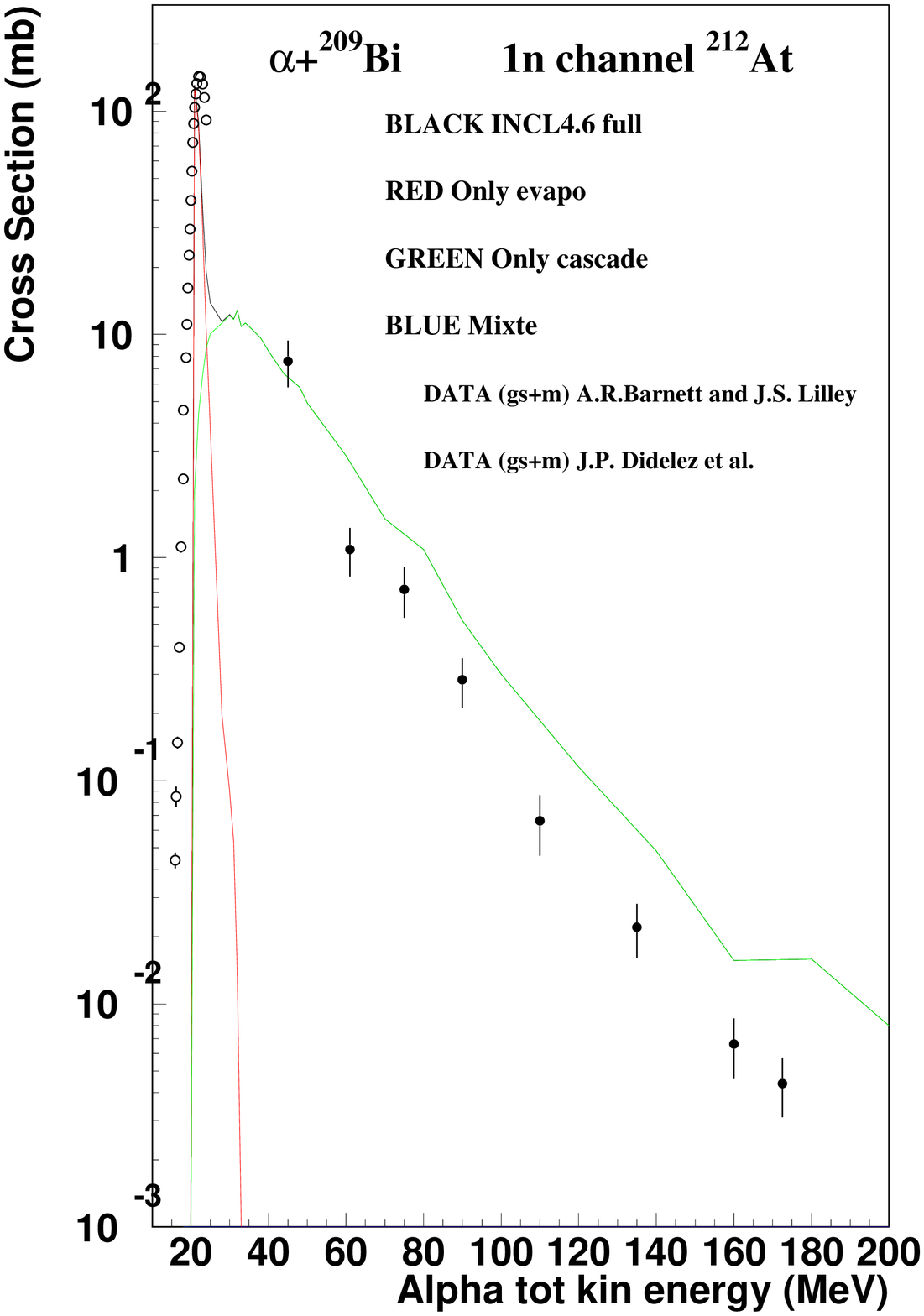}
\renewcommand{\baselinestretch}{0.5}
\end{minipage}
\hspace{1cm}
\begin{minipage}[t]{5cm}
\includegraphics[width=5cm, height=7cm]{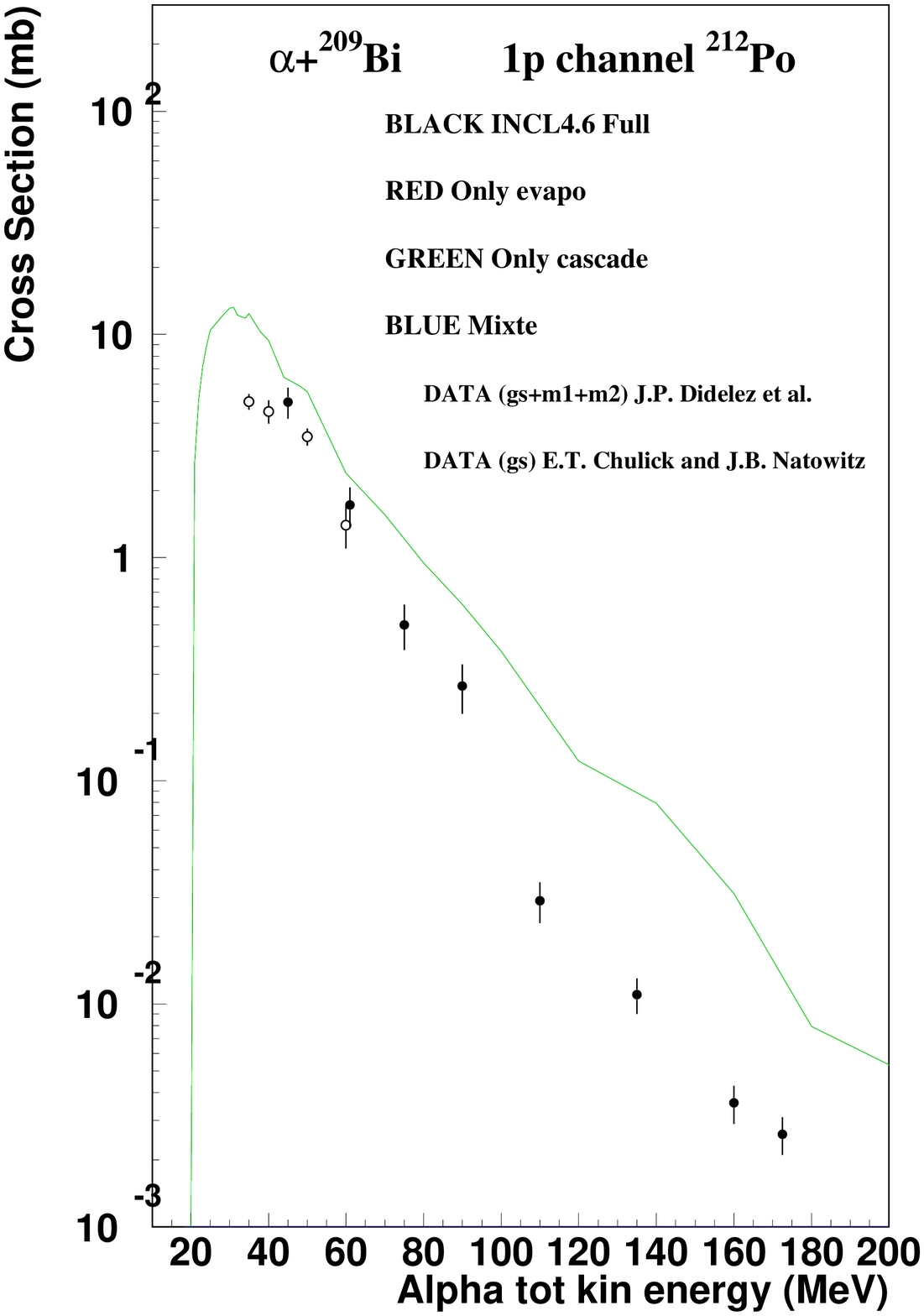}
\renewcommand{\baselinestretch}{0.5}
\end{minipage}
\hfill
\begin{minipage}[t]{5cm}
\includegraphics[height=7cm, width=5cm,  angle=0]{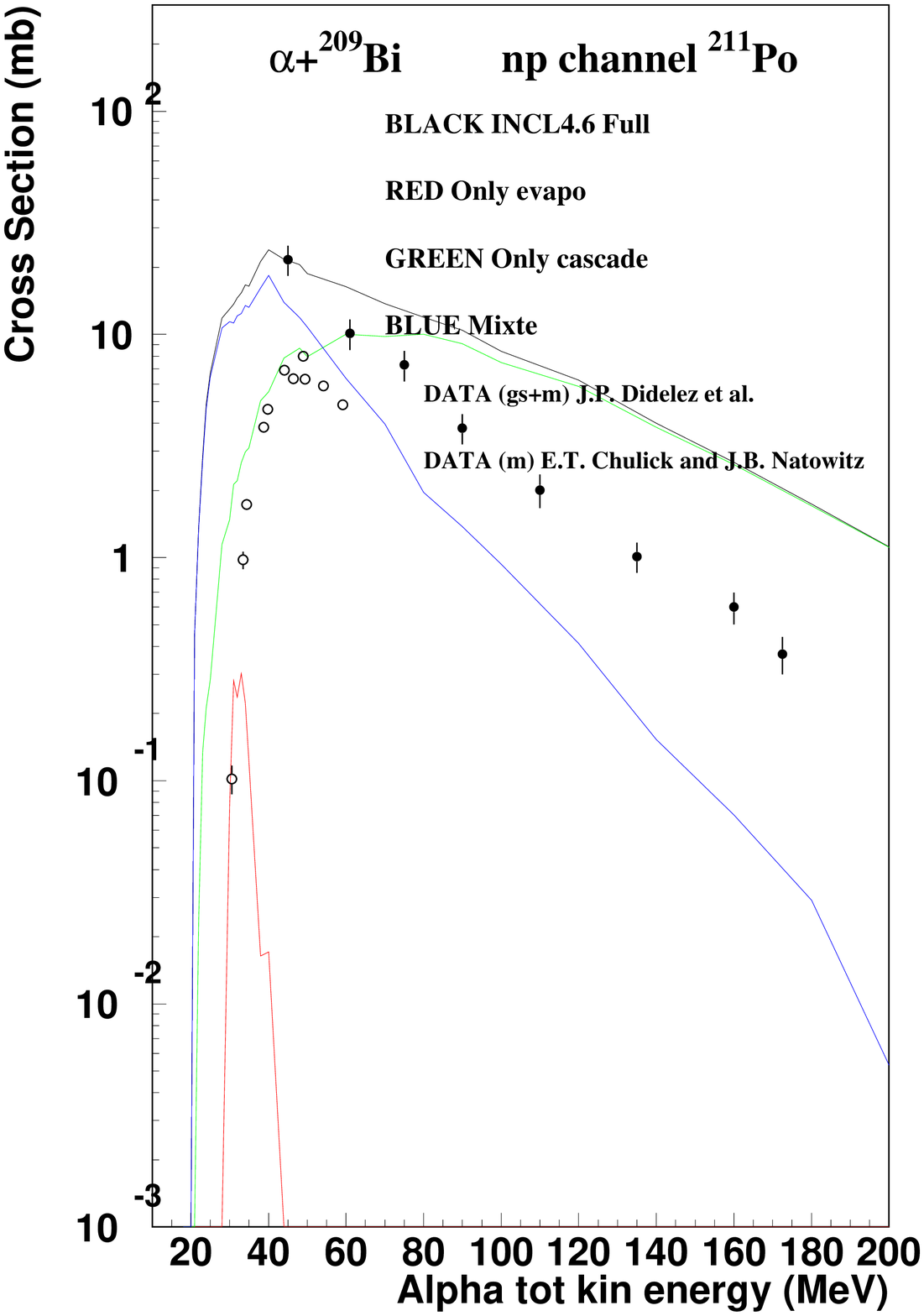}
\renewcommand{\baselinestretch}{0.5}
\end{minipage}

\caption{\small $^{209}Bi (\alpha,n) $  (left), $^{209}Bi (\alpha,p) $ (center), $^{209}Bi (\alpha,np)$ (right) cross sections, as functions of the $\alpha$ incident kinetic energy.  Each predicted cross section using  the INCL4.6-ABLA07v5 model (black lines) is splitted into a cascade component (green lines), an evaporation component (red lines) and a mixed component (blue lines). See text for detail. The black lines give the sum of the three components and are visible when they are not identical to colored lines. Experimental data are taken from Refs.~\cite{BA74,DI80} for $^{209}Bi (\alpha,n) $ and from Refs.~\cite{CH71,DI80} for the two other reactions. In the first panel, the data of Ref.~\cite{DI80} are singled out by the large error bars. }
\renewcommand{\baselinestretch}{0.5}
\label{Bian}
\end{figure*}

As a first example of our results, we show in Fig.~\ref{Biaxn} the cross sections for the production of $At$ isotopes through $^{209}Bi (\alpha, xn) $ reactions. Globally, the shapes and the magnitudes of the peaks of the cross sections are satisfactorily reproduced. The correct position of the respective maxima indicates a good description of the competition between the successively opening channels in the evaporation. On the other hand, the good evaluation of their magnitudes indicates that the total reaction cross section is correctly given by the cascade model, as shown in Fig.~\ref{adsigtot}. Fig.~\ref{Biaxn} also shows the results of a calculation using the Bass fusion model~\cite{BA74a}, coupled to the ABLA07 de-excitation model. It can be seen that in these conditions, where fusion dominates, our model yields results which compare rather well with those of a very commonly used fusion model.  It may be argued however that the Bass model is not really suitable for alpha particles.

A little bit more detail concerning the $^{209}Bi (\alpha,n) $ reaction is given in the first panel of Fig.~\ref{Bian}, where data from Ref.~\cite{DI80} at higher energy have been added to those of Ref.~\cite{BA74}. The high incident energy part of the excitation function is a good test of the cascade model. Indeed,  we have divided the theoretical cross section in Fig.~\ref{Bian} into three components: the first one (in green) corresponding to events where the particles are ejected during the cascade stage\footnote{This includes the geometrical spectators.}, the second one (in red) corresponding to events with particles emitted during the de-excitation of the target remnant and the last one (in blue) for events in which the emitted particles are coming partly from the cascade and partly from the evaporation stages. The red curves can be viewed as giving the complete fusion (+evaporation) contribution whereas the two other curves can be interpreted as corresponding to incomplete fusion and/or pre-equilibrium processes. For the $^{209}Bi (\alpha,n) $ reaction, the mixed (blue) component is identically vanishing, since a single neutron is emitted either in the evaporation of the fused
system or in the cascade stage. From the first panel of Fig.~\ref{Bian}, it can be seen that the cross section above $\sim$30 MeV incident energy is well described by our model. The emission of a single neutron is then entirely due to the cascade mechanism. The remnant, i.e. the compound nucleus, is then sufficiently excited to emit two neutrons, at least, with a large probability and thus does not significantly feed the 1$n$ channel.  The same is true for the $^{209}Bi (\alpha,p) $ reaction (shown in the central panel of Fig.~\ref{Bian}). In that case, the evaporation component is very much suppressed, even in the 20-30 MeV range, due to the Coulomb barrier.  Finally, the third panel of Fig.~\ref{Bian} displays the situation for the $^{209}Bi (\alpha,np)$ reaction. Here, above $\sim$70 MeV incident energy, the cross section is dominated by the emission of the neutron and the proton during the cascade stage. At lower incident kinetic energy, there is a very small component for the emission of the two nucleons in the evaporation stage, but at low incident kinetic energy the cross section is dominated by the mixed emission of the two particles.

\begin{figure}[h]

\includegraphics[height=10cm, width=7cm,  angle=0]{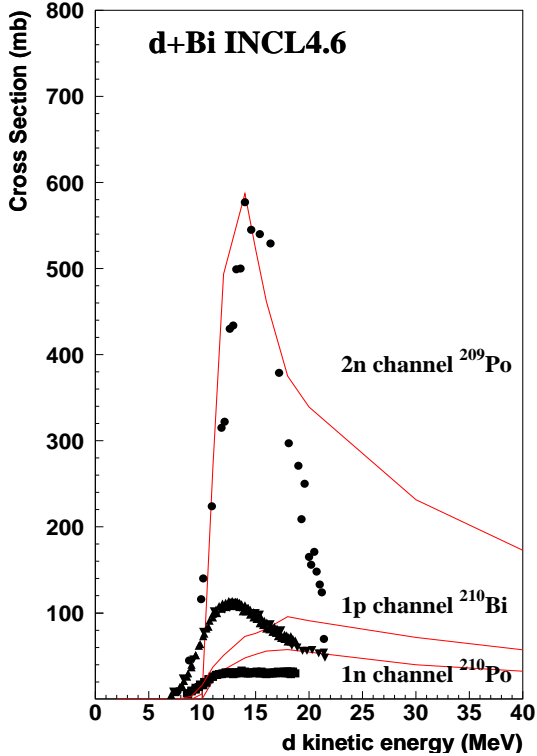}

\caption{\small Deuteron-induced reactions on $^{209}Bi$. The  predictions of the INCL4.6-ABLA07v5 model (full red lines) are compared with experimental data (symbols) relative to  $^{209}Bi (d,n) $ (squares),  $^{209}Bi (d,2n)$ (full dots) and  $^{209}Bi (d,p)$ (triangles) reactions. Cross sections are given as  functions of the deuteron incident kinetic energy. Data from Refs.~\cite{RA59,BU63,KE49}.}
\label{Bidxn}
\end{figure}

Another example is provided by Fig.~\ref{Bidxn}, which shows deuteron-induced reactions on $^{209}Bi$. The shapes of the cross sections are globally reproduced fairly well, though some details are missed. The decrease of the $(d,2n) $ cross section on the right of the peak is somehow too slow. The precise threshold behavior for $(d,n)$ and $(d,p)$ reactions is not exactly reproduced, in spite of the use of exact Q-values. Actually, the theoretical $(d,n)$ and $(d,p)$ cross sections are roughly the same.  However, the experimental $(d,n)$ cross section is roughly 4 times smaller than the $(d,p)$ cross section in the region of the maximum ($\sim$10 MeV). This discrepancy might be due to the Coulomb dissociation of the deuteron. We will discuss this point below, after the presentation of another case.

\begin{figure}[h]

\includegraphics[width=7cm, height=10cm]{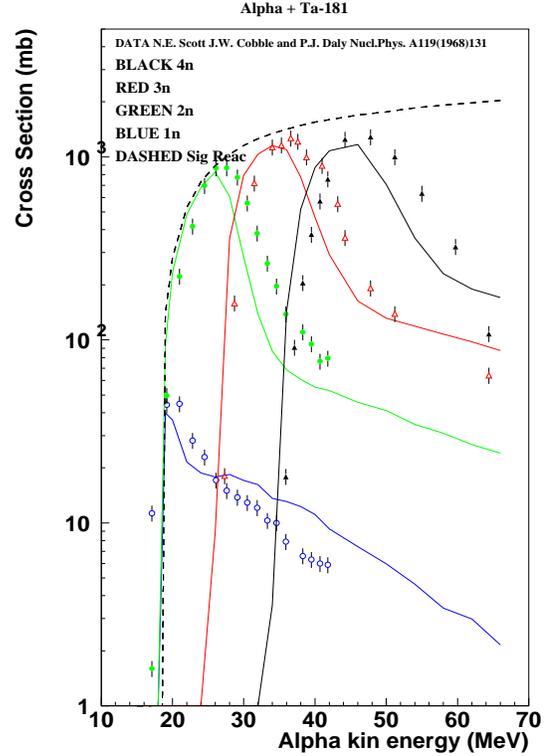}
\caption{\small $^{181}Ta (\alpha,xn) $ cross sections  for $x$=1 (blue), 2 (green), 3 (red) and 4 (black), respectively, as functions of the $\alpha$ incident kinetic energy. The full lines correspond to the predictions of the INCL4.6+ABLA07 model. Data (symbols) are taken from  Ref.~\cite{SC68}. The dashed line gives the theoretical total reaction cross section. }

\label{Taaxn1}
\end{figure}

\begin{figure}[h]

\begin{minipage}[t]{4cm}
\includegraphics[width=4cm, angle=0]{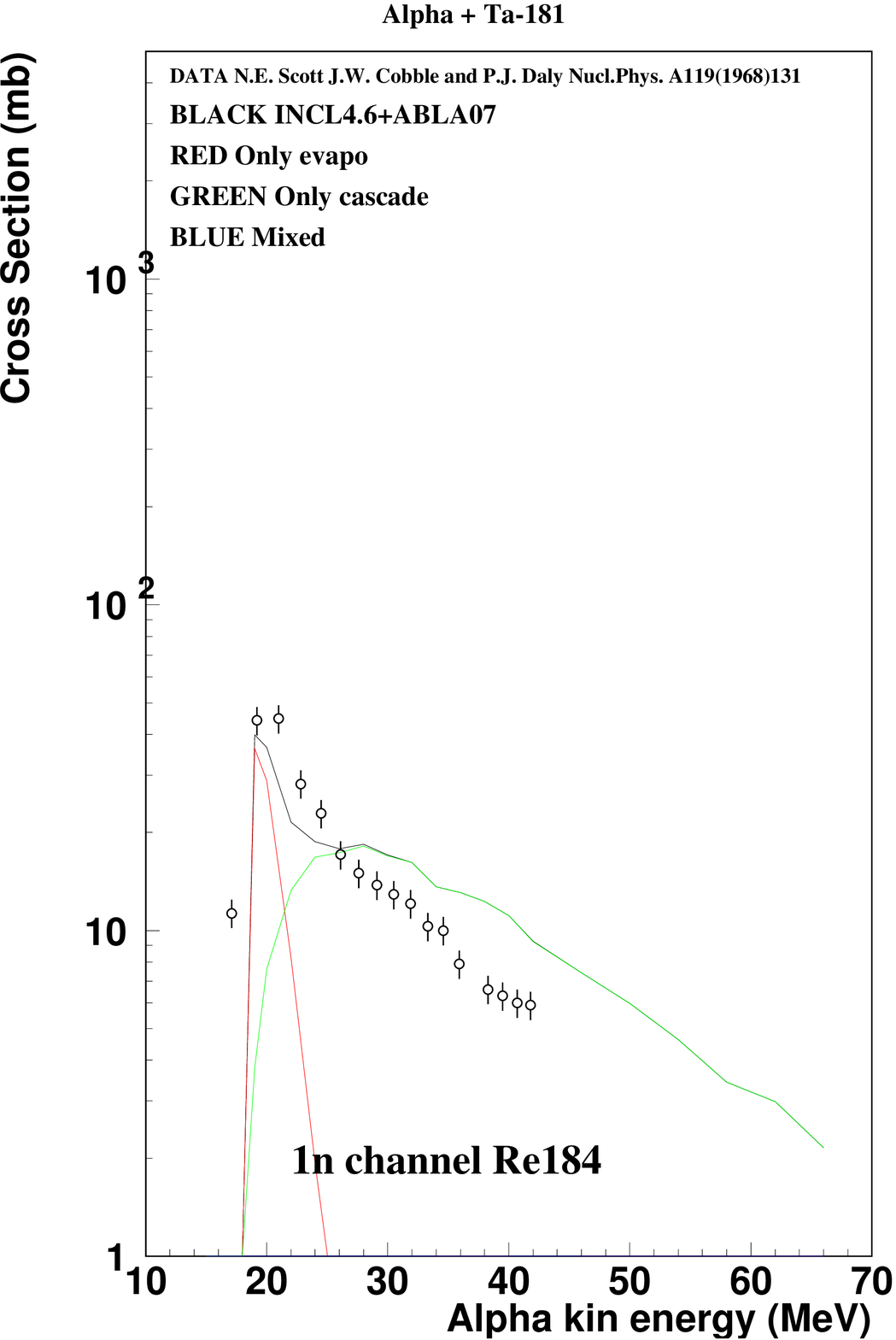}
\renewcommand{\baselinestretch}{0.5}
\end{minipage}
\begin{minipage}[t]{4cm}
\includegraphics[width=4cm]{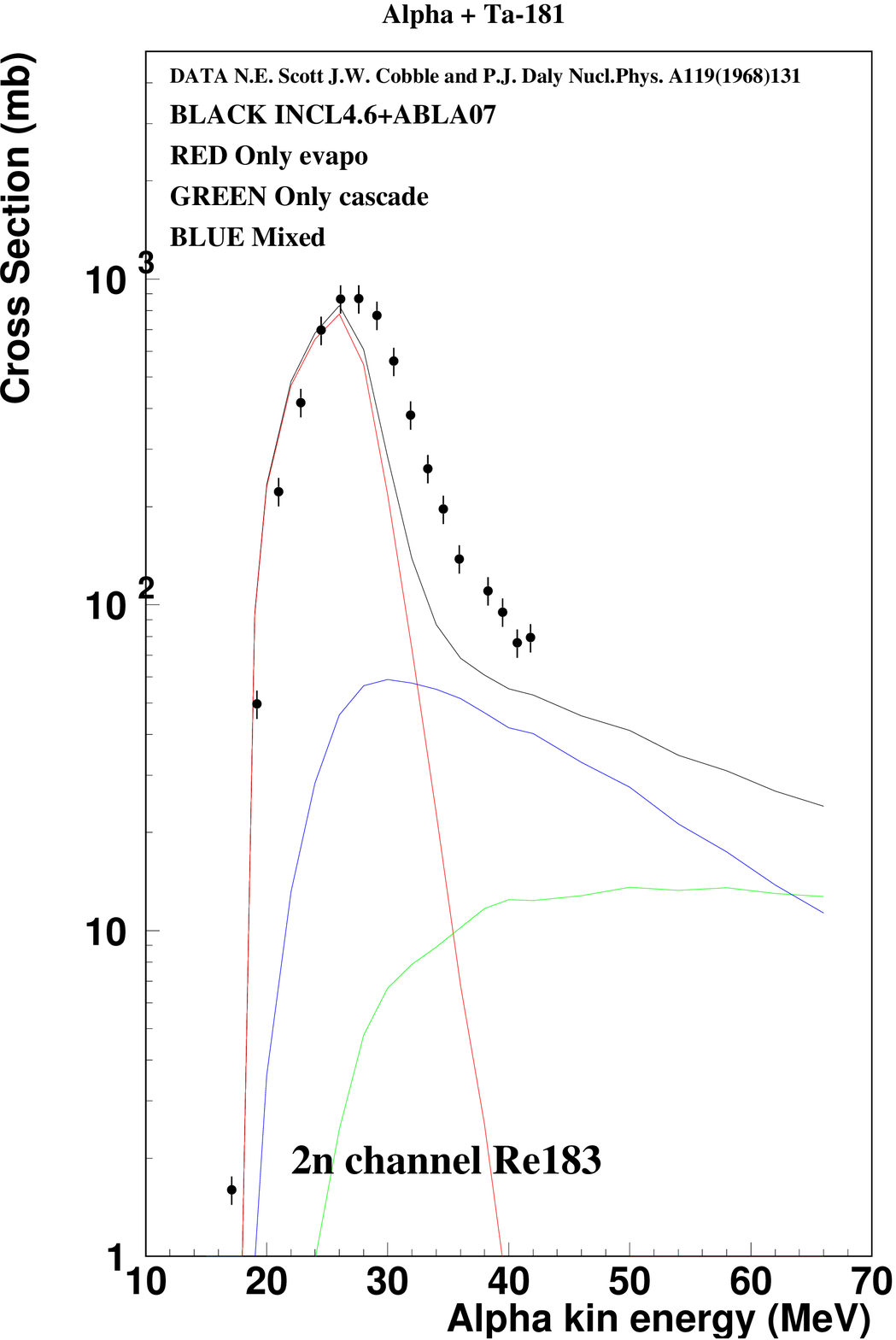}
\renewcommand{\baselinestretch}{0.5}
\end{minipage}
\\
\begin{minipage}[t]{4cm}
\includegraphics[width=4cm]{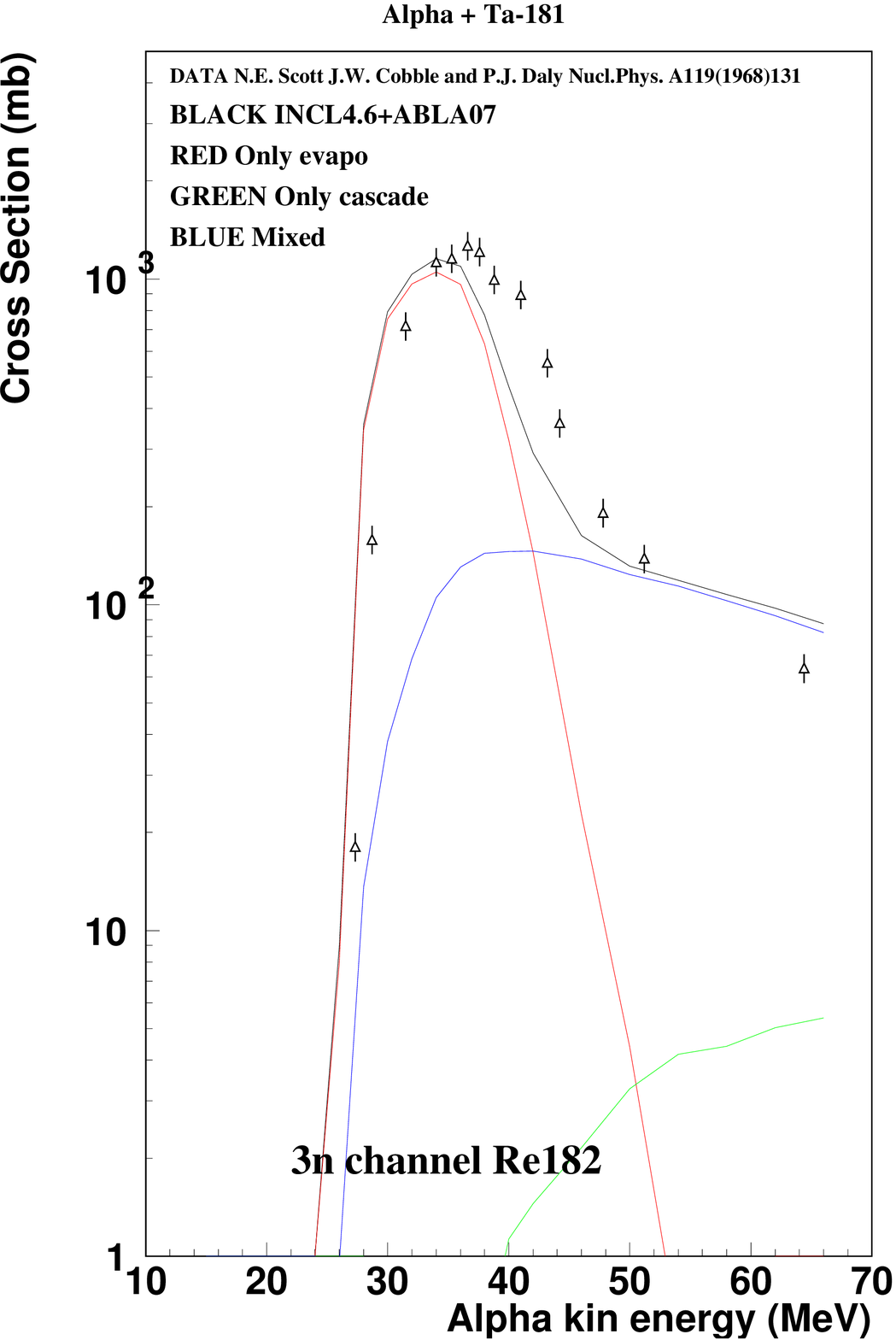}
\renewcommand{\baselinestretch}{0.5}
\end{minipage}
\begin{minipage}[t]{4cm}
\includegraphics[width=4cm]{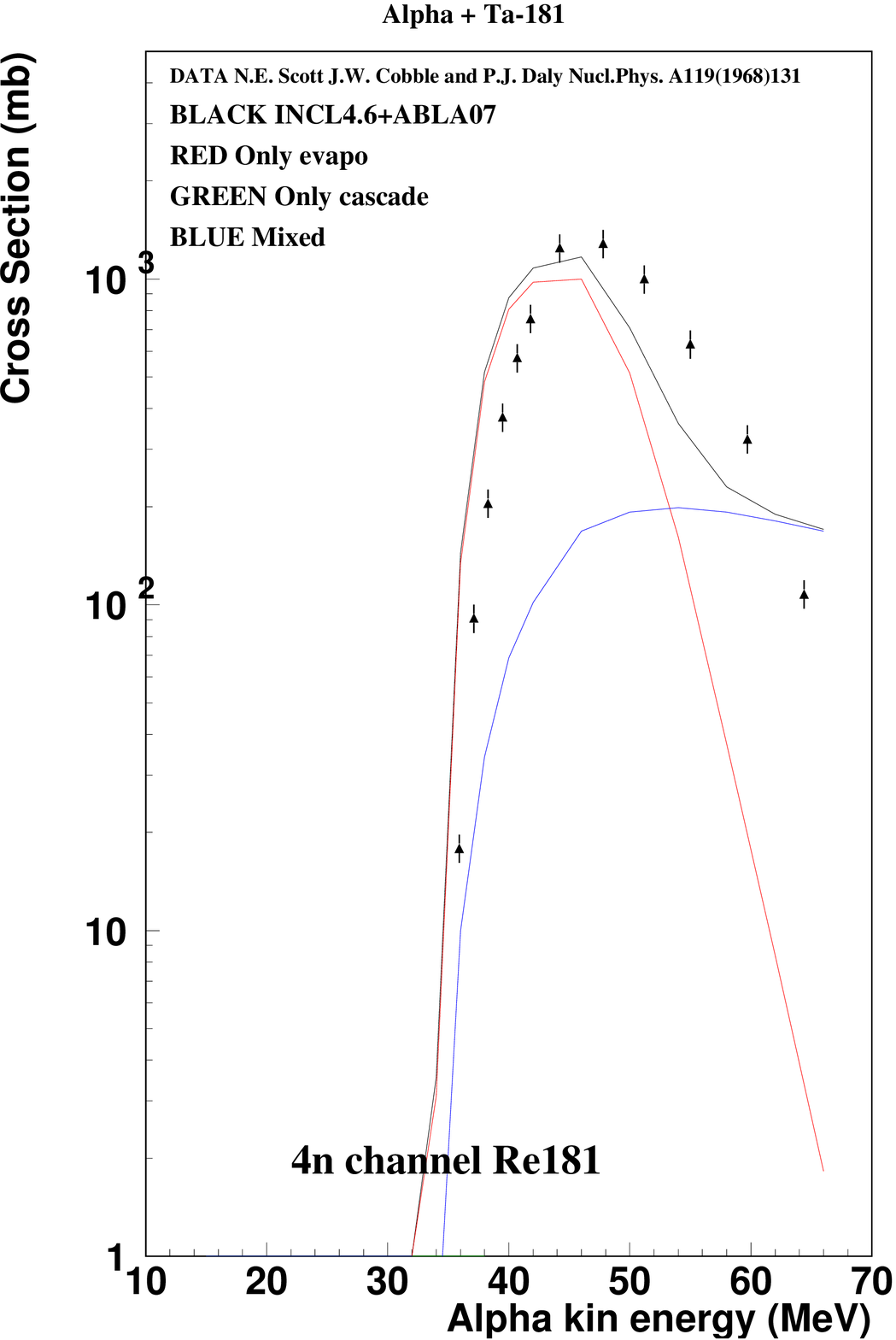}
\renewcommand{\baselinestretch}{0.5}
\end{minipage}

\caption{\small Splitting of the calculated $^{181}Ta (\alpha,xn) $ cross sections (black lines) for $x$=1, 2, 3  and 4, respectively, as functions of the $\alpha$ incident kinetic energy, into the various contributions: cascade (green), evaporation (red) and mixed (blue). See text for detail. Data (symbols) are taken from  Ref.~\cite{SC68}. }
\renewcommand{\baselinestretch}{0.5}
\label{Taaxn2}
\end{figure}

Results concerning another heavy target are reported on in Figs.~\ref{Taaxn1} and \ref{Taaxn2}. The first one shows the various $^{181}Ta (\alpha,xn) $ cross sections as functions of the $\alpha$ incident kinetic energy. One can see a remarkable agreement between the predictions of our model and the experimental data.  We also notice that the sum of the $(\alpha,xn)$ cross sections nicely exhausts the total reaction cross section.

The splitting of each of the $ (\alpha,xn) $ theoretical cross sections into three components as defined in Fig.~\ref{Bian} is displayed in Fig.~\ref{Taaxn2}. The bump of each cross section is  mainly due to evaporation. For the $(\alpha,n)$ reaction, the decreasing part of the cross section above the bump is entirely due to cascade. There is, of course, no mixed component in this case. Starting from $ (\alpha,2n)$ and going to more and more emitted neutrons, the mixed component grows and become dominant above the bump region. This simply means that when the incident energy starts to correspond to this region, the excitation energy of the compound system becomes sufficient to emit $x+1$ neutrons with a high probability, the $(\alpha, xn)$ cross section is sizably reduced and the underlying reaction mechanism resembles more and more to the standard spallation reaction mechanism, involving a cascade stage, possibly resulting in incomplete fusion, and followed by a de-excitation stage, both stages emitting particles. It is remarkable that both the shape and the magnitude of the various bumps and the trend of  decreasing part of the cross  sections are well described by our model. This means that our model is catching the main features of the compound nucleus formation at low energy and its progressive change into a mechanism involving  so-called pre-equilibrium features.  

\begin{figure}[h]

\includegraphics[width=8cm, height=10cm]{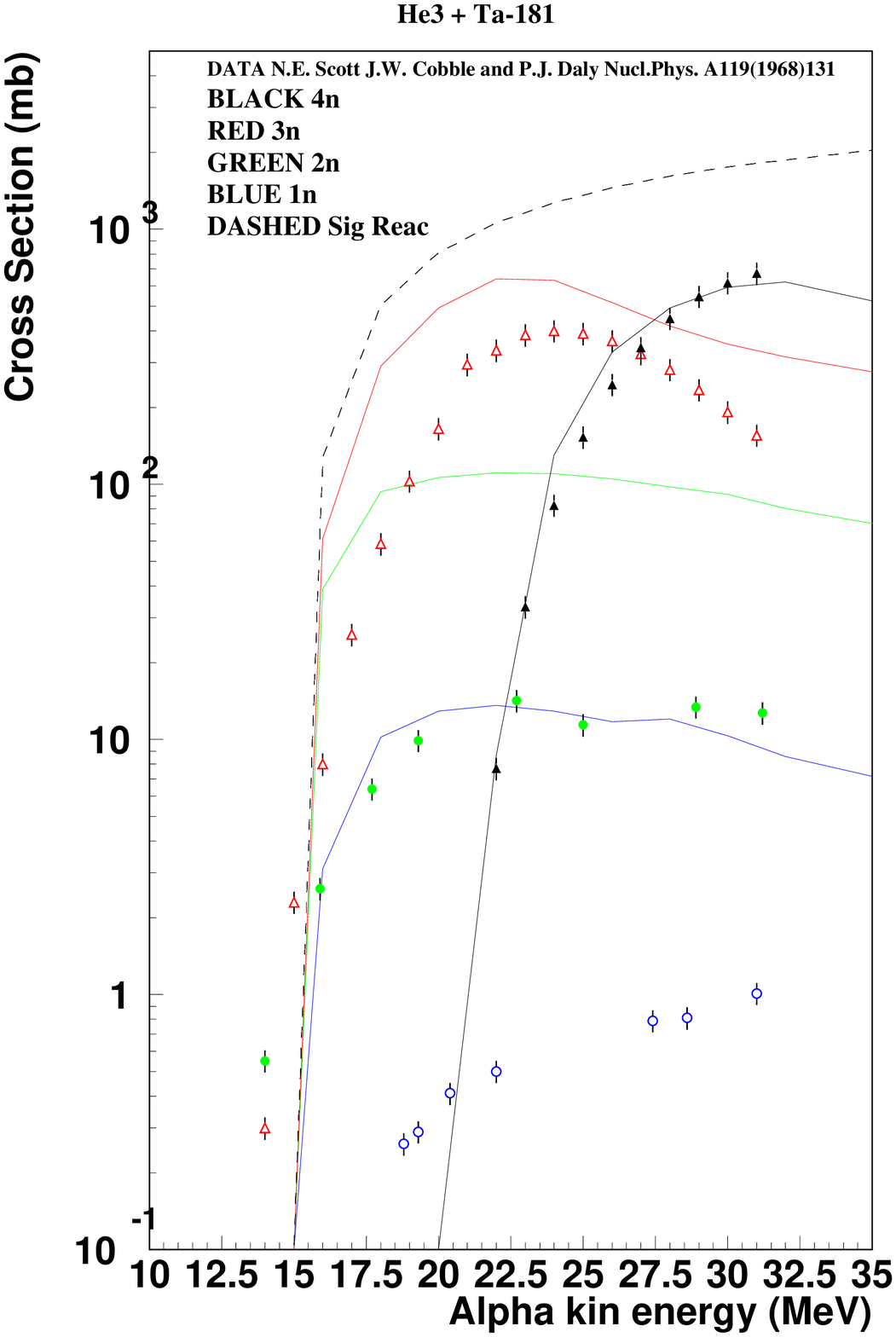}
\caption{\small Same as Fig.~\ref{Taaxn1} for $^{181}Ta (^3He,xn)$ reaction. Data (symbols) are taken from  Ref.~\cite{SC68}. The dashed line gives the theoretical total reaction cross section.}

\label{Tahe3xn1}
\end{figure}

\begin{figure}[h]

\begin{minipage}[t]{4cm}
\includegraphics[width=4cm]{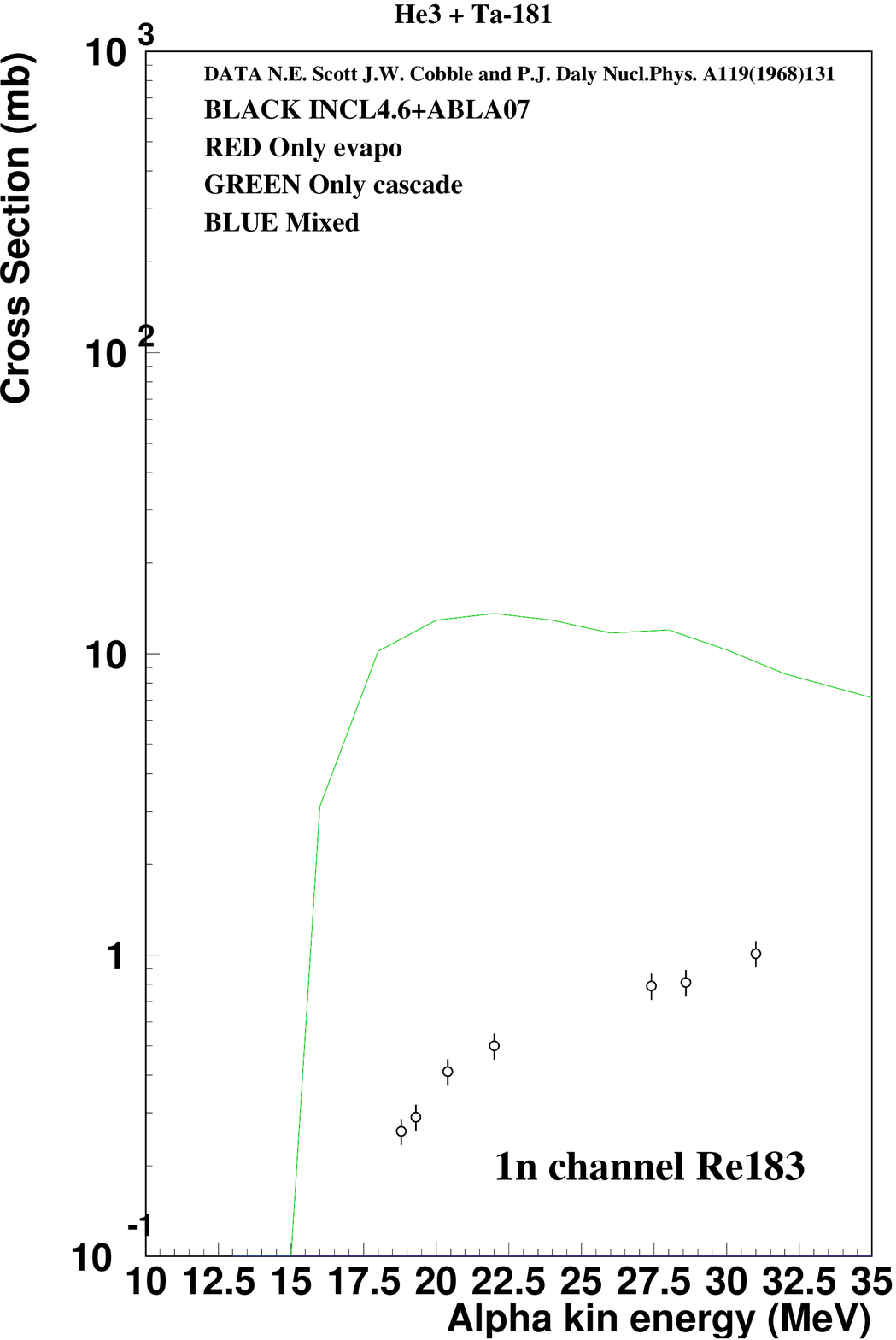}
\renewcommand{\baselinestretch}{0.5}
\end{minipage}
\begin{minipage}[t]{4cm}
\includegraphics[width=4cm]{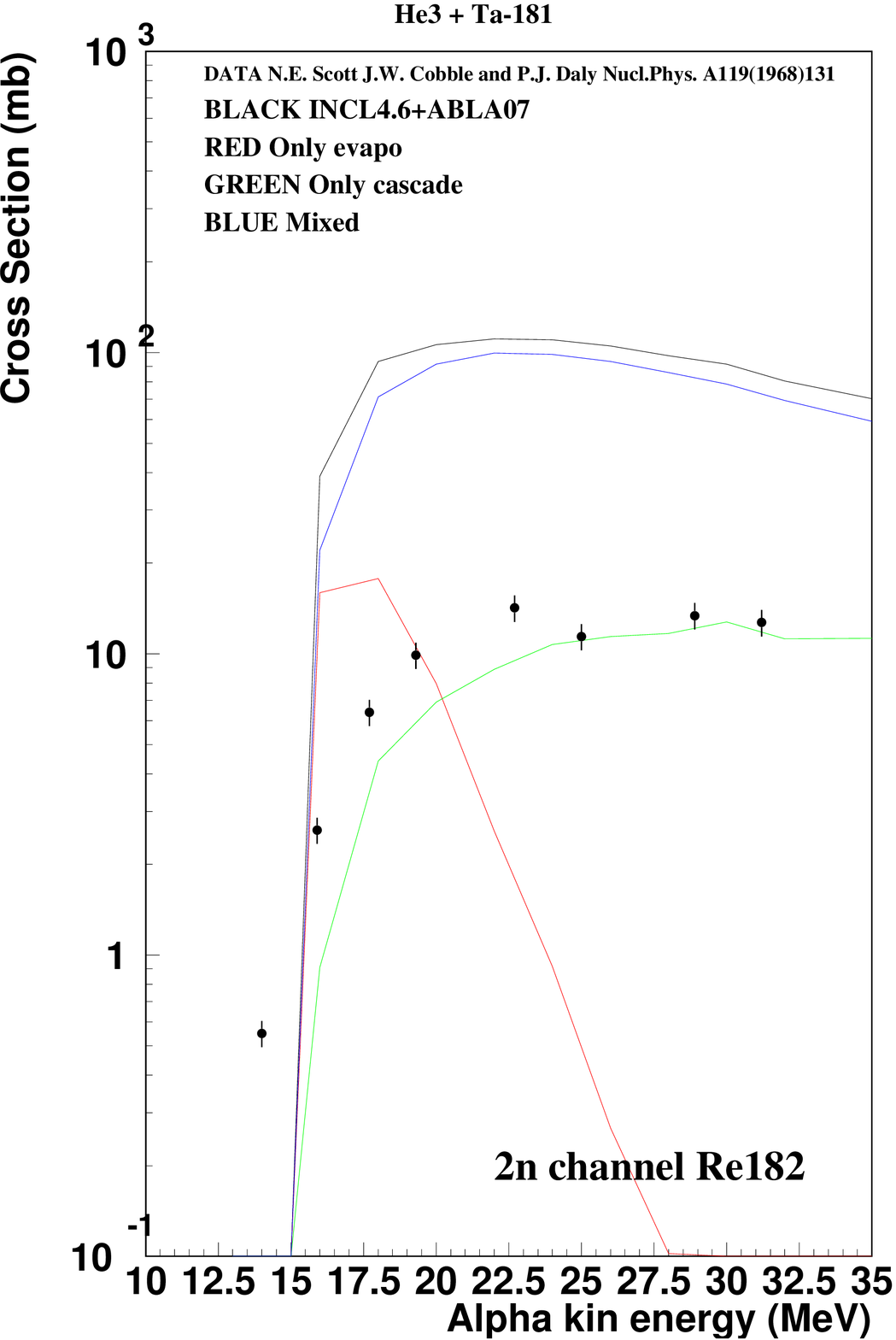}
\renewcommand{\baselinestretch}{0.5}
\end{minipage}
\\
\begin{minipage}[t]{4cm}
\includegraphics[width=4cm]{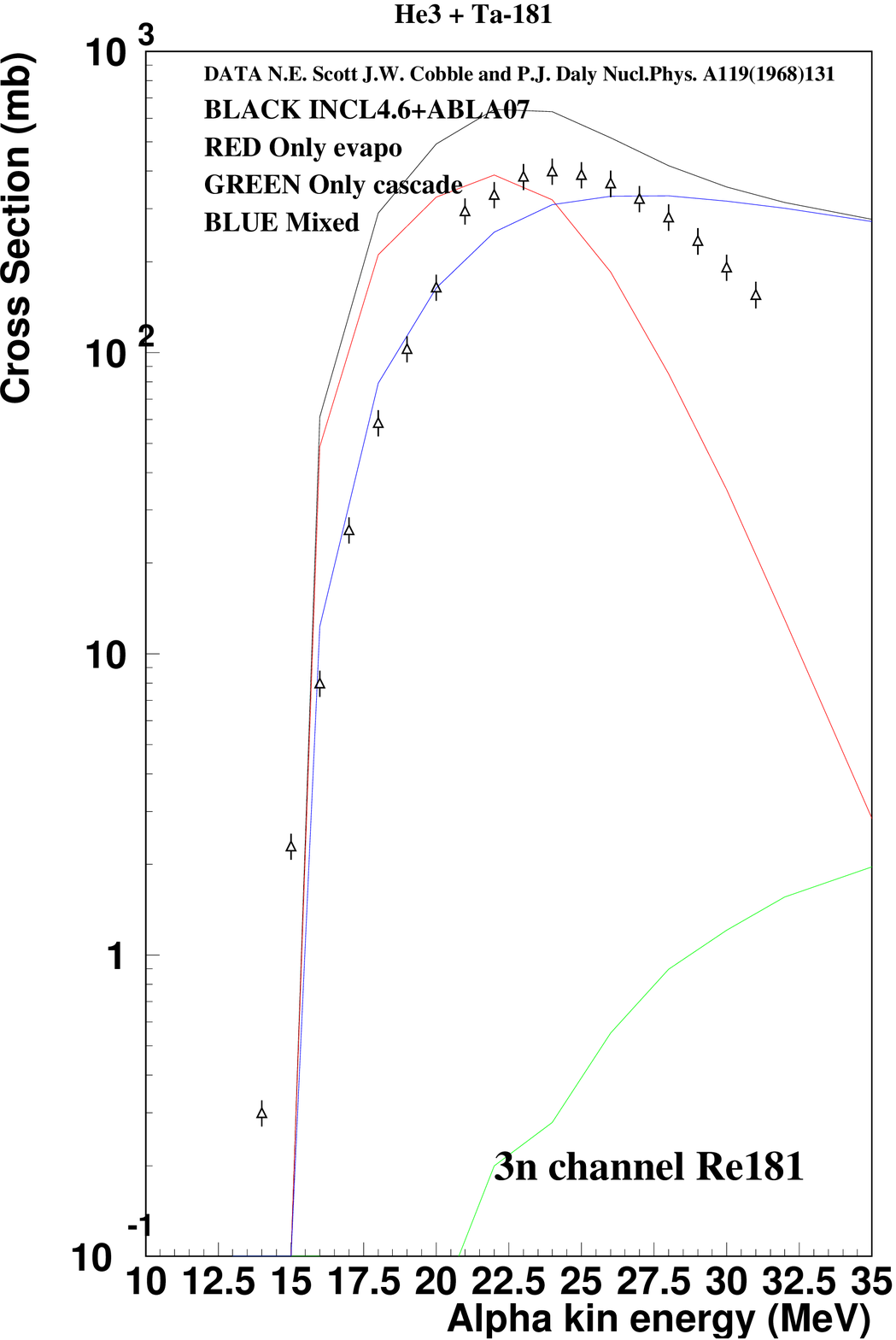}
\renewcommand{\baselinestretch}{0.5}
\end{minipage}
\begin{minipage}[t]{4cm}
\includegraphics[width=4cm]{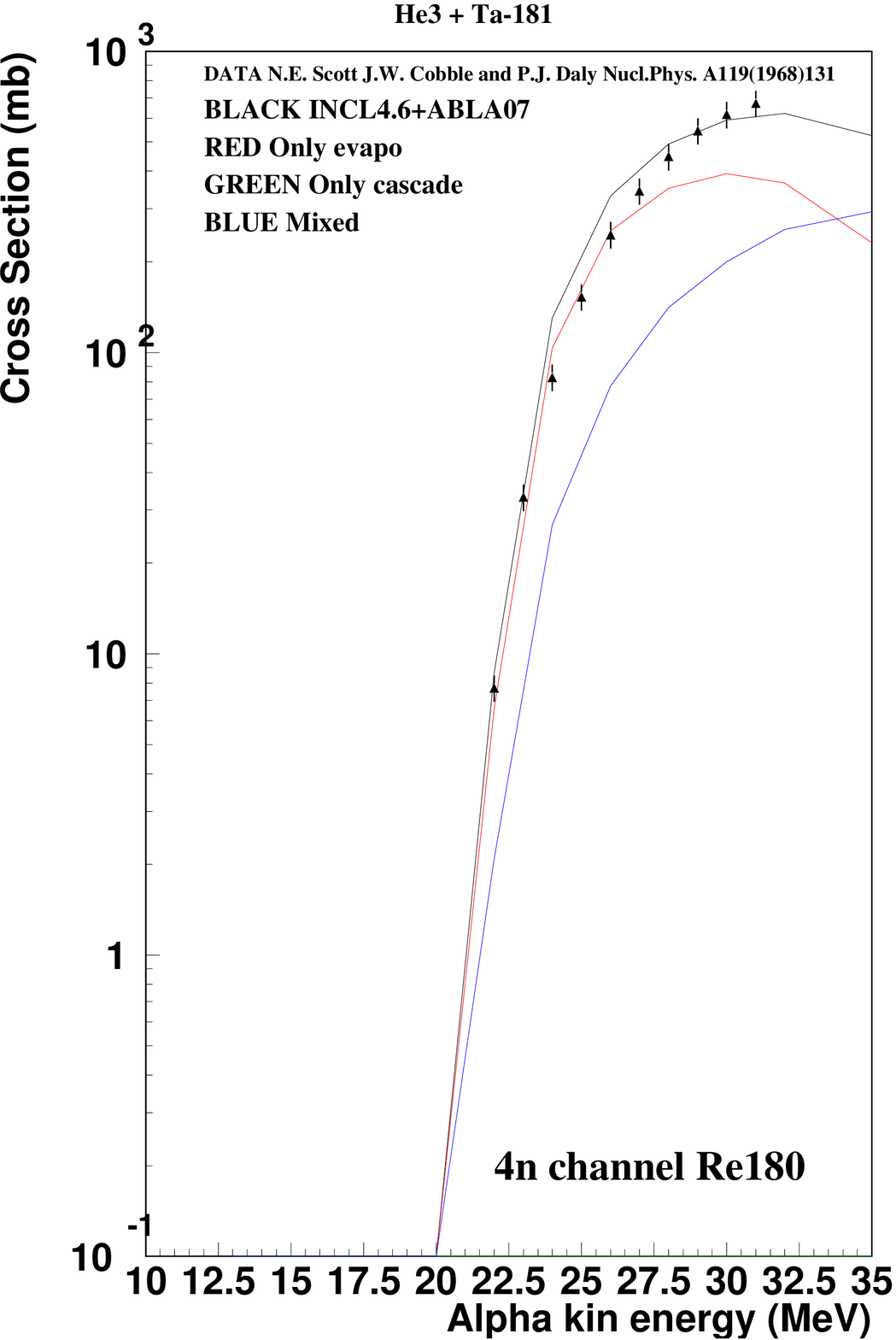}
\renewcommand{\baselinestretch}{0.5}
\end{minipage}

\caption{\small Same as Fig.~\ref{Taaxn2} for   $^{181}Ta (^3He,xn) $ . Data (symbols) are taken from  Refs.~\cite{SC68}. }
\renewcommand{\baselinestretch}{0.5}
\label{Tahe3xn2}
\end{figure}

We show in Figs.~\ref{Tahe3xn1} and \ref{Tahe3xn2} the results of our calculation for $^3He$-induced reactions on $^{181}Ta$. If the trends are correctly accounted for, the agreement is much less satisfactory than for the $^{181}Ta (\alpha,xn)$ reactions. The cross sections for $x$=1 and $x$=2 are substantially overestimated, and  the one for $x=$3 is still too large by a factor 2. 

The  $^{181}Ta (^3He,xn) $ cross sections are split in the various contributions in Fig.~\ref{Tahe3xn2}. Several features have to be noticed: there is no evaporation component for $x=$1, the mixed component is increasing very sharply from the very opening of the cross section for $x=$2 and $x=$3. This component is dominating for $x=$2. 

Before interpreting our results, we want to comment on the shapes of the experimental cross sections. For $(\alpha,xn) $ reactions, both for $^{181}Ta$ and $^{209}Bi$, the cross sections show the typical pattern of raising and decreasing stages, replacing the $(x-1)n$ cross section and then giving place to the $(x+1)n$ cross section. This is due to the positive threshold values (the opposite of the $Q$-value), increasing with $x$. For example, for the $^{209}Bi (\alpha, xn)$ reactions, the threshold values are located at 15.56, 20.62, 28.43 and 35.61 MeV, for $x=$1, 2, 3 and 4, respectively. On the other hand, for the $^{181}Ta (^3He,xn)$ reactions, the threshold values are -10.83, 3.88, 11.16 and 19.88 MeV. At the Coulomb barrier (roughly 15 MeV), the $(^3He,n)$, $(^3He,2n)$ and $(^3He,3n)$ channels are already open. That is why the cross sections, at least for $x=$1 and 2,  do not show a typical bump (as for $^{209}Bi (\alpha, xn)$), but rather a steady increase followed by a plateau. A compound nucleus formed at the Coulomb barrier has already an excitation energy of 25 MeV and has thus few chance to decay to the $1n$ exit channel and even to the $2n$ exit channel. 

We now come back to our numerical results for $^{181}Ta (^3He,xn)$.In accordance with the last remark, in our calculations (see Fig.~\ref{Tahe3xn2}, the $(^3He,xn)$ channel is fed by the cascade only. According to Section~\ref{cir}, the three nucleons of $^3He$ are lying above the Fermi energy, at the beginning of the reaction. Very likely, the neutron is emitted freely or quasi freely (after a soft collision for instance), and the two protons are kept inside the target with a small excitation energy, otherwise a neutron would be evaporated. The $(^3He,2n)$ channel is largely dominated by a mixed emission: presumably, one of the neutrons is emitted (almost) freely, whereas the two protons initiate a cascade process leading to a remnant sufficiently excited to emit a neutron. The $(^3He,3n)$ channel presumably corresponds to a transition toward a sequence of emissions by evaporation. 

It seems to us that the overestimation in our model of the $(^3He,n)$ and  $(^3He,2n)$, and perhaps $(^3He,3n)$ cross sections indicates that the separation between compound nucleus and cascade regimes in our model, explained in Section \ref{cir}, is probaly too crude, somehow underestimating the fusion and overestimating the cascade cross section.  

Another possible interpretation calls for the Coulomb dissociation of the projectile. One can imagine that,  at low energy, below or slightly above the Coulomb barrier,  the $^3He$ is dissociated, say in $d+p$, by the Coulomb field before it really hits the nucleus. If the deuteron escapes, the accompanying proton ``sees'' a Coulomb barrier which is twice as small as the one seen by the $^3He$, but its kinetic energy is three times as small.The proton is thus repelled. This phenomena would shift a part of the  $(^3He,n)$ and  $(^3He,2n)$ cross section to other channels. An argument in favour of this interpretation, quoted in Ref.~\cite{SC68}, comes from the fact that the sum of the experimental $^{181}Ta (^3He,xn)$ cross sections does not exhaust the total reaction cross section. This effect, based on Coulomb dissociation, is expected to be roughly controlled by  a  parameter that can be losely defined as $\eta=(Zr)^2/B$, where $Z$, $r$ and $B$ are the charge, the radius and the binding energy of the cluster, i.e. increasing with the electric polarisability of the projectile and inversely proportional to its ``stability''~\cite{MR87,TR83,CU98,HE05}. On these grounds, one expects that the effects encountered for $^3He$ are largely reduced for $\alpha$-particles, as one may indeed observe above. However,the effect should be even more pronounced for deuterons. But this does not seem to be the case for the example of Fig.~\ref{Bidxn}. Of course, the effect is expectedly reduced for light targets. 

\begin{figure}[h]

\vspace{1cm}
\includegraphics[width=7cm, height=10cm, angle=0]{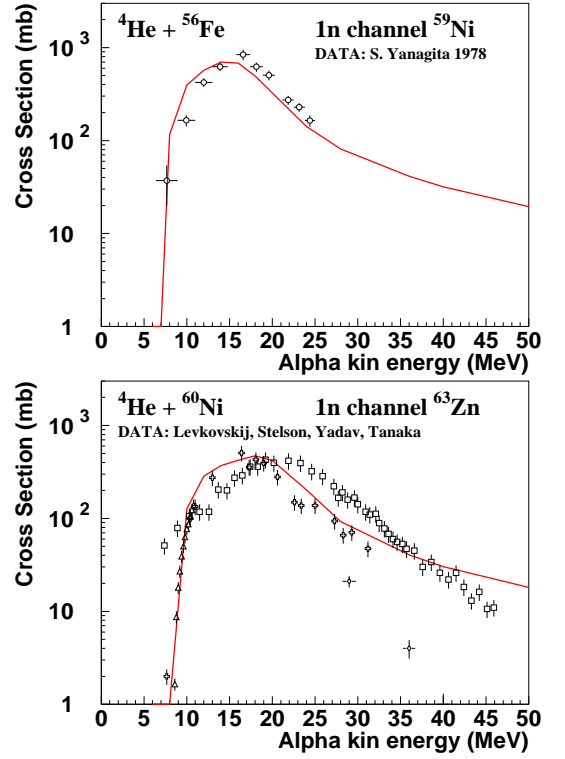}
\caption{\small $^{56}Fe (\alpha,n)$ (upper panel) and $^{60}Ni (\alpha,n)$ (lower panel) cross sections. The full curves correspond to the predictions of the INCL4.6+ABLA07 model and the symbols represent the experimental data.  Data are from Refs.~\cite{YA78,LE91,ST64,YA08a,TA60}.}

\label{FeNian}
\end{figure}

We now turn to illustrative results for light targets. Fig.~\ref{FeNian} shows a comparison of our predictions with experiment concerning $(\alpha,n)$ reactions on $^{56}Fe$ and $^{60}Ni$ targets. The agreement for the $^{56}Fe (\alpha,n)$ cross section is quite good. There are two sets of data points for the reaction $^{60}Ni (\alpha,n)$. Our predictions are close to the data set of Ref.~\cite{ST64}. In addition, one can notice that our model is able to respect the difference between the cross sections for the two targets, that are in fact rather similar to each other. The reason is that our model properly takes account of the different Coulomb penetrabilities and, especially,  of the real $Q$-values. The latter are -5.053 and -7.938 MeV, for $^{56}Fe (\alpha,n)$ and $^{60}Ni (\alpha,n)$ reactions, respectively.

\begin{figure*}[t]

\begin{minipage}[t]{5cm}
\includegraphics[width=5cm, height=7cm]{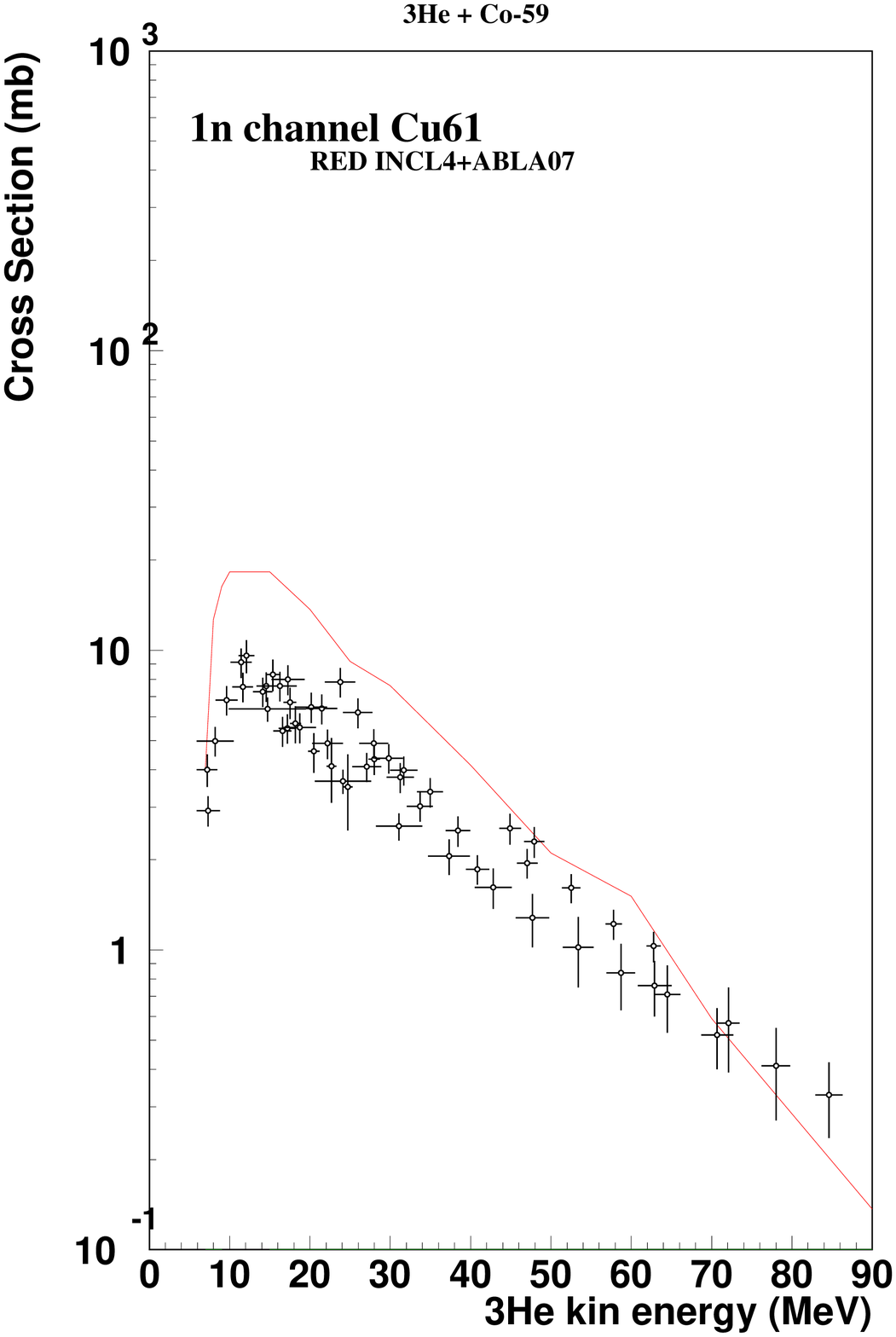}
\renewcommand{\baselinestretch}{0.5}
\end{minipage}
\hspace{1cm}
\begin{minipage}[t]{5cm}
\includegraphics[width=5cm, height=7cm]{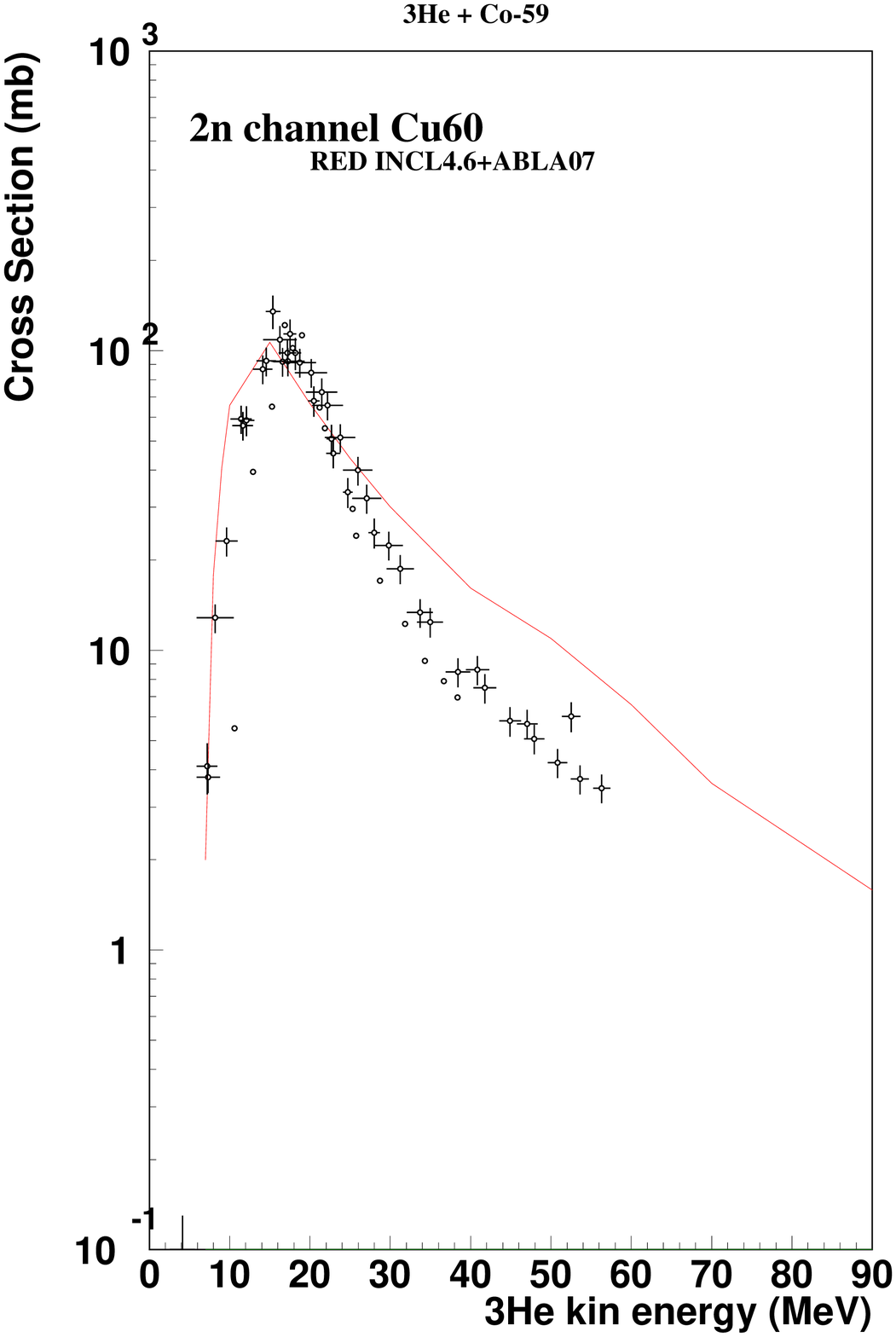}
\renewcommand{\baselinestretch}{0.5}
\end{minipage}
\hfill
\begin{minipage}[t]{5cm}
\includegraphics[height=7cm, width=5cm,  angle=0]{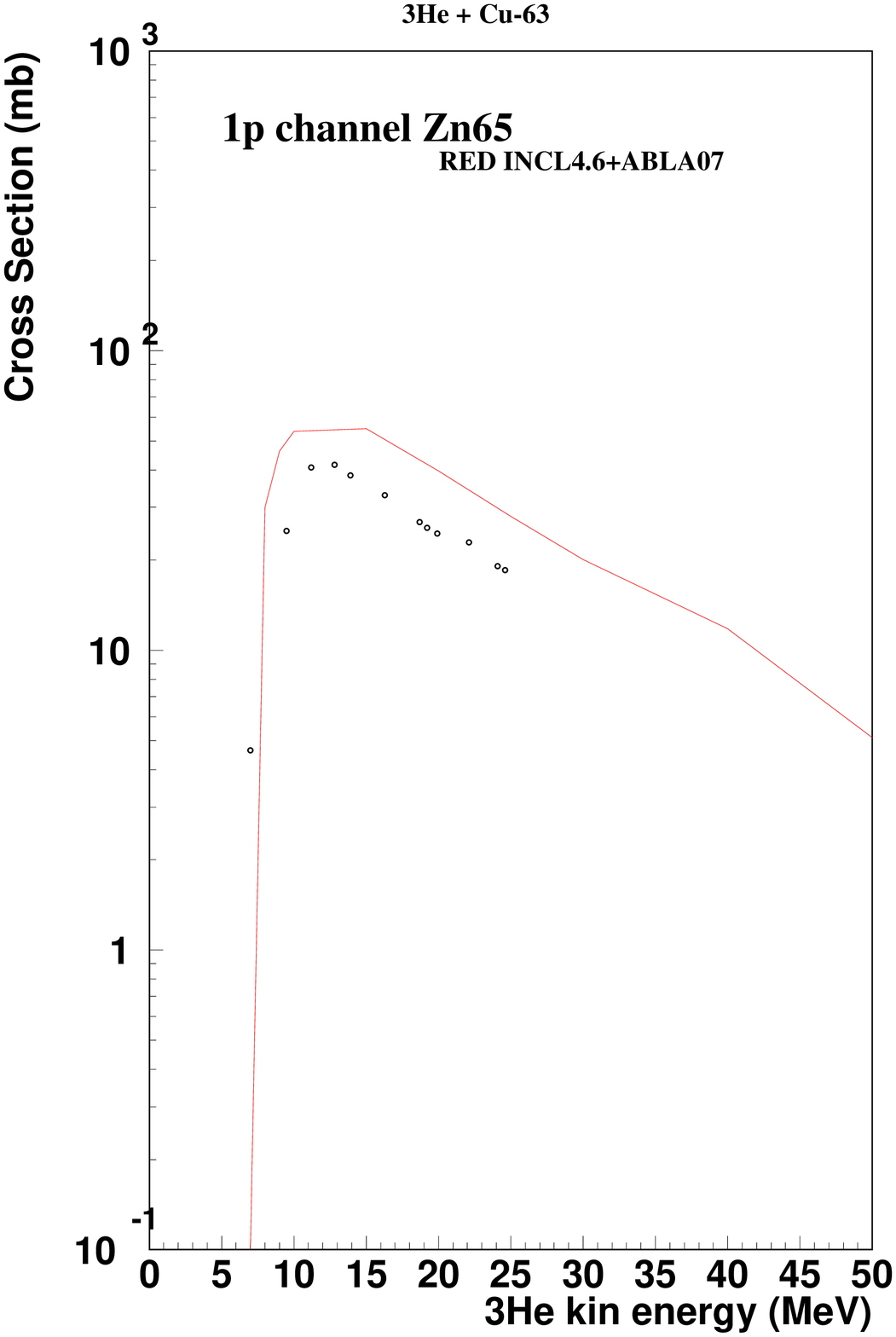}
\renewcommand{\baselinestretch}{0.5}
\end{minipage}

\caption{\small $^{59}Co (^3He,n) $  (left), $^{59}Co (^3He,2n) $ (center), $^{59}Co (^3He,p)$ cross sections, as functions of the $^3He$ incident kinetic energy.  Comparison of the INCL4.6-ABLA07v5  predictions (red lines) with experimental data from Refs.~\cite{FE04,SZ04,BR63}. }
\renewcommand{\baselinestretch}{0.5}
\label{Cohe3xn}
\end{figure*}

Another example is provided by Fig.~\ref{Cohe3xn}. In contrast to the  $^3He + ^{181}Ta$ case, illustrated in Fig.~\ref{Tahe3xn2}, the $(^3He,n)$ and $(^3He,2n)$ are now satisfactorily reproduced. Like for the previous case, these reaction channels are open at the Coulomb barrier, though not ``widely'' open. The thresholds for the two reactions are 3.41 and 5.07 MeV, respectively, and the Coulomb barrier lies around 8 MeV. The fact that we reproduce $(^3He,n)$ and $(^3He,2n)$ cross sections in this case and not in the $^{181}Ta$ case is not inconsistent with the explanation  in terms of Coulomb dissociation of the $^3He$ in this latter case. Indeed, the probability of this dissociation is expected to  be very small for a $^{59}Co$ target. Fig.~\ref{Cohe3xn} also shows an $(^3He,p)$ cross section for $^{63}Cu$ (in view of lacking data for  $^{59}Co$ a slightly different target is considered), for which our calculations reach a reasonable agreement with the data.

\begin{figure}[h]

\includegraphics[width=7cm, height=10cm, angle=0]{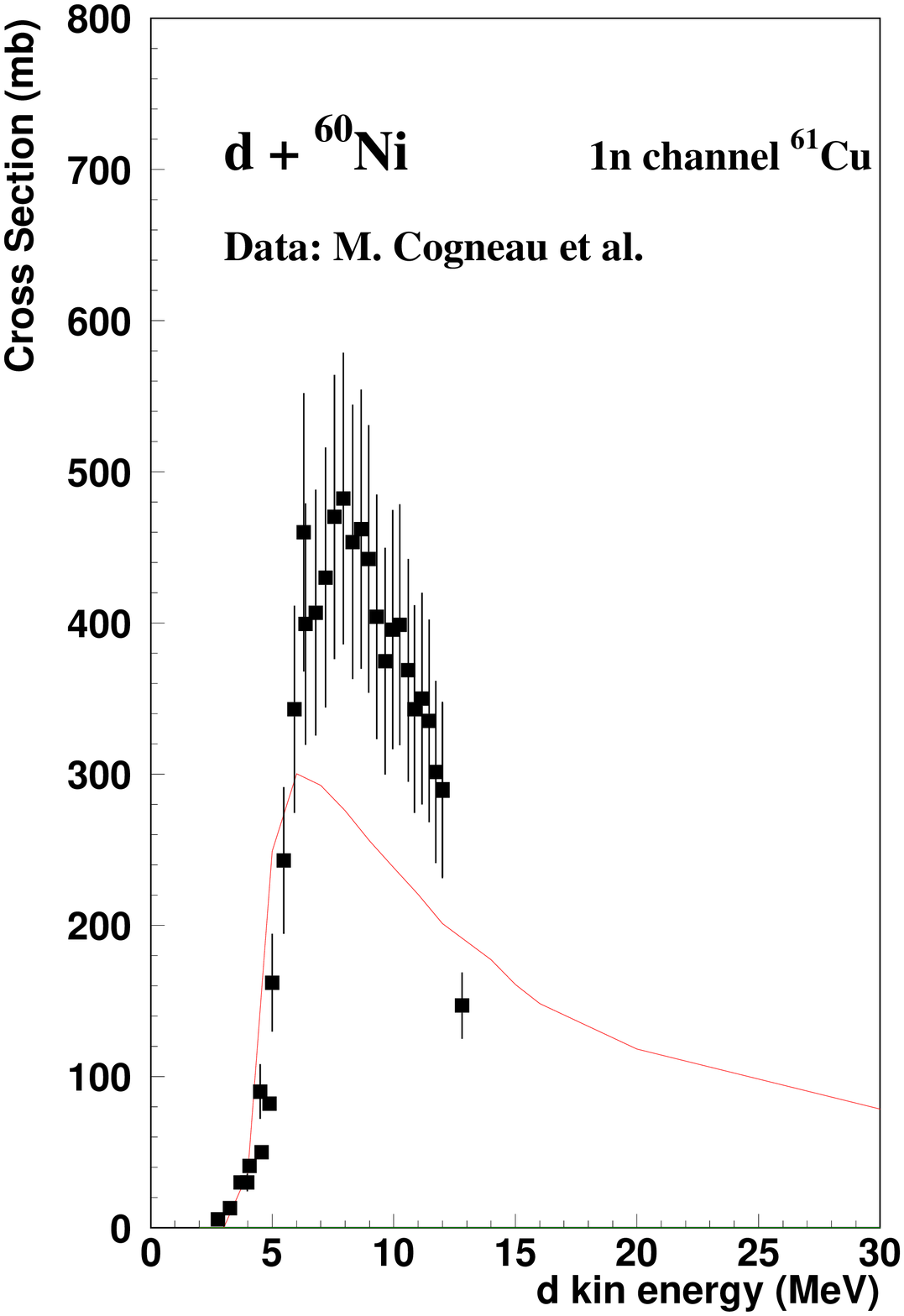}
\caption{\small $^{60}Ni (d,n)$ cross section, as a function of the deuteron incident kinetic energy. Comparison of the predictions of  the INCL4.6+ABLA07 model (red curve) with  the experimental data of Ref.~\cite{CO67}.}

\label{Nidn}
\end{figure}

The last case refers to the $^{60}Ni (d,n)$ reaction, illustrated in Fig.~\ref{Nidn}. Although the opening of the channel is well described in our model, the cross section is noticeably underestimated, in similarity with the  $^{209}Bi$ case (see Fig.~\ref{Bidxn}). This observation may be consistent with the explanation in terms of the Coulomb dissociation. However, this effect should in principle be much smaller $^{60}Ni$, as stated above. 

In summary, these examples and similar others that are not displayed here show that our model yields reasonable results for total reaction cross section and residue production cross sections at low energy. This observation seems to indicate that the model efficiently describes the probability of forming a compound nucleus and the progressive appearence of pre-compound emission, despite the crudeness of its ingredients. The model can even describe threshold behaviour satisfactorily, owing to the use of experimental $Q$-values. 

\section{Discussion. Conclusion}

We have presented here the updated and improved version (INCL4.6) of our standard INCL4.2 model. We recall that the philosophy at the origin of this latter model leads to include as much known microscopic physics as possible, without relying on parameters. Of course, this model relies on assumptions, as reminded in the Introduction.

The extension of INCL4.2 has been realized in two steps, giving birth to the INCL4.5 and INCL4.6 versions, described in Sections~\ref{features4.5} and \ref{features4.6}. The results obtained by INCL4.5 being available owing to the Intercomparison organized by the IAEA~\cite{IAEA}, we have put here the emphasis on the results obtained by INCL4.6. We will not compare either with other INC models, since such a comparison has been done in Ref.~\cite{IAEA}, concerning the INC (or QMD) models described in Refs.~\cite{BE63,BE69,YA79,YA81,MA83,MA00,IQMD,bic,JQMD,BUU,casi}). It shows that our model, coupled to ABLA07, is one of the best combinations for the description of spallation reactions, at least for the set of data proposed in the Intercomparison, basically nucleon-induced spallation reactions on key targets (maainly $Fe$ and $Pb$) for incident energy spanning from $\sim$60 MeV to 3 GeV.  See also Ref.~\cite{IAEA2} for a short analysis of the Intercomparison and Ref.~\cite{TI11} for a similar intercomparison restricted to excitation functions of residue production cross sections. The attention has here been focused  more on observables which are accessible by INCL4.6 only or which have been ignored by the Intercomparison, such as the recoil velocities and heavy cluster production.

We first want to discuss the new features of our model. Part of the extension has been realized in the spirit of INCL4.2, namely by introducing known and well-established phenomenology. This bears on the introduction of energy and isospin-dependent nucleon potentials, of pion potentials, of Coulomb deflection of charged particles and  of experimental Q-values, as discussed in Section~\ref{features4.6}. The impact of these modifications on the results is globally modest. However, they are important for some peculiar features. This holds for $(p,\pi)$ reactions~\cite{AO09} and threshold behaviour of light cluster-induced reactions leading to the emission of a small number of nucleons (see Section~\ref{rpX}). 

A second part of the extension, that we want to single out, consists in the introduction of the dynamical coalescence model for production of clusters during the cascade. It is based on the simple idea that an outgoing nucleon can carry along other nucleons which are close enough in phase space. We insist on the fact that this idea departs from usual coalescence model by two features: it deals with phase space and not the momentum space, and it depends on the instantaneous distribution of the nucleons in the course of the cascade process and not on the final distribution. The importance of this difference is discussed in Ref.~\cite{BO04}. This model is performing rather well with a single proximity parameter (valid for all light clusters) at high energy, say above 500 MeV. The model had to be refined at lower energy, basically on two points: the proximity parameter had to vary, moderately, with the type of the cluster, and tangential emission of clusters had to suppressed.  As far as the results are concerned, these modifications allow a real advance, since emission of light clusters (up to $\alpha$-particles) are well reproduced by our model, down to 60 MeV incident energy. See Ref.~\cite{IAEA} and Section~\ref{elp}. It is showed in Section \ref{ehc} that this model can be extended to emission of heavier clusters. We will comment on this point later on.

Finally, the remaining part of the extension is more founded on recipes, admittedly to remove discrepancies with experimental data, than on solidly established phenomenology or pieces of models. It is accompanied by the introduction of ad hoc parameters or assumptions. The most intrincate aspect of this remaining part of the extension is given by the new procedure for handling incident clusters. In INCL4.2, the incident cluster is considered as a collection of on-shell nucleons, gifted with a Fermi motion, and having a total energy equal to the incident total cluster energy.  This approximation, which is pretty reasonable at high  energy per nucleon of the incident cluster (compared to binding energy per nucleon and/or Fermi energy), is bound to fail at low energy.  Therefore, in our opinion, it is necessary to account for the off-shellness of the nucleons while keeping total momentum and energy to their nominal values. Another necessary feature appears to be the almost unavoidable compound nucleus formation at very low energy, at least for sufficiently small impact parameters. In addition,  incomplete fusion and/or soft excitations of the projectile should be mad epossible at grazing impact parameters. The last ingredient of a satisfactory reaction model should  allow the progressive growth of the pre-equilibrium processes. 

The surprising result of our study is that all these requirements can be met in a model using as basic quantities the 4-momenta of all nucleons and leading to results in reasonable agreement with experiment. Even more, our model, whatever its ad hoc features, possesses the unique property of describing the gradual move from compound nucleus to pre-equilibrium processes with a unique set of assumptions. Usually, two distinct models are used, with sometimes an additional model to ponder the relative importance of compound nucleus and pre-equilibrium. Actually, we have achieved, in some sense, for cluster-induced reactions, the same objective as in our cascade model for nucleon-induced reactions, namely describing absorpion and so-called pre-equilibrium processes with the same cascade tools (propagation and binary collisions of individual nucleons).  It should be stressed, however, that this objective has been realized on a rather empirical basis. 

Let us now turn to a discussion of the predictive power of our new model. We first first on nucleon, pion and residue production cross sections, either differential or global, for nucleon-induced reactions. These observables were already accessible by our previous model INCL4.2. We remind that some of the modifications included in INCL4.6 (even restricted to those which have an influence on  the observables under consideration) may appear to have some ad hoc character, like the treatment of soft collisions or the ``local E'' trick, described in Section \ref{features4.5}. As it is often the case, when ad hoc or empirical modifications are added to an already performing model, some of the results are improved and some others are slightly less good than before. For instance, the cascade part of the neutron spectra are slightly less good with INCL4.6 than with INCL4.2, which gives an excellent reproduction of the data (see Fig.~\ref{neutron}). On the other side, INCL4.6  generally gives more excitation energy than INCL4.2. As a consequence, the residue mass spectra have noticeably been improved (see Section~\ref{resmc}). A considerably important improvement brought by INCL4.6 concerns the prediction of the total reaction cross sections, which has also allowed an improvement of the results at low incident energy, well below 200 MeV, the usually quoted limit of validity of the INC models. 

We now turn to the new  potentialities of our improved model, which, in our opinion, outpace the improvements mentioned above and others discussed in the paper. We will shortly comment on each of them.    
\newcounter{marker2}
\begin{list}{\arabic{marker2}.}{\usecounter{marker2} \setlength{\parsep}{0.3ex}}
\item Our improved model allows the production of clusters, owing to the implementation of a dynamical coalescence model. We have presented here (and in Ref.~\cite{IAEA}) results for double differential production of light clusters, up to $\alpha$ particles, which are in rather good agreement with experimental data, for a wide range of incident energy and target mass. We have in addition showed that the production of heavier clusters with a kinetic energy larger than the typical evaporation energies can also be generated by the same mechanism. In our simulations, which requires the production of all potential clusters, we are limited, by computational time, to A$\leq$8 clusters. We have shown very promising results in Fig.~\ref{hc}, but not all cases show a similar agreement. Nevertheless, we are convinced that our implementation of the coalescence module is flexible and that a more involved search of the better parameters will give eventually an overall satisfying description.     
\item In Section~\ref{rec}, devoted to the recoil of the residues, we have shown perhaps the nicest of our results. Recoil energies are small, but their knowledge is of uttermost importance for technological applications, since they determine the damage to materials under irradiation. We think it is an important result, because the agreement with experiment is impressive, but also because the distribution of the recoil velocity shows the fingerprint of a diffusion process, which is in concordance with the description of the reaction process as a succession of independent binary collisions. 
\item In Section~\ref{CIR}, we have set out our model for cluster-induced reactions. If at high incident energy the cluster can be viewed, in first approximation, as a collection of individual nucleons (with perhaps a procedure to handle spectators), it is crucial to introduce departures from this picture at low energy: off-shellness of the nucleons, geometrical spectators and participants, formation of compound nucleus and progressive appearance of pre-equilibrium processes. We have shown that all these features can be taken care of with individual nucleons and by using criteria involving geometry and the 4-momenta of the nucleons. Even more, the results displayed in Section~\ref{CIR} testify of the success  of this procedure. Of course,  rather global quantities have been considered and the model should be tested also on double-differential cross sections. But the  results accumulated up to now  are very encouraging. We want to draw the attention on the importance of this potentiality of the model for technical applications. In a thick target bombarded by a high energy beam, many secondary reactions will be induced by clusters produced in primary interactions.          
\end{list}
    
We hope to have shown that the improvement of our INCL model, from INCL4.2 to INCL4.6, has generated a powerful tool for the description of  nucleon and light-cluster induced reactions in a large domain of incident energy spanning from a few tens of MeV to ~3 GeV. We stress that this description is based on a single microscopic cascade model, even if some ad hoc prescriptions have been added, as we have discussed in detail. This addition has been made to improve the predictive power of the model in view of applications to thick spallation targets. Actually, the model is already included as an option in the  Geant4~\cite{Geant4}, MCNPX~\cite{MCNPX} and PHITS~\cite{PHITS} transport codes. Nevertheless, we think that the model is still prone to further improvement. We  mentioned in this paper several results which are not sufficiently satisfactory, like for instance the yield of isotopes close to the target, which have large cross sections and whose accurate predictions are  important for applications. We are currently working on such topics.
   
\section{Appendix. Parametrization of $R_{Coul}$} 
The quantity $R_{Coul}$ (in fm), defined in Section~\ref{cir}, is parametrized as
\begin{equation}
R_{Coul}=\frac{1.44 z Z_T}{a A_T^{2/3}-b} - c,
\label{pRCoul} 
\end{equation}
where $z$ is the charge of the incident particle. The values of the quantities $a$, $b$ and $c$ are provided in Table~\ref{abc}.

This is valid for target mass number $A_T \geq 10$. For $A_T < 10$, $R_{Coul}=R_{max}$  (see Eq. \ref{Rmax}). The last equality also holds for incident protons and pions.

\begin{table}[h]
\caption{\it Values of the parameters a, b, c.}
\vspace{4mm}
\renewcommand{\baselinestretch}{1.5}
\begin{tabular}{|c|c|c|c|}
\hline
incident particle & a  & b & c \\
\hline
$d$ &   0.2565 & 0.78 & 2.5 \\
$t$ &  0.2504 & 0.58  & 0.5  \\
$^3He$ & 0.5009 & 1.16 & 0.5  \\
$^4He$ & 0.5939 & 1.64 & 0.5  \\
\hline
\end{tabular}
\renewcommand{\baselinestretch}{0.5}
\label{abc}
\end{table}

\begin{acknowledgments} This work has been partly done in the frame of the EU IP EUROTRANS (European Union Contract No. FI6W-CT-2004-516520) and EU ANDES (FP7-249671) projects. We acknowledge the EU financial support. We are grateful to our ANDES colleagues for interesting discussions and particularly to Drs A. Keli\'{c}-Heil and M. V. Ricciardi for valuable enlightments concerning the use of their evaporation code. 
\end{acknowledgments}


\begin{thebibliography}{99} 
\bibitem{FI10}~D. Filges and F. Goldenbaum, {\it Handbook of Spallation Research
Theory, Experiments and Applications}, Wiley VCH, Berlin, 2010.
\bibitem{GU99}~W. Gudowski, Nucl. Phys. {\bf{A654}}, 436c (1999).
\bibitem{AB10}~H. Abderrahim, P. Baeten, D. De Bruyn, J. Heyse, P. Schuurmans and J. Wagemans, {\it MYRRHA,  a Multipurpose Hybrid Research Reactor
for High-end Applications}, in Nuclear Physics News 20, 24 (2010).
\bibitem{Geant4}~Geant4 collaboration, "Geant4 - a simulation toolkit," Nucl. Instr. Meth. Phys. Res. A506, 250 (2003). 
\bibitem{TA89}~I. Tanihata, On the possible use of secondary radioactive beams,
in {\it Treatise on Heavy-Ion Science}, vol. 8, ed. by D. A. Bromley, Plenum Press, 1989.
\bibitem{DU02}~M. Durante: Radiation protection in space, Riv. Nuovo Cimento 25 (8), 1 (2002). 
\bibitem{LO97}~M. Longair, {\it High Energy Astrophysics}, vol. 1 and 2, Cambridge University Press, 1997.
\bibitem{KR90}~G. Kraft: The radiobiological and physical basis for radiotherapy with protons and heavier ions, Strahlenther. Onkol., 166, 10 (1990). 
\bibitem{HIN}~J.-P. Meulders et al, eds., ``HINDAS Detailed final report'',\\
http://www.theo.phys.ulg.ac.be/wiki/index.php/Cugnon\_Joseph (2005).
\bibitem{EUR}~EUROTRANS/NUDATRA project EU Contract FI6W-CT-2004-516529, http://nuklear-server.ka.fzk.de/eurotrans/
\bibitem{BO02} A. Boudard, J. Cugnon, S. Leray and C. Volant, Phys. Rev. C {\bf{66}}, 044615 (2002). 
\bibitem{GA91}~J.-J. Gaimard and K.-H. Schmidt, Nucl. Phys. {\bf{A531}}, 709 (1991).
\bibitem{JU98}~A. R. Junghans, M. de Jong, H. G. Clerc, A. V. Ignatyuk, G. A. Kudyaev and K.-H. Schmidt, Nucl. Phys. {\bf{A629}}, 635 (1998).
\bibitem{BE98}~J. Benlliure, A. Grewe, M. de Jong, K.-H. Schmidt and S. Zhdanov, Nucl. Phys. {\bf{A628}}, 458 (1998).
\bibitem{BO04}~A. Boudard, J. Cugnon, S. Leray and C. Volant, Nucl. Phys. {\bf{A740}},195 (2004).
\bibitem{RA06}~B. Rapp, J.-C. David, V. Blideanu, D. Dor\'e, D. Ridikas, N. Thiolli\`ere, in: Proceedings of International Workshop on Shielding Aspects of Accelerators, Targets and Irradiation Facilities (SATIF-8), EURISOL DS/Task5/TN-06-04, Pohang, South Korea, May 22-24, 2006.
\bibitem{LE12}~J.-C. David {\it et al.}, to be published.
\bibitem{IAEA} Benchmark of Spallation Models, organized by the IAEA,\\  http://www-nds.iaea.org/spallations/
\bibitem{CU10} J. Cugnon, A. Boudard, S. Leray and D. Mancusi,  Proc. of Int. Topical Meeting on Nuclear Research Applications and Utilization of Accelerators (AccApp09), IAEA, Vienna, 2009, ISBN 978-92-0-150410-4, SM/SR-02, IAEA Publications, Vienna, 8p. (2010). 
\bibitem{IAEA2}~J.-C. David, D. Filges, F. Gallmeier, M. Khankader, A. Konobeyev,S.  Leray, G. Mank, A. Mengoni, R. Michel, N. Otuka and Y. Yariv, Progress in NUCLEAR SCIENCE and TECHNOLOGY, Vol. 2, 942 (2011). 
\bibitem{BO86} W. Botermans and R. Malfliet, Phys. Lett. {\bf{B171}}, 22 (1986).
\bibitem{BO90} W. Botermans and R. Malfliet, Phys. Rep. {\bf{198}}, 115 (1990). 
\bibitem{BU85} V. E. Bunakov and G. V. Matvejev, Z. Physik A {\bf{322}}, 511 (1985). 
\bibitem{WE89}~G. Welke, R. Malfliet, C. Gr\'{e}goire, M. Prakash and E. Suraud, Phys. Rev. C{\bf{40}}, 2611 (1989). 
\bibitem{CU11} J. Cugnon, Proc. of the workshop "30 years of strong interactions", Spa, Belgium, 6-8 April 2011, Few-Body Syst {\bf{53}} 181 (2012). 
\bibitem{CU03} J. Cugnon and P. Henrotte, Eur. Phys. J. A {\bf{16}}, 393 (2003). 
\bibitem{HO94}~P. E. Hodgson, {\it The Nucleon Optical Potential}, World Scientific, Singapore, 1994.
\bibitem{JE76}~J.-P. Jeukenne, A. Lejeune, C. Mahaux, Phys. Rep.  {\bf{25}}, 83 (1976).
\bibitem{JE91}~J.-P. Jeukenne, C. Mahaux, R. Sartor, Phys. Rev. {\bf{43}}, 2211 (1991).
\bibitem{AO04}~Th. Aoust and J. Cugnon, Eur. Phys. J. A {\bf{21}}, 79 (2004). 
\bibitem{AO09}~Th. Aoust and J. Cugnon, Nucl. Phys. {\bf{A828}}, 52 (2009) 
\bibitem{LA62}~A. M. Lane, Phys. Rev. Lett. {\bf{8}}, 171 (1962).
\bibitem{AO06}~Th. Aoust and J. Cugnon, Phys. Rev. C {\bf{74}}, 064607 (2006).
\bibitem{CU11a}~J. Cugnon, A. Boudard, J.-C. David, A.  Keli\'c-Heil, S. Leray, D. Mancusi and M. V. Ricciardi, Proc. of the INPC Conference 2010, Vancouver, July 2010, J. Phys.: Conf. Ser. {\bf{312}}, 082019 (2011).
\bibitem{HE05}~P. Henrotte, PhD thesis, University of Li\`ege, 2005. 
\bibitem{BO08}~A. Boudard, Proc. of the Int. Conf. on Nuclear Data for Science and Technology 2007, ed. by O. Bersillon et al, EDP Sciences, 2008, pp 1103-1108.
\bibitem{YA79}~Y. Yariv and Z. Fraenkel, Phys. Rev. C {\bf{20}}, 2227 (1979).
\bibitem{YA81}~Y. Yariv and Z. Fraenkel, Phys. Rev. C {\bf{24}}, 488 (1981).
\bibitem{YA08}~Y. Yariv, Th. Aoust, A. Boudard, J. Cugnon, J.-C. David, S. Lemaire and S. Leray, Proc. of the Int. Conf. on Nuclear Data for Science and Technology 2007, ed. by O. Bersillon et al, EDP Sciences, 2008, pp. 1125-1128.
\bibitem{BO10}~A. Boudard, J. Cugnon, P. Kaitaniemi, S. Leray and D. Mancusi,
Proc. of Int. Topical Meeting on Nuclear Research Applications and Utilization of Accelerators (AccApp09), IAEA, Vienna, 2010, ISBN 978-92-0-150410-4, contribution AP/IE-08, 8p.
\bibitem{KA10}~P. Kaitaniemi, A. Boudard, S. Leray, J. Cugnon and D. Mancusi, on behalf of the Geant4 	collaboration, Joint International Conference on Supercomputing in Nuclear Applications and Monte Carlo 2010 (SNA + MC2010), Hitotsubashi Memorial Hall, Tokyo, Japan, October 	17-21, 2010, in Progress in NUCLEAR SCIENCE and TECHNOLOGY, Vol. 2, Atomic Energy Society of Japan, pp.788-793 (2011).
\bibitem{GO74}~A. S. Goldhaber, Phys. Lett. B {\bf{53}}, 306 (1974).
\bibitem{FE73}~H. Feshbach and K. Huang, Phys. Lett. B {\bf{47}}, 300 (1973).
\bibitem{HU81}~J. H\"{u}fner  and M. C. Nemes, Phys. Rev.  C {\bf{23}}, 2538 (1981).
\bibitem{AO00}~Y. Aoki, N. Okumura, T. Joh, N. Takahashi and Y. Honkyu, Nucl. Phys. {\bf{A673}}, 189 (2000). 
\bibitem{BE63}~H. W. Bertini, Phys. Rev.  {\bf{131}}, 1801 (1963).
\bibitem{BE69}~H. W. Bertini, Phys. Rev.  {\bf{188}}, 1711 (1969).
\bibitem{MA83}~K. K. Gudima, S. G. Mashnik  and V. D. Toneev, Nucl. Phys. {\bf{A401}}, 329 (1983).
\bibitem{MA00}~S. G. Mashnik and A. J. Sierk, invited talk to the Int. Conf. on Nuclear Data for Science and Technology, Tsukuba (Japan), Oct.7-12, 2001 and Los Alamos preprint LANL Report LA-UR-01-5390.
\bibitem{BUU}~Z. Rudy and A. Kowalczyk, IAEA INDC (NDS)-1530, IAEA Publications, Vienna, Austria, 2008, pp.53-64. 
\bibitem{IQMD}~C. Hartnack {\it et al.},  Eur. Phys. J. A {\bf{1}}, 151 (1998).
\bibitem{JQMD}~K. Niita, S. Chiba, Toshiki Maruyama, Tomoyuki Maruyama, H. Takada, T. Fukahori, Y. Nakahara, A. Iwamoto, Phys. Rev. C {\bf{52}}, 2620 (1995).
\bibitem{FI08}~D. Filges, S. Leray, Y. Yariv, A. Mengoni, A. Stanculescu and G. Mank, IAEA INDC (NDS)-1530, IAEA Publications, Vienna, Austria, 2008.
\bibitem{KE08}~A. Keli\'{c}, M. V. Ricciardi  and K.-H. Schmidt, in ``Joint ICTP-IAEA Advanced Workshop on Model Codes for Spallation Reactions'', ed. by D. Filges {\it et al.}, IAEA INDC (NDS)-1530, IAEA Publications, Vienna, Austria, 2008, pp.181-222.
\bibitem{PR97}~R. E. Prael and M. B. Chadwick, preprint  Los Alamos National 
Laboratory LA-UR-97-1744, 1997.
\bibitem{BA93}~B. C. Barashenkov, {\it Cross-Sections of Interactions of Particles and Nuclei with Nuclei}, JINR publications, Dubna, 1993.
\bibitem{CA96}~R. F. Carlson, Atomic Data and Nuclear Data Tables {\bf{63}}, 93 (1996).
\bibitem{TR01}~R. K. Tripathi, J. W. Wilson and F. A. Cucinotta, Nucl. Instr. Meth. Phys. Res.  B {\bf{173}}, 391 (2001).
\bibitem{LE02}~S. Leray    {\it et al.}, Phys. Rev.  C {\bf{65}}, 044621 (2002).
\bibitem{LE99}~X. Ledoux {\it et al.}, Phys. Rev.  Lett. {\bf{82}}, 4412 (1999).\bibitem{HJ96}~E. L. Hjort  {\it et al.}, Phys. Rev.  C {\bf{53}}, 237 (1996).
\bibitem{GU05}~A. Guertin et al., Eur. Phys. J. A {\bf{23}}, 49 (2005).
\bibitem{BE73}~F. E. Bertrand and R. W. Peelle, Phys. Rev.  C {\bf{8}}, 1045  (1973).
\bibitem{EXFOR}~EXFOR/CSISRS, National Nuclear Data Center, 
http://www.nndc.bnl.gov/exfor/exfor00.htm
\bibitem{BU09}~A. Budzanowski {\it et al.}, Phys.Rev. C {\bf{80}}, 054604 (2009). 
\bibitem{CH80}~R. E. Chrien {\it et al.}, Phys. Rev. C {\bf{21}}, 1014 (1980).
\bibitem{Mc84}~J. A. McGill, G. W. Hoffmann, M. L. Barlett, R. W. Fergerson,
E. C. Milner, R. E. Chrien, R. J. Sutter, T. Kozlowski and R. L. Stearns, Phys. Rev. C {\bf{29}}, 204 (1984).
\bibitem{BU08}~A. Budzanowski {\it et al.}, Phys.Rev. C {\bf{78}}, 024603 (2008).
\bibitem{FR90}~J . Franz {\it et al.}, Nucl. Phys. {\bf{A828}}, 774 (1990).
\bibitem{CO96}~A. A. Cowley et al., Phys. Rev. C {\bf{54}}, 778 (1996). 
\bibitem{BU07}~A. Bubak {\it et al.}, Phys.Rev. C {\bf{76}}, 014618 (2007). 
\bibitem{LET02}~A. Letourneau {\it et al.}, Nucl. Phys. {\bf{A712}}, 133 (2002). 
\bibitem{HE06}~C.-M. Herbach {\it et al.}, Nucl. Phys. {\bf{A765}}, 426 (2006). 
\bibitem{LE10}~S. Leray, A. Boudard, J. Cugnon, J.-C. David, A. Keli\`{c}-Heil, D. Mancusi, M. V. Ricciardi,  Nucl. Instr. and Meth. in Phys. Res. B {\bf{268}}, 581 (2010). 
\bibitem{MA06}~H. Machner {\it et al.}, Phys.Rev. C {\bf{73}}, 044606 (2006). 
\bibitem{DA11}~J.-C. David, A. Boudard, J. Cugnon, S. Leray and D. Mancusi, FP7\_ANDES report/Task 4.1/Deliverable 4.1, 2011,  available at http://www.andes-nd.eu/ 
\bibitem{CO72}~D. R. F. Cochran {\it et al.}, Phys. Rev. D {\bf{6}}, 3085 (1972).
\bibitem{EN01}~T. Enqvist {\it et al.}, Nucl. Phys. {\bf{A686}}, 481 (2001).
\bibitem{KE11}~A. Keli\'{c}-Heil, private communication
\bibitem{VI07}~C. Villagrasa-Canton {\it et al.}, Phys. Rev. C {\bf{75}}, 044603 (2007).
\bibitem{FE05}~B. Fern\'{a}ndez-Dom\'{i}nguez {\it et al.}, Nucl. Phys.{\bf{A747}}, 227 (2005).
\bibitem{AU06}~L. Audouin {\it et al.}, Nucl. Phys.  {\bf{A768}}, 1 (2006).
\bibitem{TA03}~J. Ta\"{i}eb {\it et al.}, Nucl. Phys. {\bf{A724}}, 413 (2003).
\bibitem{BE03}~M. Bernas {\it et al.}, Nucl. Phys. {\bf{A725}}, 213 (2003).
\bibitem{BE06}~M. Bernas {\it et al.}, Nucl. Phys. {\bf{A765}}, 197 (2006).
\bibitem{RI06}~M. V. Ricciardi {\it et al.},  Phys. Rev. C {\bf{73}}, 014607 (2006).
\bibitem{MA11}~D. Mancusi, A. Boudard, J. Cugnon, J.-C. David, T. Gorbinet and S. Leray, Phys. Rev. C {\bf{84}}, 064615 (2011). 
\bibitem{TA08}~Y. Tall et al., Proceedings of the International Conference on Nuclear Data for Science and Technology, April 22-27, 2007, Nice, France, ed. by O. Bersillon, F. Gunsing, E. Bauge, R. Jacqmin, and S. Leray, EDP Sciences, 2008, p1069.
\bibitem{MA06a}~S. G. Mashnik {\it et al.}, J. Phys.: Conf. Ser. {\bf{41}}, 340 (2006).
\bibitem{RI99}~D. Ridikas, W. Mittig and J. A. Tostevin, Phys. Rev. {\bf{C59}}, 1555 (1999).
\bibitem{TO98}~J. A. Tostevin {\it et al.}, Phys. Rev. C {\bf{57}}, 3225 (1998). 
\bibitem{OK94}~H. Okamura {\it et al.}, Phys. Lett. B {\bf{325}}, 308 (1994). 
\bibitem{AV12}~M. Avrigeanu, V. Avrigeanu and A. J. Koning, Phys. Rev. C {\bf{85}}, 034603 (2012). 
\bibitem{SI11}~E. \v{S}ime\v{c}kov\'a {\it et al.}, Phys. Rev. C {\bf{84}}, 014605 (2011).  
\bibitem{BA74}~A. R. Barnett and J. S. Lilley,  Phys. Rev. C {\bf{9}}, 2010 (1974). 2010-2027
\bibitem{HE05a}~A. Hermanne, F. T\'{a}rk\'{a}nyi, S. Tak\'{a}cs, Z. Sz\"{u}cs, Yu. N. Shubin, A. I. Dityuk, Appl. Radiat.  Isot., Vol. 63(1), 1 (2005). 
\bibitem{RI90}~I. A. Rizvi, M. K. Bhardwaj, M. Afzal Ansari, A. K. Chaubey, Appl. Radiat.  Isot., Vol. 41(2), 215 (1990). 
\bibitem{BA74a}~R. Bass, Nucl. Phys. {\bf{A231}}, 45 (1974). 
\bibitem{DI80}~J.-P. Didelez, R. M. Lieder, H. Beuscher, D. R. Haenni, H. Machner, M. M\"{u}ller-Veggian and C. Mayer-B\"{o}ricke, Nucl. Phys. {\bf{A341}}, 421 (1980). 
\bibitem{CH71}~E. T. Chulick and J. B.  Natowitz, Nucl. Phys. {\bf{A173}}, 487 (1971). 
\bibitem{RA59}~W. J. Ramler, J. Wing, D. J. Henderson and J. R. Huizenga, Phys. Rev.  {\bf{114}}, 154 (1959). 
\bibitem{BU63}~A. Budzanowski, L. Freindl, K. Grotowski, M. Rzeszutko, M. Slapa, J. Szmider and P. E. Hodgson, Nucl. Phys. {\bf{A49}}, 144 (1963).
\bibitem{KE49}~E. L. Kelly and E. Segr\`e, Phys. Rev.  {\bf{75}}, 999 (1949). 
\bibitem{SC68}~N. E. Scott, J. W. Cobble and P. J. Daly, Nucl. Phys. {\bf{A119}}, 131 (1968).
\bibitem{MR87}~S. Mr\'{o}wczy\'{n}ski, Phys. Rev. D {\bf{36}}, 1520 (1987).
\bibitem{TR83}~D. Trautmann, G. Baur and F. R\"{o}sel, J. Phys. B {\bf{16}}, 3005 (1983).
\bibitem{CU98}~J. Cugnon and D. Vautherin, Proc. of the Int. Workshop ``Hadronic atoms and positronium in the standard model'', ed. by M. A. Ivanov, A. Arbuzov, E. Kuraev, V. Lyubovitskij and A. Rusetsky, JINR publications, Dubna, 1998, ISBN 5-85165-514-3, pp. 128-134.
\bibitem{YA78}  S. Yanagita, K. Yamakoshi, R. Gensho, Nucl. Phys. {\bf{A303}}, 254 (1978).  
\bibitem{LE91}~V. N. Levkovskij, {\it Activation Cross Sections by Protons and Alphas}, Moscow, 1991. 
\bibitem{ST64}~P. H. Stelson and F. K. McGowan,  Phys. Rev. {\bf{133}}, B911 (1964). 
\bibitem{YA08a}~A. Yadav {\it et al.}, Phys. Rev. C {\bf{78}}, 044606 (2008). 
\bibitem{TA60}~S. Tanaka, J. Phys. Soc. Jpn. {\bf{15}}, 2159 (1960). 
\bibitem{FE04}~A. Fenyvesia, F. T\'{a}rk\'{a}nyia, S.-J. Heselius, Nucl. Instr. Meth. Phys. Res. B {\bf{222}}, 355 (2004). 
\bibitem{SZ04}~F. Szelecs\'{e}nyia, Z. Kov\'{a}csa, K. Suzukib, K. Okadab, T. Fukumurab and K. Muka, Nucl. Instr. Meth. Phys. Res. B {\bf{222}}, 364 (2004). 
\bibitem{BR63}~E. A. Bryant, D. R. F. Cochran, and J. D. Knight, Phys. Rev. {\bf{130}}, 1512 (1963).  
\bibitem{CO67}~M. Cogneau, L. J. Gilly and  J. Cara, Nucl. Phys. {\bf{A99}}, 686 (1967). 
\bibitem{bic}~G. Folger, V. N. Ivanchenko and H. P. Wellisch, Eur. Phys. J. A {\bf{21}}, 407 (2004).  
\bibitem{casi}~H. Kumawat and V. S. Barashenkov, Euro. Phys. J. A  {\bf{26}}, 61 (2005).
\bibitem{TI11}~Yu. E. Titarenko {\it et al.}, Phys. Rev. C {\bf{84}}, 064612 (2011). 
\bibitem{MCNPX}~D. B. Pelowitz {\it et al.}, MCNPX 2.7.B extensions,
Los Alamos Report LA-UR-09-04130 (2009).
\bibitem{PHITS}~K. Niita {\it et al.}, JAEA-Data/Code 2010-022 (2010).

\end{thebibliography}
\end{document}